\documentclass[a4paper,twoside]{article}
\usepackage{tikz}
\usepackage[utf8]{inputenc}
\usepackage[ukrainian,english]{babel}
\usepackage{subcaption} 

\usepackage{eejp}

\begin{document}

\title[Benjamin–-Feir Instability of in a Two-Layer Fluid. Part II]
{Benjamin--Feir Instability of Interfacial Gravity--Capillary Waves in a Two-Layer Fluid. Part II. Surface-Tension Effects}

\author{0000-0002-7960-1436}{Olga Avramenko}{a*,b}
\author{0000-0001-5187-8831}{Volodymyr Naradovyi}{c}
\authors[O. Avramenko, V. Naradovyi]%
{Olga Avramenko, Volodymyr Naradovyi}


\BeginPaper 
\begin{center}
\AuthorPrint 
\affiliation{a}{National University of Kyiv-Mohyla Academy, 2 Skovorody,
        Kyiv, 04070, Ukraine}
\affiliation{b}{Vytautas Magnus University, 58 K.~Donelaičio g.,
        Kaunas 44248, Lithuania}
\affiliation{c}{Volodymyr Vynnychenko Central Ukraine State University, 1 Shevchenka, Kropyvnytskyi, 25006, Ukraine}
\email{o.avramenko@ukma.edu.ua}

\data{Received Month XX, 20XX; revised Month XX, 20XX; accepted December Month XX, 20XX}
\end{center}



\newcommand{\ep}{\varepsilon}
\newcommand{\eps}[1]{{#1}_{\varepsilon}}



\begin{abstract}
This second part of the study develops a complete geometric and asymptotic
description of how surface tension governs the modulational stability of
interfacial waves in a two-layer fluid. Extending the analytical framework of
Part~I, surface tension is treated as a freely adjustable parameter, making it
possible to trace the nonlinear and dispersive properties of the system across
the full range of depth ratios and density contrasts. Using the nonlinear
Schrödinger reduction together with long-wave asymptotics, the mechanisms that
shape the boundaries between stable and unstable regimes are identified and
their dependence on surface tension is quantified.

The long-wave structure is controlled by two special density values that mark
the bases of the loop and the corridor on the stability diagrams. Their
ordering switches at a threshold that exists only when the lower layer is
deeper, and loop-type structures occur only in this regime. A second organising
parameter is the classical Bond threshold, at which the dispersive and
nonlinear singularities coincide. When surface tension exceeds this value and
the upper layer is sufficiently deep, the interaction between resonant and
dispersive effects produces a capillary cut that replaces the corridor and
characterises strongly capillary, upper-layer-dominated configurations.

To unify these observations, the full three-dimensional critical surfaces that
separate different types of nonlinear and dispersive behaviour are computed. The
familiar loop, corridor, and cut appear as planar sections of these surfaces,
and their transitions follow directly from the deformation of the intersection
between the resonant and dispersive sheets. Two depth ratios correspond to
genuine geometric degeneracies: equal layer depths, where the intersection
reduces to a straight line, and the golden-ratio configuration, where the
critical surface becomes horizontally tangent at the Bond threshold.

Overall, Part~II completes the geometric and physical classification of
modulational stability in two-layer interfacial waves and provides a framework
for future extensions incorporating shear, external forcing, flexible
boundaries, or variable bathymetry.

\key{modulational instability; interfacial gravity--capillary waves;
two-layer fluid; surface tension; Benjamin--Feir instability}

\end{abstract}


\section{Introduction}
\label{sec_Intro}

In Part~I~\cite{AvramenkoNarad2025_P1}, the modulational instability of interfacial waves in a two-layer fluid was examined for different depth ratios and density contrasts under the assumption of a fixed or characteristic interfacial tension. The present paper extends that analysis by investigating the effects of varying surface tension $T$ on the stability and topology of modulational–stability domains.

The theoretical foundation of modulational instability originates from the classical works of Benjamin and Feir~\cite{Benjamin1967} and Zakharov~\cite{Zakharov1968}, with extensions to interfacial waves by Grimshaw and Pullin~\cite{Grimshaw85} and by Christodoulides and Dias~\cite{Christodoulides1995}. Dullin, Gottwald, and Holm further developed a unified long-wave shallow-water asymptotic framework linking several equivalent model equations (KdV, fifth-order KdV/Kawahara, and Camassa–Holm), clarifying the role of surface tension through the Bond-number parameter and identifying the parameter regimes where each model applies~\cite{Dullin_Gottwald_Holm_2003,Dullin_Gottwald_Holm_2004}. Building on this foundation, more recent studies have explored the influence of surface tension in both surface and interfacial systems.

Sun and Wahlén~\cite{Sun2025} performed a rigorous spectral analysis of periodic gravity–capillary waves and showed that positive $T$ excludes spectral instabilities at high-frequency crossings, yielding regions of spectral stability.In their analysis based on the Evans function method, Hur and Yang~\cite{Hur2023} examined the spectral properties of gravity–capillary waves, identifying regimes of Wilton‐ripple instability and establishing the precise criteria for their onset. Ward et~al.~\cite{Ward2019} examined the Faraday instability in a system with two interfaces, demonstrating experimentally and theoretically that surface tension governs excitation thresholds.

For interfacial waves in two-layer fluids, several recent studies have derived nonlinear Schrödinger (NLS) reductions and constructed stability diagrams. Li et~al.~\cite{Li_Cao_Song_Yu_Chen_2020} obtained the NLS equation for arbitrary depths and showed that stronger density contrast and thinner layers compress the modulational–instability region, while finite $T$ shifts the instability boundary toward shorter waves. Li et~al.~\cite{Li22} considered a linear shear profile and demonstrated that the combined effects of shear and $T$ alter the sign of the nonlinear coefficient, changing the envelope–stability type. Murashige and Choi~\cite{Murashige2022} performed a two-dimensional stability analysis of finite-amplitude waves and found competition between modulational and Kelvin–Helmholtz instabilities: surface tension suppresses the latter but alters the width of the modulational–instability band. Pal and Dhar~\cite{Pal22,Pal24} produced stability maps for oceanographic parameters and emphasized the sensitivity of instability boundaries to $T$. Boral, Ni, and Korobkin~\cite{Boral2025} studied the interaction between interfacial and flexural–gravity waves in the presence of a discontinuous background current, while Halder et al.~\cite{Halder2025} analyzed the effect of constant vorticity and showed that vorticity and surface tension jointly modify nonlinear corrections and stability.

Other studies have explored situations where capillarity interacts with external influences. Goldobin et~al.~\cite{Goldobin2015} investigated interfacial waves in a two-layer system subjected to horizontal vibrations and showed that surface tension \(T\) influences the onset and stability of long-wave oscillatory modes.
Doak et~al.~\cite{Doak2020} considered gravity–capillary waves at the boundary between dielectric fluids under an external electric field, where surface tension modifies both the dispersion relation and the stability condition. Chow, Chan, and Grimshaw~\cite{Chow2019} analyzed long internal waves in smoothly stratified shallow fluids—although surface tension was absent, their approach is directly transferable to interfacial configurations. Liang, Zareei, and Alam~\cite{Liang2016} demonstrated resonant harmonic generation in internal waves, producing narrow instability windows analogous to those induced by capillarity. Boral, Sahoo, and Stepanyants~\cite{Boral2021} investigated surface waves under wind forcing and identified $T$ as a key control parameter for the transition between stability and instability. Andreeva, Bulavin, and Tkachenko~\cite{Andreeva2020} investigated the Rayleigh--Plateau dissipative instability with viscous effects in both contacting fluids and showed that surface tension, together with dissipation, controls the transition between stable and unstable regimes.

Overall, existing research provides rigorous spectral criteria for gravity–capillary waves at $T>0$, modulational–instability maps for two-layer systems with shear and currents, and analyses of auxiliary factors such as vorticity, vibrations, electric fields, wind forcing, and stratification. However, a consistent topological classification of stability diagrams in the $(\rho,k)$ plane under varying $T$ has not been established. The mechanisms governing the emergence and disappearance of corridors, loops, and cuts, as well as the limiting configurations HS–La, La–HS, and HS–HS (as defined in Part I), remain insufficiently understood. These questions are addressed in the present work.

\section{Problem formulation, preliminary background, and limiting cases}
\label{sec_ProblemLimit}

\subsection{Mathematical formulation and background assumptions}
\label{subsec_2.1}

All variables and scaling follow the notation of Part~I~\cite{AvramenkoNarad2025_P1}.

We consider the same two-layer inviscid, incompressible fluid system as in Part~I, composed of
a lower layer $\Omega_1=\{(x,z)\,|\, -h_1<z<0\}$ and an upper layer
$\Omega_2=\{(x,z)\,|\, 0<z<h_2\}$, separated by the interface $z=\eta(x,t)$ with surface tension $T$.
The fluid densities are $\rho_1$ and $\rho_2$, and the density ratio is $\rho=\rho_2/\rho_1$.
In Part~I, the analysis was carried out for a fixed interfacial tension corresponding to $T=1$.
Here, $T$ is treated as a free control parameter governing the capillary contribution.

Following the nondimensionalization adopted previously, all quantities are scaled with
\( L=(T_0/(\rho_1 g))^{1/2} \), \( t_0=(L/g)^{1/2} \), and \( m_0=\rho_1 L^3 \),
where $T_0$ is the reference interfacial tension.
The dimensionless coefficient $T=T/T_0$ measures the relative magnitude of surface tension effects.

The governing equations for the velocity potentials $\varphi_j(x,z,t)$
and the interface displacement $\eta(x,t)$ are identical to system~(1) of Part~I.
The dependent variables are expanded in powers of the small steepness parameter
$\alpha=a/l$, where $a$ is the maximum interface displacement and $l$ the wavelength:
\[
(\eta,\varphi_j)
 = \sum_{n=1}^{3} \alpha^{n-1} (\eta_n,\varphi_{jn}) + O(\alpha^3),
\qquad
x_n=\alpha^n x, \quad t_n=\alpha^n t.
\]

From the first-order approximation, one obtains the dispersion relation
\begin{equation}
\omega^2 =
\frac{k\,(1-\rho+Tk^2)}
{\coth k h_1 + \rho\,\coth k h_2},
\label{eq:dispersion}
\end{equation}
in which $T$ explicitly controls the curvature of $\omega(k)$.

At third order, the solvability condition gives the nonlinear Schrödinger equation
\begin{equation}
iA_t + i\omega' A_x + \tfrac12\,\omega'' A_{xx}
 = -\alpha\,\omega^{-1} J\,|A|^2 A,
\label{eq:NLS}
\end{equation}
where $\omega'=\partial\omega/\partial k$ and
$\omega''=\partial^2\omega/\partial k^2$ are the first and second derivatives
of the carrier frequency.
The coefficient \(J\) is the Benjamin–Feir index,
which depends on the system geometry and on the second-harmonic correction
\(\Lambda\) (defined in Eq.~(3) of Part~I):
\begin{align}
J &= -\frac{1}{16(1-\rho)\,[\,\rho\,\coth k h_2 + \coth k h_1\,]}
  \bigg\{
  2k\omega^2(1-\rho)\Lambda
  [ -3\rho\,\coth^2 k h_2 + 3\,\coth^2 k h_1 - 1 + \rho ] \nonumber\\
&\quad\quad\quad
 - 4k\omega^4
   [ \rho(\coth^2 k h_2 - 1) - (\coth^2 k h_1 - 1) ]^2 \nonumber\\
&\quad\quad\quad
 - 4k^2\omega^2(1-\rho)
   [ \rho\,\coth^3 k h_2 + \coth^3 k h_1 - 2\rho\,\coth k h_2 - 2\,\coth k h_1 ] - 3T k^5(1-\rho)
  \bigg\}.
\label{eq:J}
\end{align}

A purely temporal solution of Eq.~\eqref{eq:NLS},
\(
A(t) = a\,\exp\!\left(i\alpha a^2\omega^{-1}J t\right),
\)
with constant envelope amplitude \(a\),
is modulationally stable when
\begin{equation}
J\omega'' < 0,
\label{eq:stability}
\end{equation}
while \(J\omega'' > 0\) signals the onset of instability.
In what follows, we examine how variations of $T$
alter the loci $J=0$ and $\omega''=0$ in the $(\rho,k)$ plane,
revealing new transitions between stable and unstable regimes
as surface tension modifies both the nonlinear and dispersive
characteristics of the interfacial mode.

\subsection{Limiting case \(\rho=1\) for equal layer depths}
\label{subsec_2.2}

We now consider the symmetric configuration of two layers with equal thickness
\(h_{1}=h_{2}=h\) and equal densities \(\rho=1\).
In this limit the interface vanishes, and the system formally reduces to a single
homogeneous fluid bounded by two free surfaces of equal curvature.
Nevertheless, this configuration provides a convenient reference for analysing the
nonlinear coefficient \(J\) and the structure of the Benjamin–Feir index near the symmetry line.

For \(\rho=1\) and \(h_{1}=h_{2}=h\),
substituting the dispersion relation~\eqref{eq:dispersion},
\[
    \omega^{2}
    =
    \frac{T k^{3}}{2\,\coth kh},
\]
into the general expression~\eqref{eq:J} for \(J\),
one obtains, after simplification,
\[
    J(h,T,1,k)
    =-\frac{1}{32}\,T\,k^{5}
    \Bigl(\coth kh-\frac{5}{\cosh kh\,\sinh kh}\Bigr).
\]
The factor \(T\) appears only as a multiplier and therefore does not affect
the location of the zero of \(J\).
The condition \(J=0\) yields \(\cosh^{2}(kh)=5\),
and hence the characteristic wavenumber at which \(J\) changes sign is
\begin{equation}
    k_{\mathrm{ch}}(h)
    =\frac{1}{2h}\ln(9+4\sqrt{5})
    \approx \frac{1.4436}{h}.
    \label{eq:kstar}
\end{equation}

The independence of \(k_{\mathrm{ch}}\) from the surface-tension coefficient \(T\)
is a notable feature of this symmetric configuration.
While \(T\) scales the dispersion relation, \(\omega \propto \sqrt{T}\),
so that the frequency increases as the square root of the surface tension,
it does not modify the nonlinear geometry encoded in \(J\).
Consequently, the transition between focusing and defocusing nonlinearity
in the Benjamin–Feir sense depends solely on the dimensionless depth parameter \(kh\).
For \(k<k_{\mathrm{ch}}(h)\) the index \(J>0\), corresponding to one type of nonlinearity,
whereas for \(k>k_{\mathrm{ch}}(h)\) the sign reverses (\(J<0\)).
Because \(\omega''(k)\) obeys the same scaling invariance
(\(\omega(k)\propto\sqrt{T}\), and thus \(\omega''(k)\) scales identically and retains its sign),
the overall modulational-stability criterion based on the product \(J\,\omega''\)
preserves the same threshold~\eqref{eq:kstar} for all \(T>0\).

The analytical symmetry of this configuration implies that both
\(\omega(k)\) and \(J(k)\) are even functions of \(k\).
Consequently, near \(k=0\) the second derivative \(\omega''(k)\)
vanishes linearly with \(k\) for unequal densities (\(\rho\neq1\)),
while in the symmetric limit \(\rho=1\) it tends to a finite constant.
In either case, \(J\) remains regular at \(k=0\).
In the long-wave limit (\(k\to0\)),
the singular conditions \(J\to\infty\) and \(\omega''=0\)
approach each other and asymptotically coincide for all admissible
density ratios \(\rho\neq1\).
For the perfectly symmetric case \(\rho=1\),
both functions remain regular,
and this degeneracy reduces to a simple linear intersection
in the \((\rho,k)\) plane observed for \(h_1=h_2\)
in the three-dimensional representations
discussed in Sec.~\ref{sec_3D}.

To the best of our knowledge, the \(T\)-independence of the characteristic
wavenumber \(k_{\mathrm{ch}}\) in the symmetric limit
\(\rho=1\), \(h_{1}=h_{2}\) has not been explicitly noted in the literature.
This follows directly from the capillary scaling of the dispersion relation
(\(\omega\propto\sqrt{T}\)) and the corresponding invariance of \(\omega''(k)\).
For a comprehensive discussion of modulational instability in
gravity–capillary systems, see Dias and Kharif~\cite{Dias_Kharif_1999}.

\subsection{Mutual placement of the points \(\rho_{\mathrm L}\) and \(\rho_{\mathrm C}\) at \(k=0\)}
\label{subsec_2.3}

In the long-wave limit \(k\to0\), the intersection points of the conditions
\(J=0\) and \(J\to\infty\) with the density axis \(O\rho\) are
\begin{align}
\rho_{\mathrm L} &= \frac{h_2^2}{h_1^2},
\label{eq:rho_L}\\[3pt]
\rho_{\mathrm C}(T) &=
\frac{-h_{2}h_{1}^{2}+h_{1}h_{2}^{2}-3T h_{1}
+\sqrt{h_{1}^{4}h_{2}^{2}+2h_{1}^{3}h_{2}^{3}+h_{1}^{2}h_{2}^{4}
+6T h_{1}^{3}h_{2}-6T h_{1}^{2}h_{2}^{2}-12T h_{1}h_{2}^{3}
+9T^{2}h_{1}^{2}}}{2h_{1}h_{2}^{2}}.
\label{eq:rho_C}
\end{align}

Geometry fixes the reference point \(\rho_{\mathrm L}=h_2^2/h_1^2\),
while capillarity shifts \(\rho_{\mathrm C}(T)\) monotonically leftward from
\(\rho=1\) for small \(T\) and toward the geometric limit for large \(T\).
The two curves intersect at a finite surface-tension value \(T_{\times}\)
defined by \(\rho_{\mathrm C}(T_{\times})=\rho_{\mathrm L}\).
Solving this relation gives
\[
T_{\times}
=\frac{h_{1}^{4}-h_{1}^{3}h_{2}+h_{1}h_{2}^{3}-h_{2}^{4}}{3h_{1}^{2}}
=\frac{(h_{1}^{2}-h_{2}^{2})(h_{1}^{2}+h_{2}^{2}-h_{1}h_{2})}{3h_{1}^{2}},
\]
whose sign determines the admissible configurations:
\[
\operatorname{sign}(T_{\times})=
\begin{cases}
>0, & h_1>h_2,\\
=0, & h_1=h_2,\\
<0, & h_1<h_2,
\end{cases}
\qquad\text{so that }T_{\times}>0\text{ exists only if }h_1>h_2.
\]

For physically relevant \(T>0\), three cases arise:
\begin{enumerate}
\item[\textbf{(i)}] \(h_{1}>h_{2}\):\
  \(T_{\times}>0\), and
  \[
  T<T_{\times}:\ \rho_{\mathrm L}<\rho_{\mathrm C}(T),\qquad
  T=T_{\times}:\ \rho_{\mathrm L}=\rho_{\mathrm C}(T_{\times}),\qquad
  T>T_{\times}:\ \rho_{\mathrm C}(T)<\rho_{\mathrm L}.
  \]
\item[\textbf{(ii)}] \(h_{1}=h_{2}=h\):\
  \(\rho_{\mathrm L}=1\) and
  \[
  \rho_{\mathrm C}(T)=\frac{-3T+\lvert 3T-2h^{2}\rvert}{2h^{2}}
  =
  \begin{cases}
  1-\dfrac{3T}{h^{2}}, & 0\le T<\dfrac{2h^{2}}{3},\\[6pt]
  -1, & T\ge\dfrac{2h^{2}}{3},
  \end{cases}
  \]
  hence \(\rho_{\mathrm C}(T)\le 1=\rho_{\mathrm L}\) for all \(T>0\).
\item[\textbf{(iii)}] \(h_{1}<h_{2}\):\
  \(T_{\times}<0\), so for admissible \(T>0\) the equality
  \(\rho_{\mathrm C}(T)=\rho_{\mathrm L}\) cannot be reached.
  Since \(\mbox{$\rho_{\mathrm L}=h_2^2/h_1^2>1$}\)
 while \(\rho_{\mathrm C}(T)\le1\) for small \(T\),
  the inequality \(\rho_{\mathrm C}(T)<\rho_{\mathrm L}\) holds for all \(T>0\).
\end{enumerate}

For small capillarity (\(T\to 0\)), a Taylor expansion of \(\rho_{\mathrm C}\)
yields
\[
\rho_{\mathrm C}(T)=1-\frac{3T}{h_1h_2}+O(T^{2}),
\]
so the curve \(\rho_{\mathrm C}(T)\) departs from unity with slope
\(-3/(h_1h_2)\).
Hence, for \(h_1>h_2\) the crossing \(\rho_{\mathrm C}(T)=\rho_{\mathrm L}<1\)
occurs at a finite \(T_{\times}>0\), whereas for \(h_1\le h_2\)
the ordering \(\rho_{\mathrm C}(T)\le1\le\rho_{\mathrm L}\)
is maintained even for arbitrarily small \(T>0\).

In the opposite, capillarity-dominated regime (\(T\to\infty\)),
the leading terms in formula~\eqref{eq:rho_C} cancel,
yielding the simple asymptotic limit
\[
\rho_{\mathrm C}(T)
=-\frac{h_{2}}{h_{1}}
+O\!\left(\frac{1}{T}\right),
\]
so that
\[
\lim_{T\to\infty}\rho_{\mathrm C}(T)=-\frac{h_2}{h_1}<0,
\qquad
\rho_{\mathrm L}=\frac{h_2^{2}}{h_1^{2}}
\ \text{is independent of }T.
\]

Thus, for \(h_2\ge h_1\) one always has
\(\rho_{\mathrm C}(T)\le1\le\rho_{\mathrm L}\) on the entire range \(T\ge0\);
for \(h_2<h_1\), the inequality eventually reverses once \(T>T_{\times}\).

The analytical relations derived above show that the relative placement
of \(\rho_{\mathrm L}\) and \(\rho_{\mathrm C}(T)\) at \(k=0\) depends on
both geometry and surface tension:
a positive threshold \(T_{\times}\) satisfying
\(\rho_{\mathrm C}(T_{\times})=\rho_{\mathrm L}\) exists only when
\(h_1>h_2\).
This behaviour underlies the long-wave topology of neutral-stability
boundaries first noted by Grimshaw and Pullin~\cite{Grimshaw85},
where the neutral curves separating focusing and defocusing regimes
form closed contours in the \((\rho,k)\) plane.

\subsection{Long-wave critical surface tension \(T^{\ast}\) (Bond threshold \(\mathrm{Bo}=1/3\))}
\label{subsec_2.4}

To determine the long-wave threshold and introduce convenient notation, we define the Bond number
\(\mathrm{Bo}=T/h_{1}^{2}\) following the standard convention
\cite{Bond_1927,Dias_Kharif_1999}.
The corresponding critical surface tension is
\begin{equation}
T^{\ast}=\frac{h_{1}^{2}}{3},
\label{eq:Tast}
\end{equation}
which is equivalent to \(\mathrm{Bo}^{\ast}=1/3\)
\cite{Dullin_Gottwald_Holm_2003,Dullin_Gottwald_Holm_2004}.
This threshold separates the regimes of normal and anomalous dispersion in the long-wave limit:
for \(\mathrm{Bo}<\mathrm{Bo}^{\ast}\) (\(T<T^{\ast}\)) the dispersion is normal, whereas for
\(\mathrm{Bo}>\mathrm{Bo}^{\ast}\) (\(T>T^{\ast}\)) it becomes anomalous.

From the dispersion relation~\eqref{eq:dispersion}, in the limit of a light upper layer (\(\rho\to0\)),
\[
\omega^{2}
=\frac{(1+Tk^{2})\,k}{\coth(kh_{1})}
=(1+Tk^{2})\,k\,\tanh(kh_{1})
= h_{1}k^{2}\!\left[1+\!\left(\mathrm{Bo}-\tfrac{1}{3}\right)\!(kh_{1})^{2}
+O\!\left((kh_{1})^{4}\right)\!\right].
\]
Hence,
\[
\omega(k)=\sqrt{h_{1}}\,k\!\left[1+\tfrac12\!\left(T-\tfrac{h_{1}^{2}}{3}\right)\!k^{2}+O(k^{4})\right],
\qquad
\omega''(k)=3\sqrt{h_{1}}\!\left(T-\tfrac{h_{1}^{2}}{3}\right)k+O(k^{3}).
\]
At \(T=T^{\ast}\) (\(\mathrm{Bo}=\mathrm{Bo}^{\ast}\)), the curvature \(\omega''(k)\) vanishes, and the dispersion relation becomes
degenerate. This is the classical long-wave, or shallow-water, threshold where the dispersive coefficient
changes sign \cite{Dullin_Gottwald_Holm_2003,Dullin_Gottwald_Holm_2004}.

The link between this linear threshold and the nonlinear modulation follows from the asymptotics of the
Benjamin–Feir index~\eqref{eq:J}.
In the double limit \(k\to0\) and \(\rho\to0\) (long waves on a light upper layer),
\[
\lim_{\rho\to0}\lim_{k\to0} J
= \frac{9}{16\,h_{1}\,(-h_{1}^{2}+3T)},
\]
so \(J\) diverges precisely at \(T=T^{\ast}\),
corresponding to \(\mathrm{Bo}=\mathrm{Bo}^{\ast}\), in agreement with
the long-wave shallow-water theory
\cite{Dullin_Gottwald_Holm_2003,Dullin_Gottwald_Holm_2004}.
At this value, the curves \(J\to\infty\) and \(\omega''=0\)
intersect at the origin of the \((\rho,k)\) plane:
for \(T<T^{\ast}\), the \(J\to\infty\) branch is locally vertical,
whereas for \(T>T^{\ast}\) it becomes nearly horizontal.
As \(T\) increases further, the curves \(J\to\infty\) and \(\omega''=0\)
approach one another and nearly coincide,
forming the degenerate ``cut'' structure observed
in the modulational-stability diagrams.

\subsection{Geometric condition for horizontal tangency of the critical surface}
\label{subsec_2.5}

As shown in Subsec.~\ref{subsec_2.4}, the long-wave Bond threshold
\(
T^{\ast}=h_1^2/3,\qquad \mathrm{Bo}^{\ast}=1/3,
\)
marks the point where the singular conditions
\(J\to\infty\) and \(\omega''=0\) coincide at
\((\rho,k)=(0,0)\).
The geometry of the critical surface \(T(\rho,k)\) near this point
depends on the depth ratio of the layers.

Expanding the dispersion relation~\eqref{eq:dispersion} for small \(k\) yields
\[
\omega^2 =
c_2(\rho)\,k^2 + c_4(\rho;T)\,k^4 + \mathcal{O}(k^6),
\]
with
\begin{align*}
c_2(\rho)&=\frac{h_1 h_2(1-\rho)}{h_1\rho+h_2},\\[3pt]
c_4(\rho;T)&=\frac{h_1 h_2\!\left[3T(h_1\rho+h_2)
+h_1^2 h_2(\rho-1)
+h_1 h_2^2(\rho^2-\rho)\right]}%
{3(h_1\rho+h_2)^2}.
\end{align*}
The condition \(c_4(\rho;T)=0\) defines the
critical surface-tension function
\[
T^{\ast}(\rho)
=\frac{h_1 h_2\big(-h_1\rho+h_1-h_2\rho^2+h_2\rho\big)}%
{3(h_1\rho+h_2)},
\qquad
T^{\ast}(0)=\frac{h_1^2}{3}.
\]
In the long-wave limit, this degeneracy condition coincides with the
singularity set \(J\to\infty\); hence, the same function \(T^{\ast}(\rho)\)
locally describes both the linear dispersion surface
\(\omega''=0\) (green) and the nonlinear singular surface
\(J\to\infty\) (blue) in the vicinity of \((\rho,k)=(0,0,T^{\ast})\).

A horizontal tangency of \(T(\rho,k)\) to the plane \(T=T^{\ast}\)
at \((\rho,k)=(0,0)\) requires
\(\partial T^{\ast}/\partial\rho|_{\rho=0}=0\), which gives
\begin{equation}\label{eq:h2_phi_h1_short}
h_1^2+h_2(h_1-h_2)=0,
\qquad
\frac{h_2}{h_1}=\frac{1+\sqrt{5}}{2}\approx1.618.
\end{equation}
Hence, the critical surface is horizontally tangent at the origin
when the depth ratio satisfies~\eqref{eq:h2_phi_h1_short}.
This ratio corresponds to the \emph{golden ratio}
\(\varphi=(1+\sqrt{5})/2\),
representing a neutral balance between the inertial contributions
of the two layers in the long-wave limit.
For \(h_2/h_1<\varphi\),
\(\partial T^{\ast}/\partial\rho|_{\rho=0}<0\),
indicating a gravity-dominated regime;
for \(h_2/h_1>\varphi\),
the derivative is positive,
corresponding to a capillarity-dominated regime,
while the golden-ratio configuration
\(\,h_2/h_1=\varphi\,\)
defines the boundary between these two regimes.

\section{Stability diagrams: the influence of surface tension}
\label{sec_Matrix}

\subsection{General remarks and reference configuration}
\label{subsec_3.1}

In Part~I, modulational–stability maps were constructed for two representative sets of layer–depth combinations,
\((h_1,h_2)\in\{1,2,3,4\}\) and \((h_1,h_2)\in\{1,5,9,13\}\),
which together illustrated the main topological types of stable and unstable regions
and their evolution with varying density ratio~\(\rho\) and wavenumber~\(k\).
In the present part, attention is focused on the second, broader matrix
\((h_1,h_2)\in\{1,5,9,13\}\),
which serves as a reference configuration for analysing the effect of surface tension.
Other geometric and physical parameters of the system will be examined in the following subsections.

\begin{figure}
\centering
\vspace{-1.5em}

\hspace{-3ex}
\begin{subfigure}[b]{0.26\textwidth}
  \includegraphics[width=\linewidth]{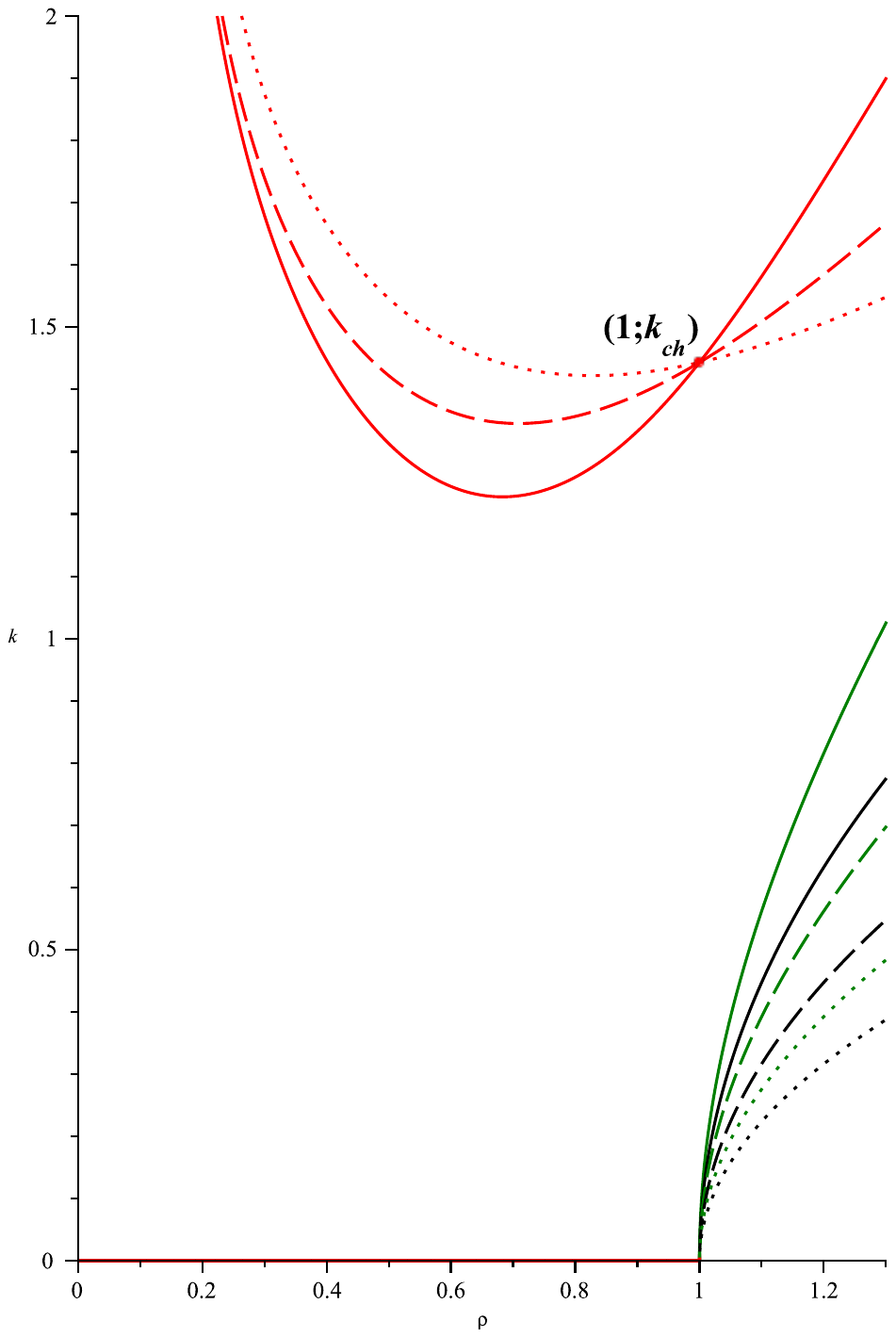}
  \caption{$h_1\!=\!1,\, h_2\!=\!1$}
  \label{fig:Fig1a}
\end{subfigure}\hspace{-2ex}
\begin{subfigure}[b]{0.26\textwidth}
  \includegraphics[width=\linewidth]{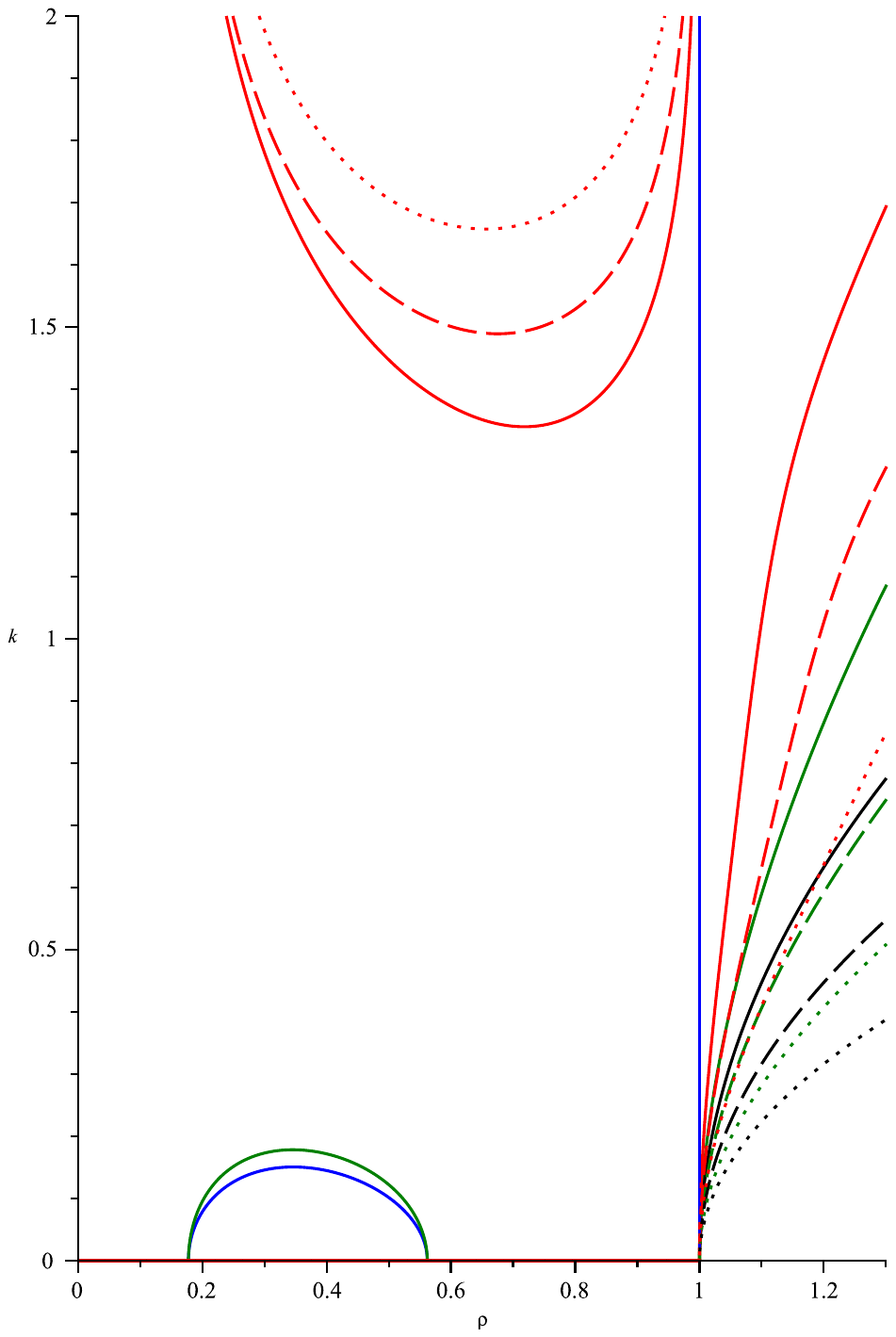}
  \caption{$h_1\!=\!1,\, h_2\!=\!5$}
  \label{fig:Fig1b}
\end{subfigure}\hspace{-2ex}
\begin{subfigure}[b]{0.26\textwidth}
  \includegraphics[width=\linewidth]{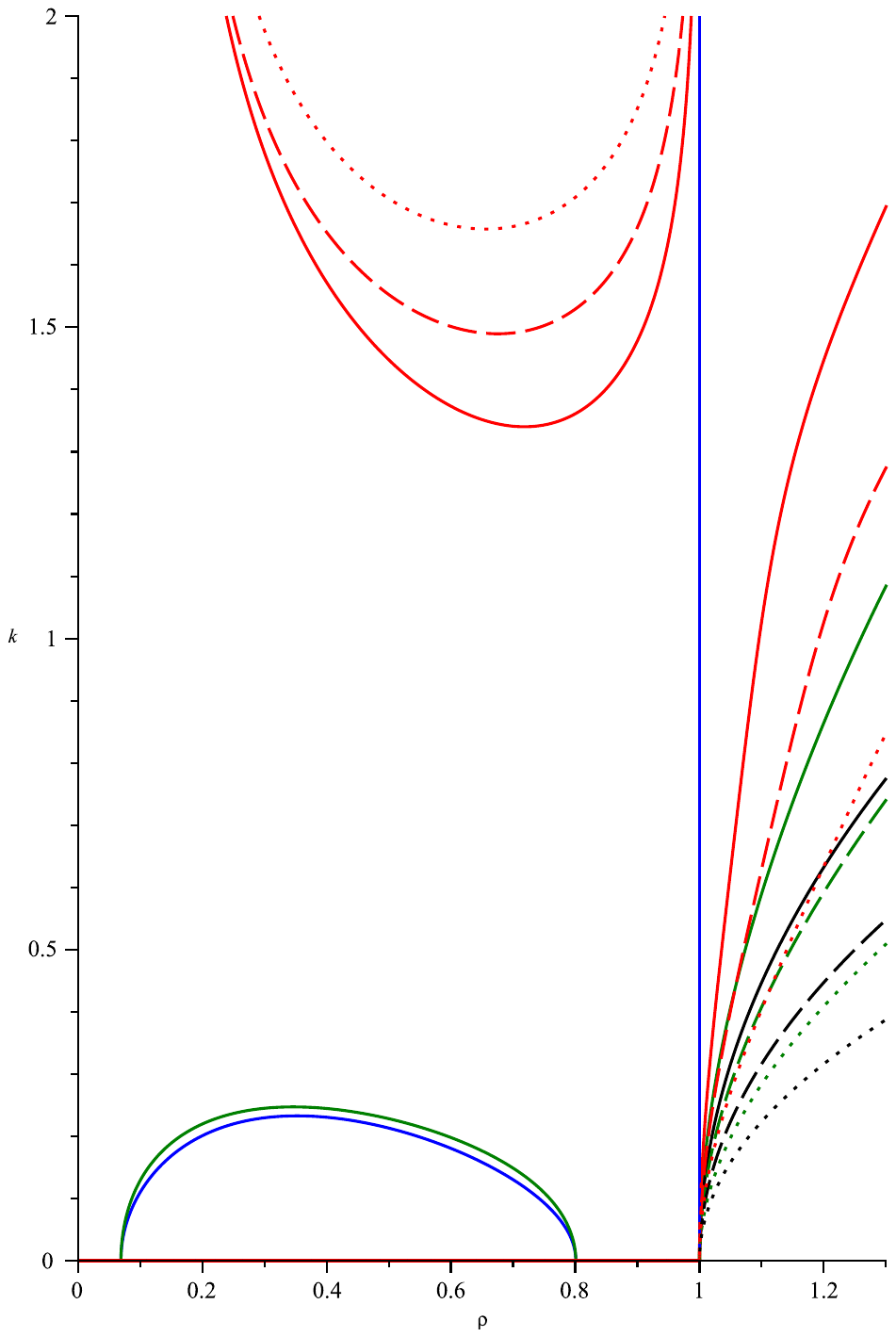}
  \caption{$h_1\!=\!1,\, h_2\!=\!9$}
  \label{fig:Fig1c}
\end{subfigure}\hspace{-2ex}
\begin{subfigure}[b]{0.26\textwidth}
  \includegraphics[width=\linewidth]{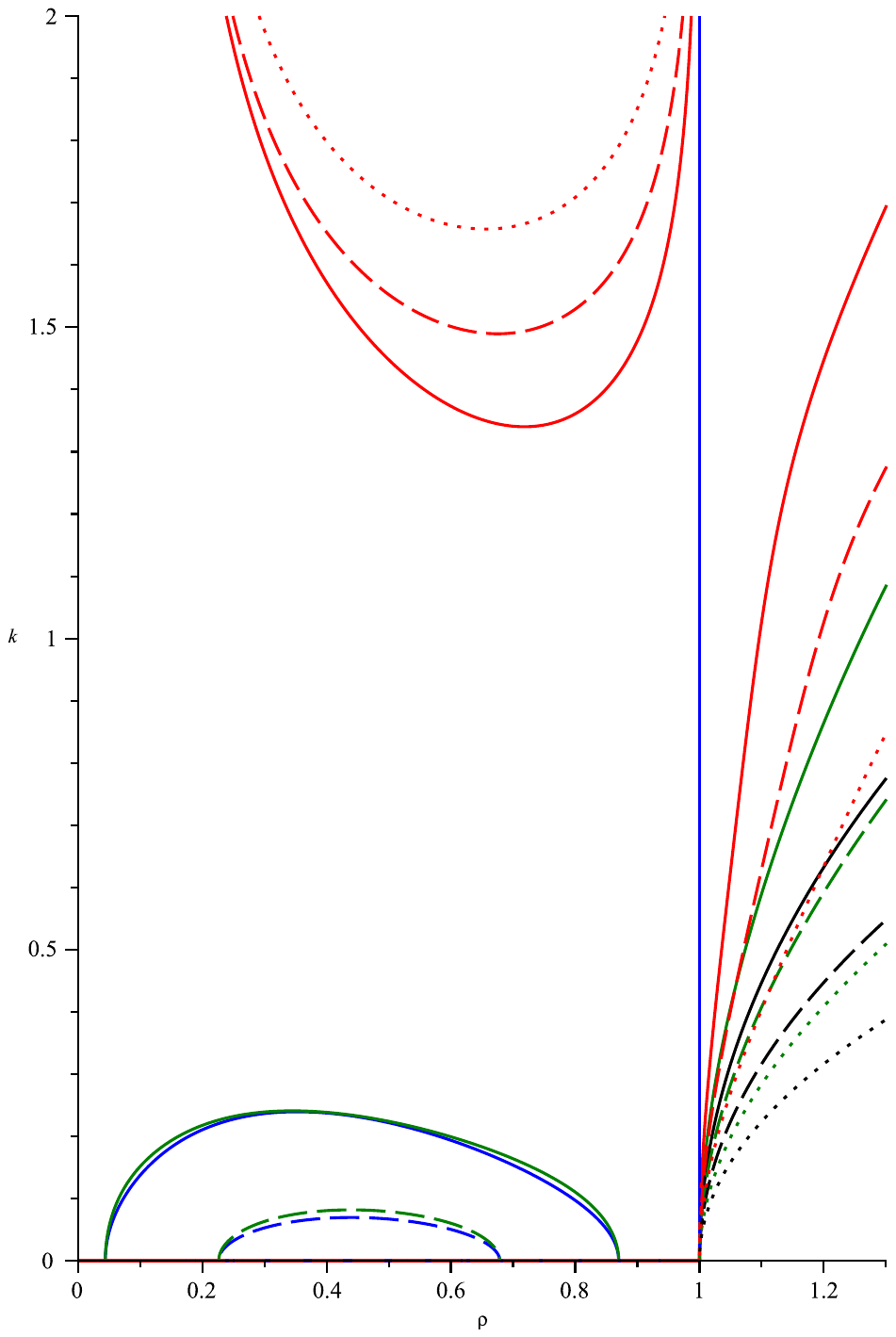}
  \caption{$h_1\!=\!1,\, h_2\!=\!13$}
  \label{fig:Fig1d}
\end{subfigure}

\vspace{-0.5em}

\hspace{-3ex}
\begin{subfigure}[b]{0.26\textwidth}
  \includegraphics[width=\linewidth]{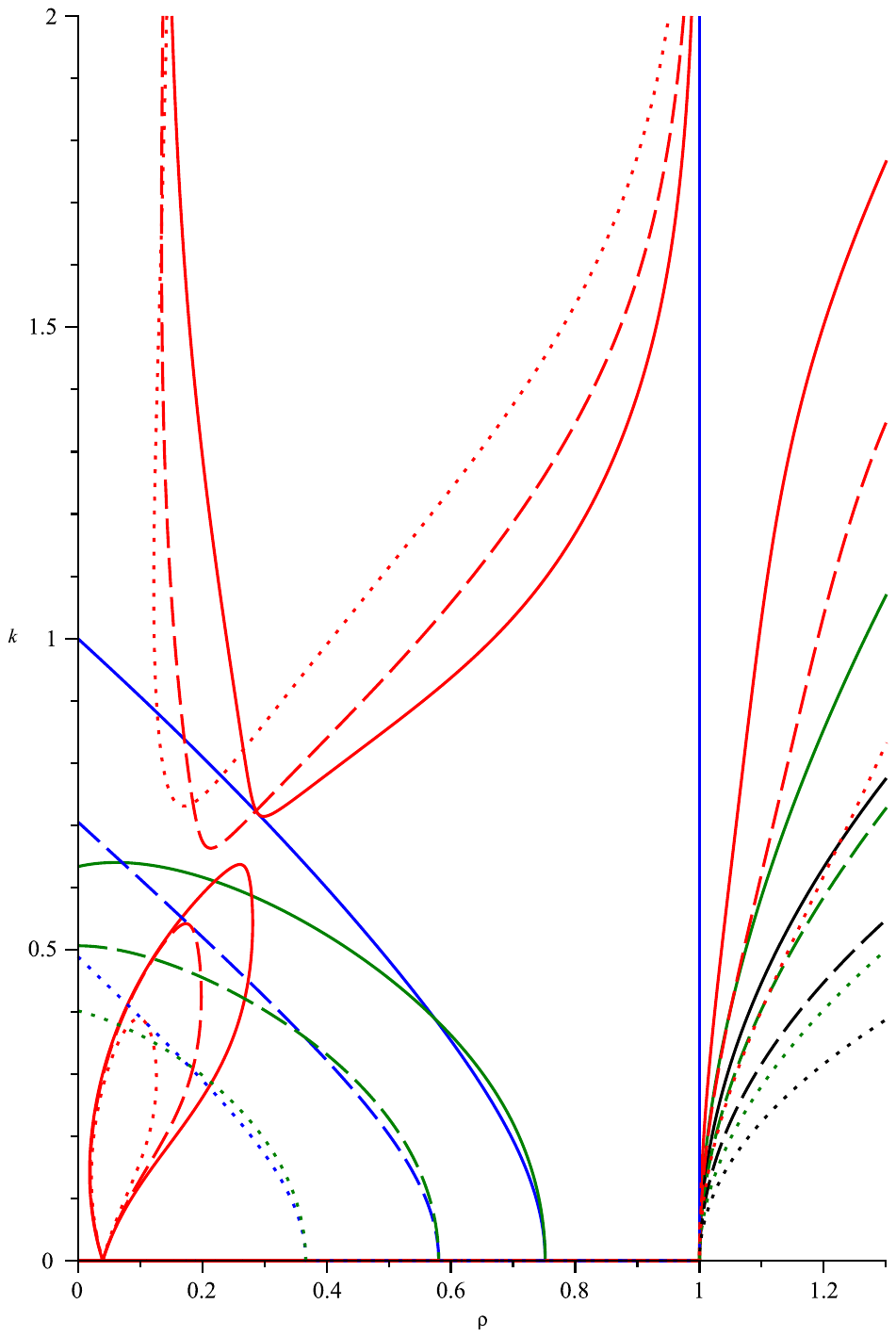}
  \caption{$h_1\!=\!5,\, h_2\!=\!1$}
  \label{fig:Fig1e}
\end{subfigure}\hspace{-2ex}
\begin{subfigure}[b]{0.26\textwidth}
  \includegraphics[width=\linewidth]{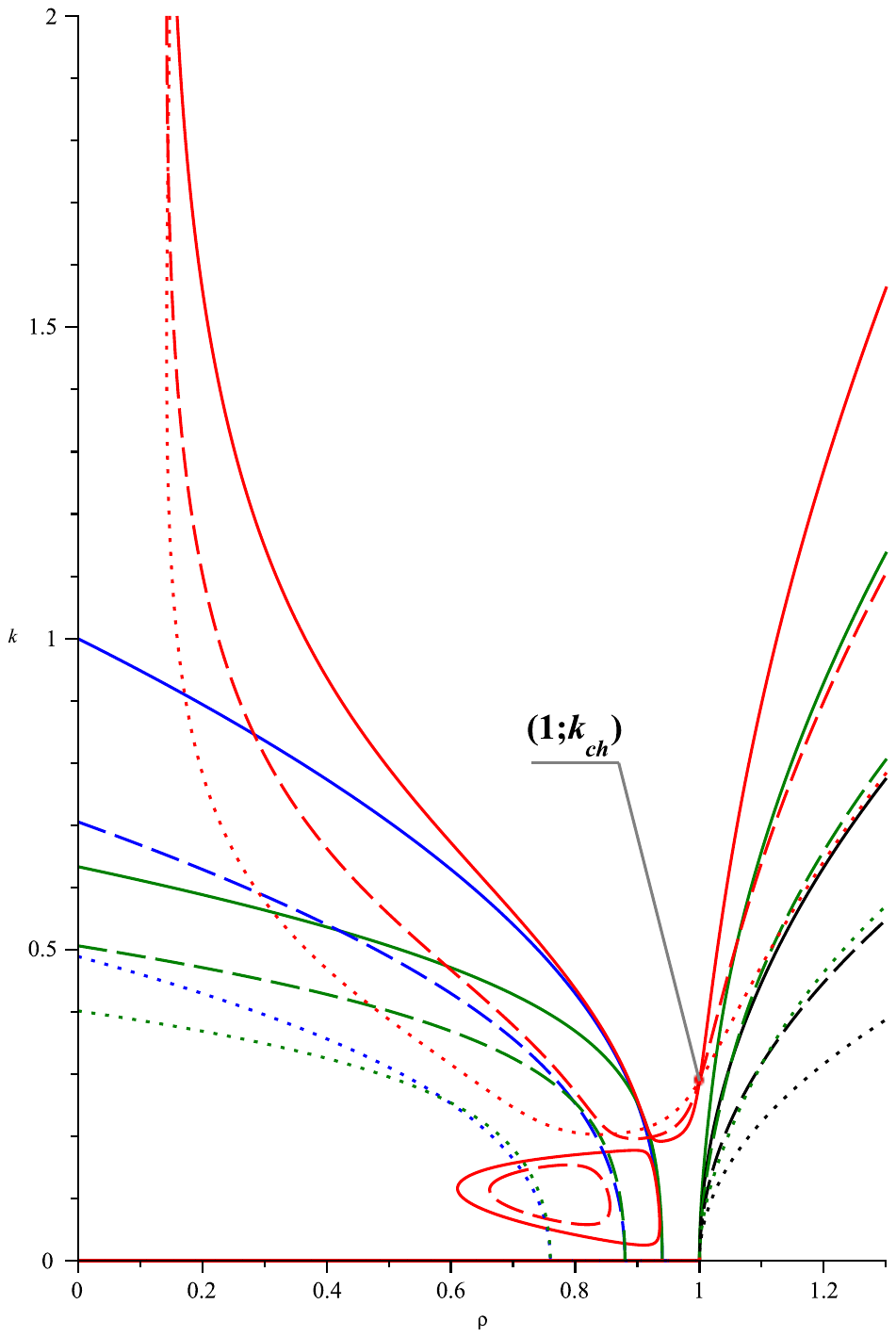}
  \caption{$h_1\!=\!5,\, h_2\!=\!5$}
  \label{fig:Fig1f}
\end{subfigure}\hspace{-2ex}
\begin{subfigure}[b]{0.26\textwidth}
  \includegraphics[width=\linewidth]{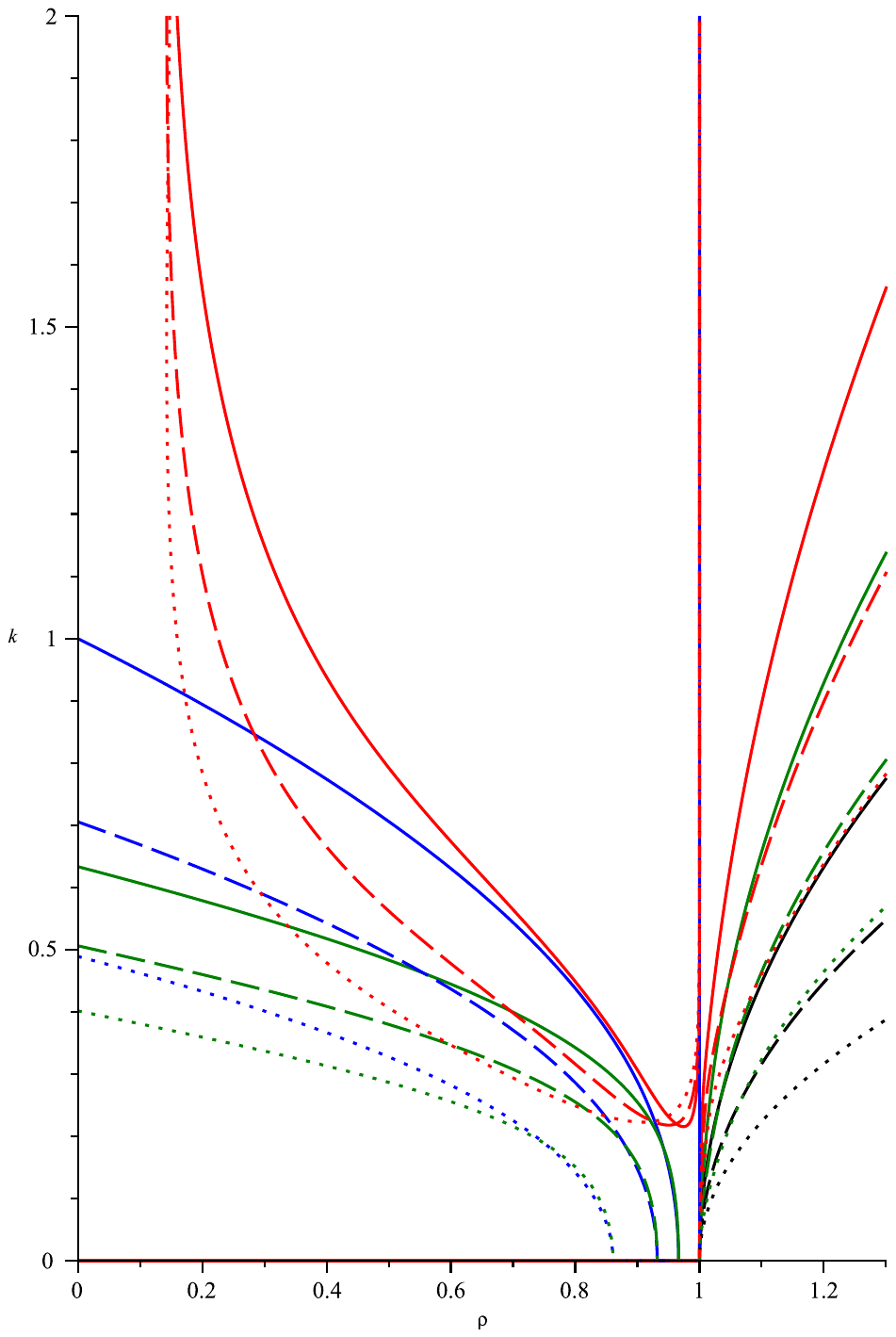}
  \caption{$h_1\!=\!5,\, h_2\!=\!9$}
  \label{fig:Fig1g}
\end{subfigure}\hspace{-2ex}
\begin{subfigure}[b]{0.26\textwidth}
  \includegraphics[width=\linewidth]{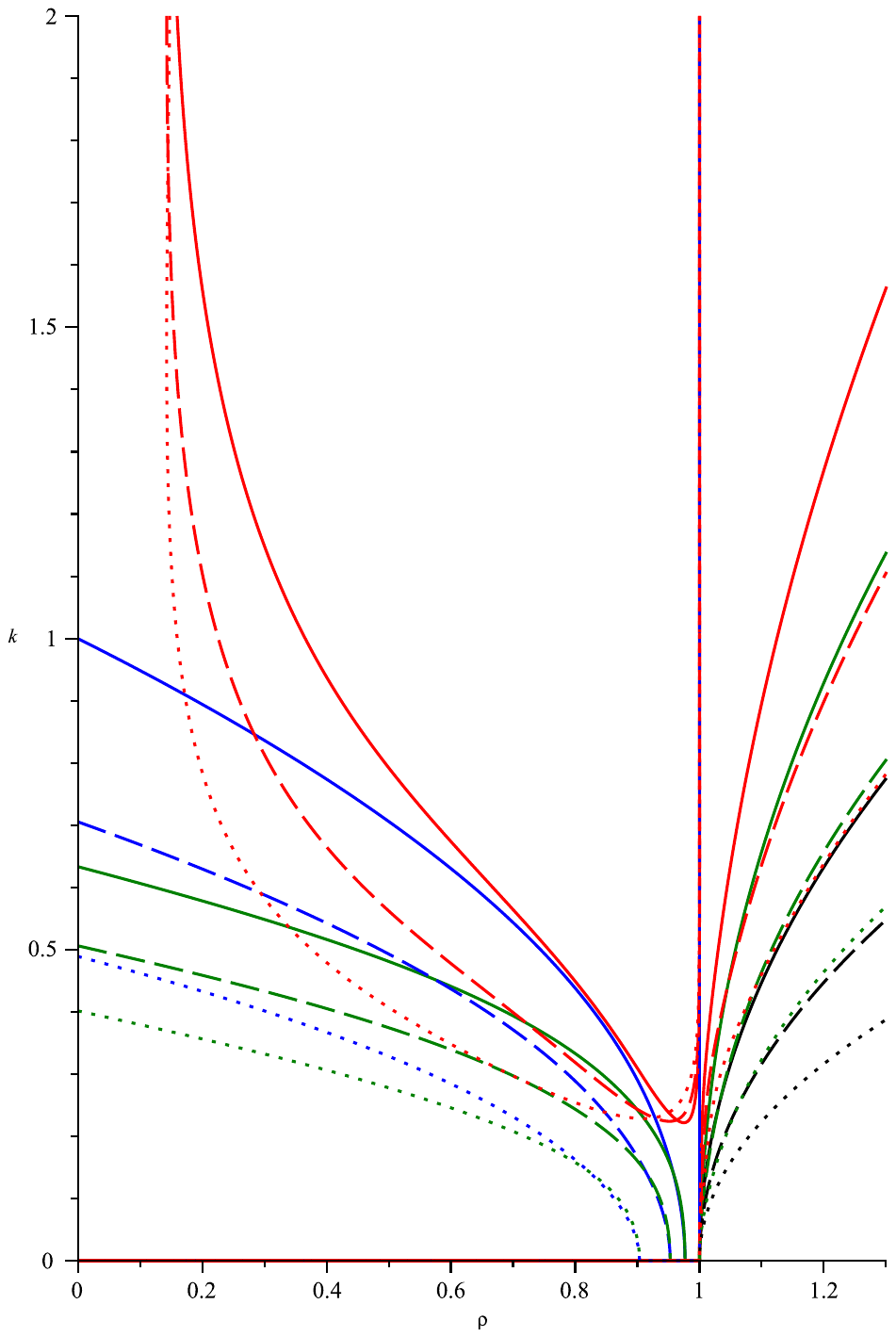}
  \caption{$h_1\!=\!5,\, h_2\!=\!13$}
  \label{fig:Fig2h}
\end{subfigure}

\vspace{-0.5em}

\hspace{-3ex}
\begin{subfigure}[b]{0.26\textwidth}
  \includegraphics[width=\linewidth]{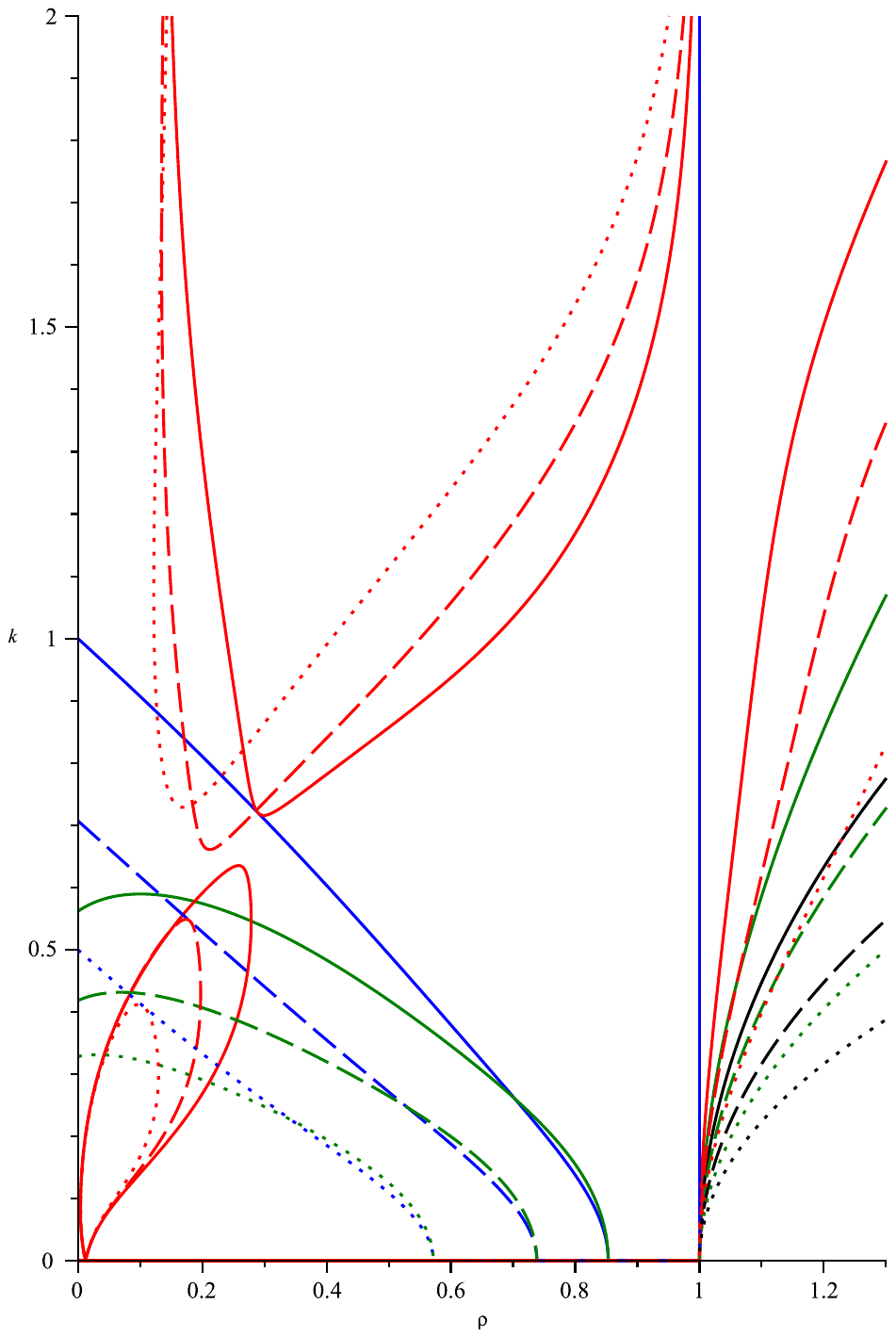}
  \caption{$h_1\!=\!9,\, h_2\!=\!1$}
  \label{fig:Fig1i}
\end{subfigure}\hspace{-2ex}
\begin{subfigure}[b]{0.26\textwidth}
  \includegraphics[width=\linewidth]{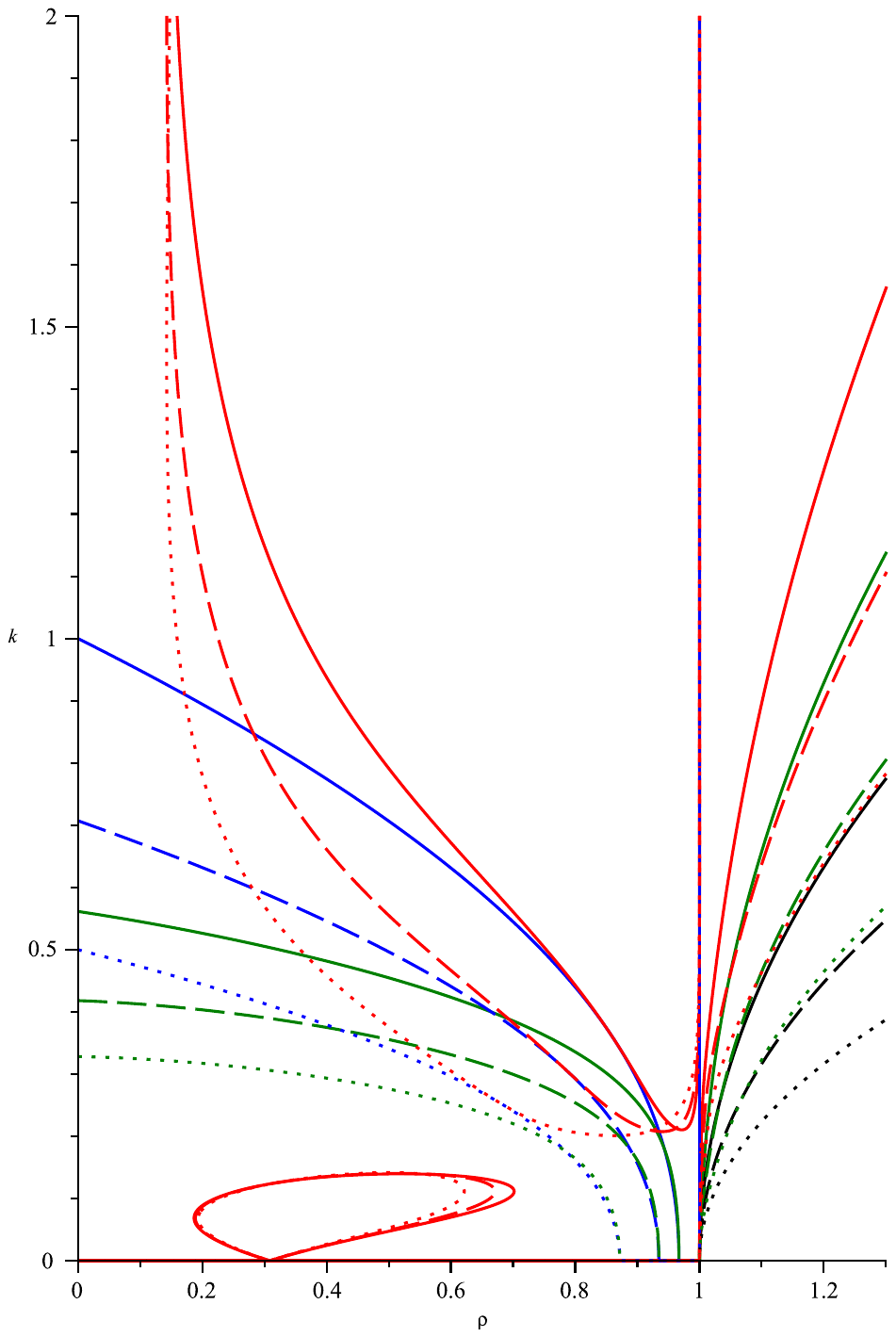}
  \caption{$h_1\!=\!9,\, h_2\!=\!5$}
  \label{fig:Fig1j}
\end{subfigure}\hspace{-2ex}
\begin{subfigure}[b]{0.26\textwidth}
  \includegraphics[width=\linewidth]{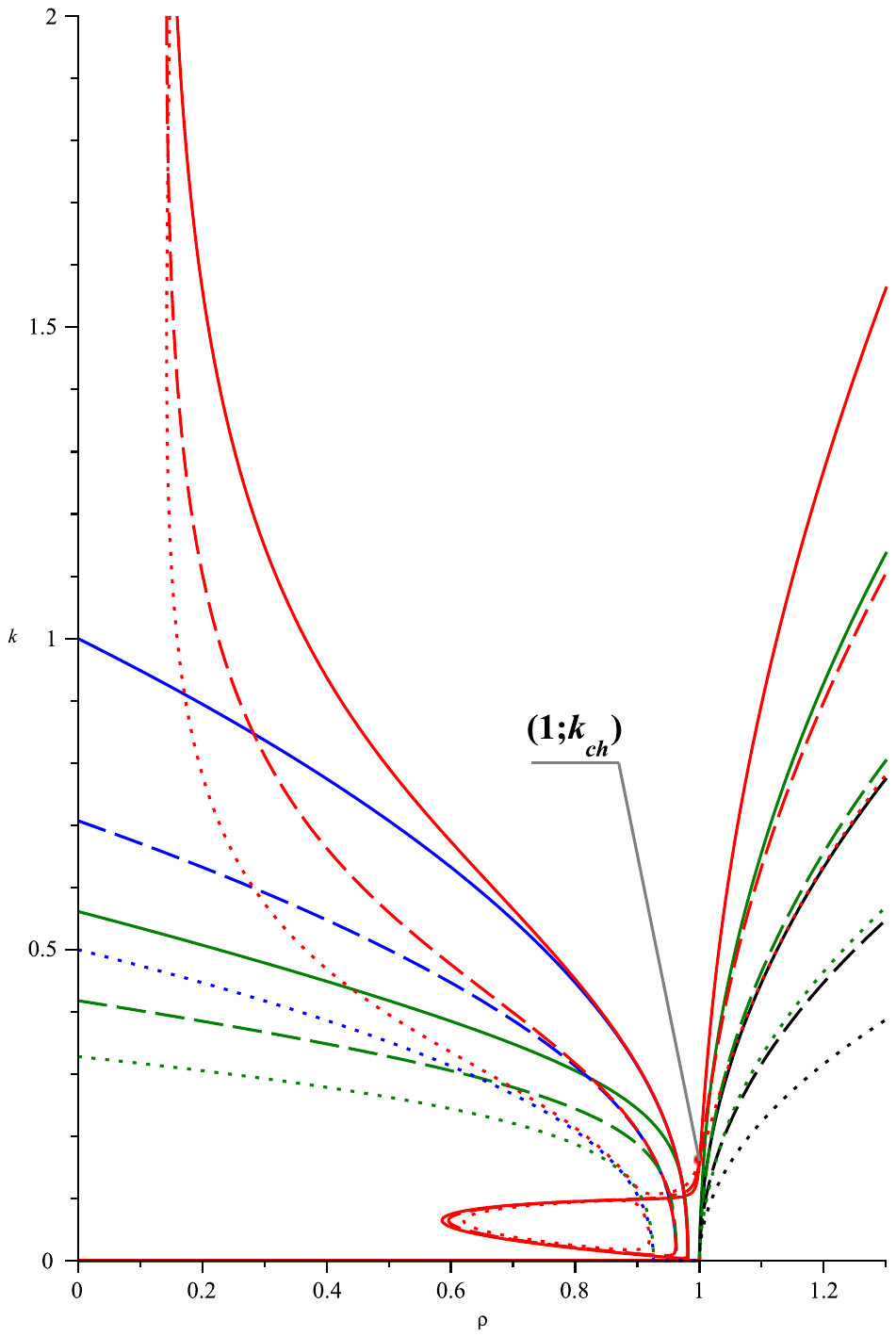}
  \caption{$h_1\!=\!9,\, h_2\!=\!9$}
  \label{fig:Fig1k}
\end{subfigure}\hspace{-2ex}
\begin{subfigure}[b]{0.26\textwidth}
  \includegraphics[width=\linewidth]{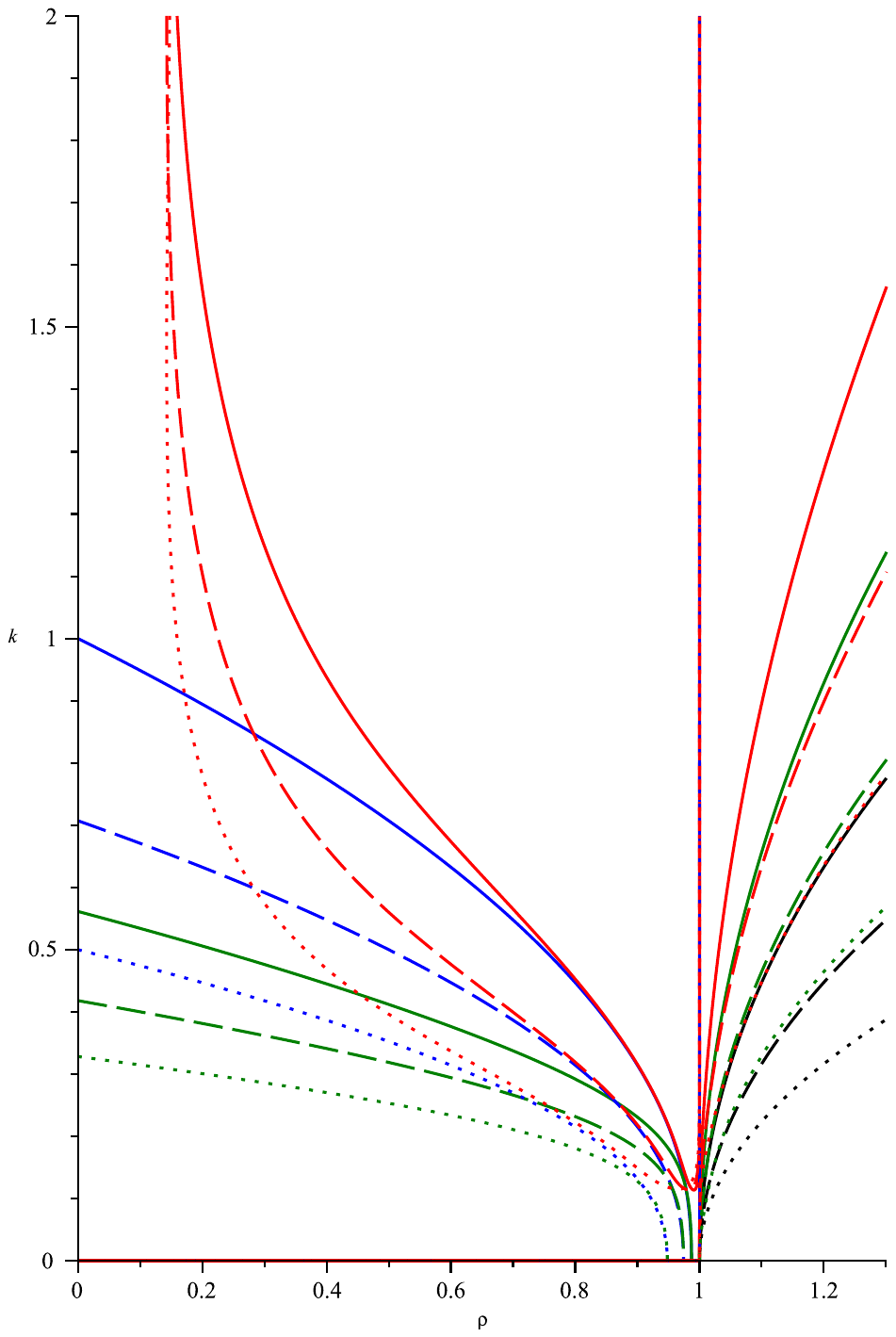}
  \caption{$h_1\!=\!9,\, h_2\!=\!13$}
  \label{fig:Fig1l}
\end{subfigure}

\vspace{-0.5em}

\hspace{-3ex}
\begin{subfigure}[b]{0.26\textwidth}
  \includegraphics[width=\linewidth]{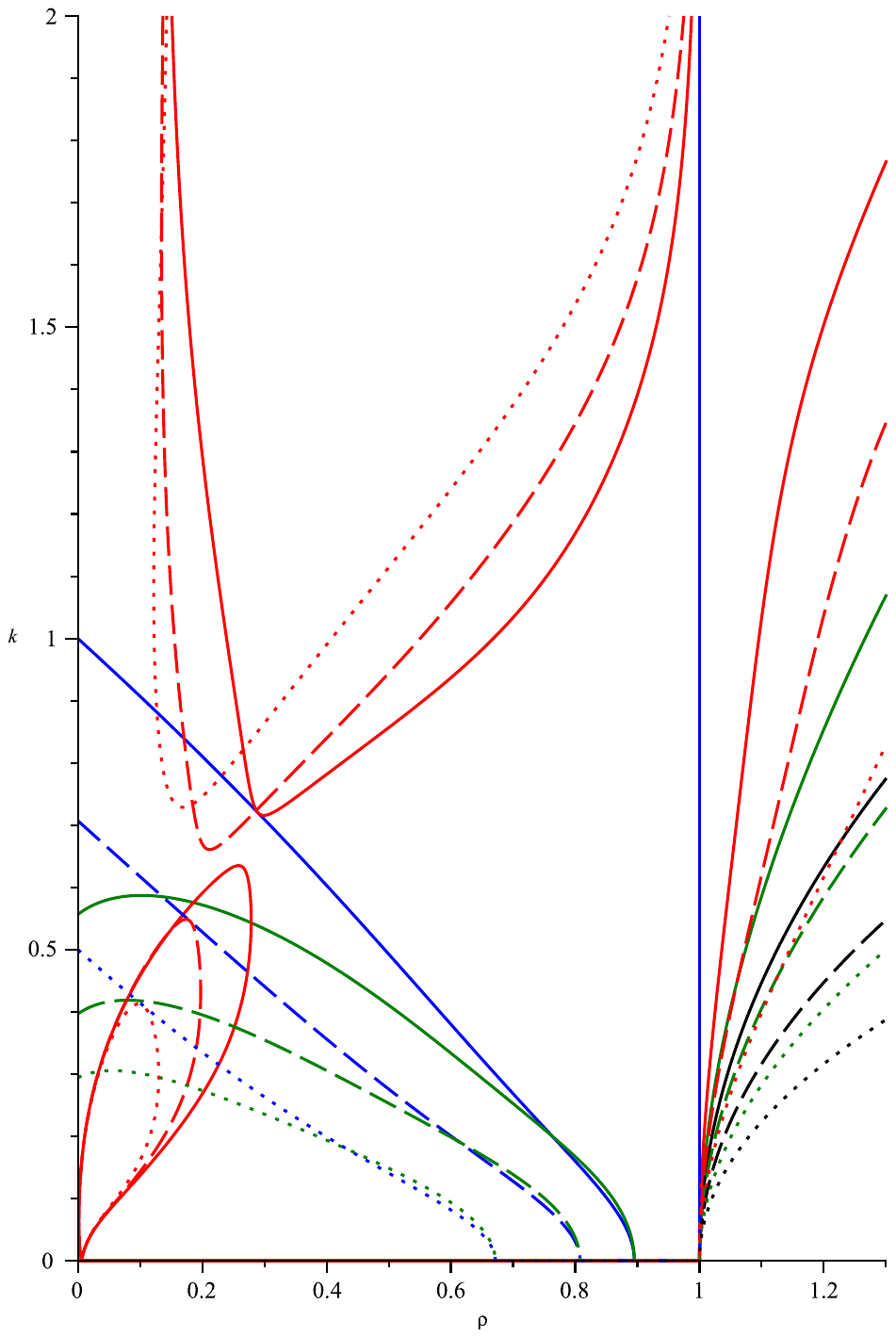}
  \caption{$h_1\!=\!13,\, h_2\!=\!1$}
  \label{fig:Fig1m}
\end{subfigure}\hspace{-2ex}
\begin{subfigure}[b]{0.26\textwidth}
  \includegraphics[width=\linewidth]{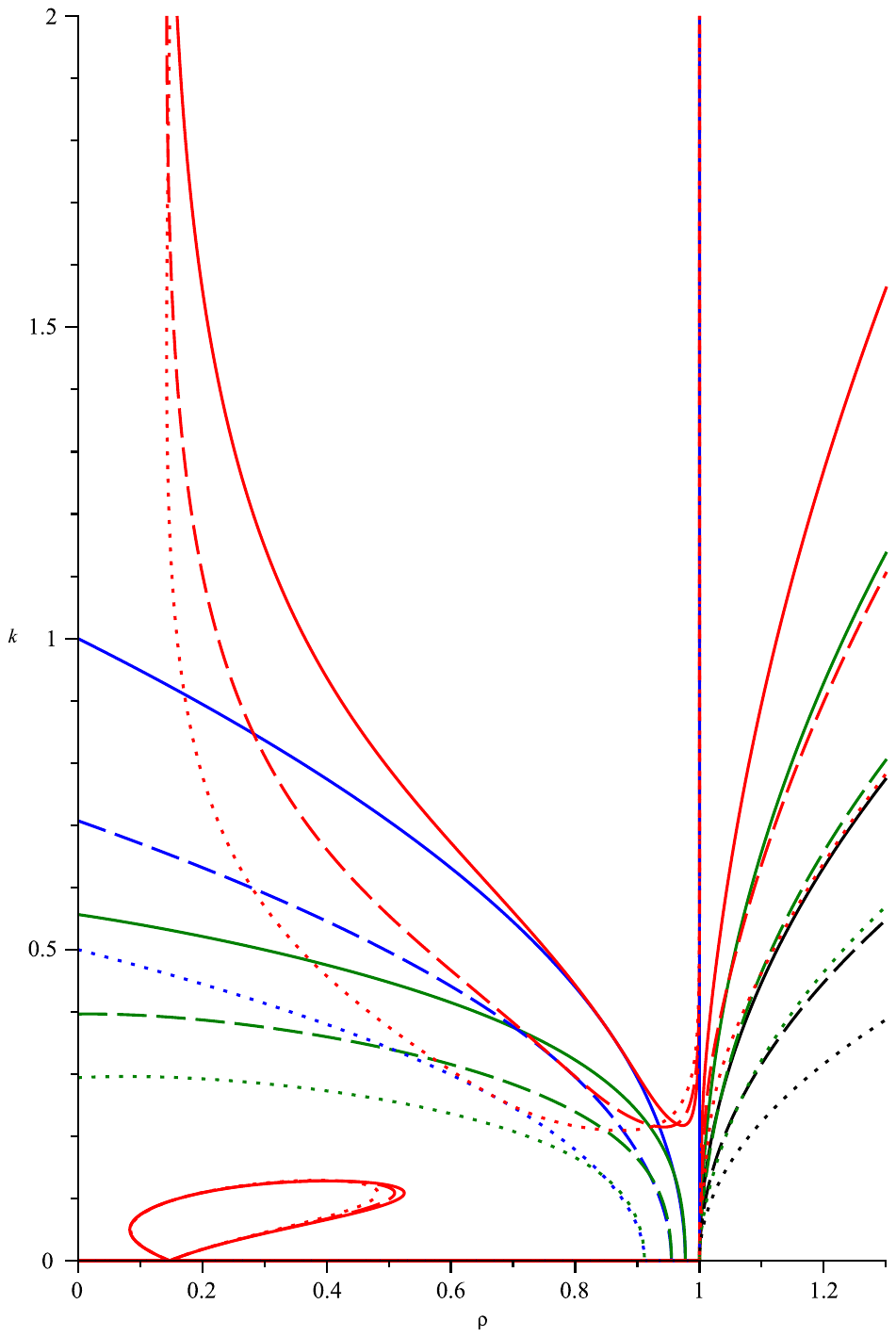}
  \caption{$h_1\!=\!13,\, h_2\!=\!5$}
  \label{fig:Fig1n}
\end{subfigure}\hspace{-2ex}
\begin{subfigure}[b]{0.26\textwidth}
  \includegraphics[width=\linewidth]{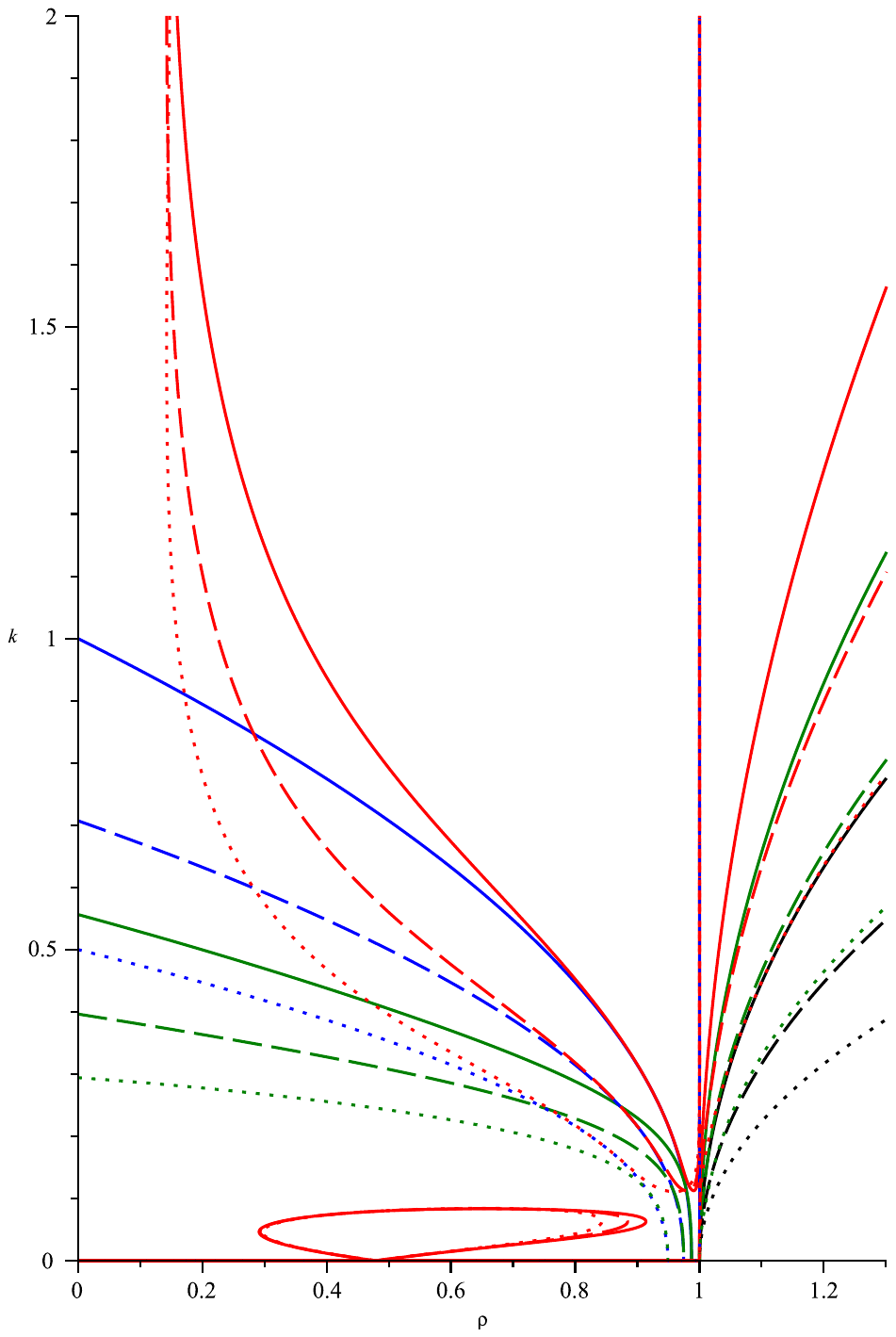}
  \caption{$h_1\!=\!13,\, h_2\!=\!9$}
  \label{fig:Fig1o}
\end{subfigure}\hspace{-2ex}
\begin{subfigure}[b]{0.26\textwidth}
  \includegraphics[width=\linewidth]{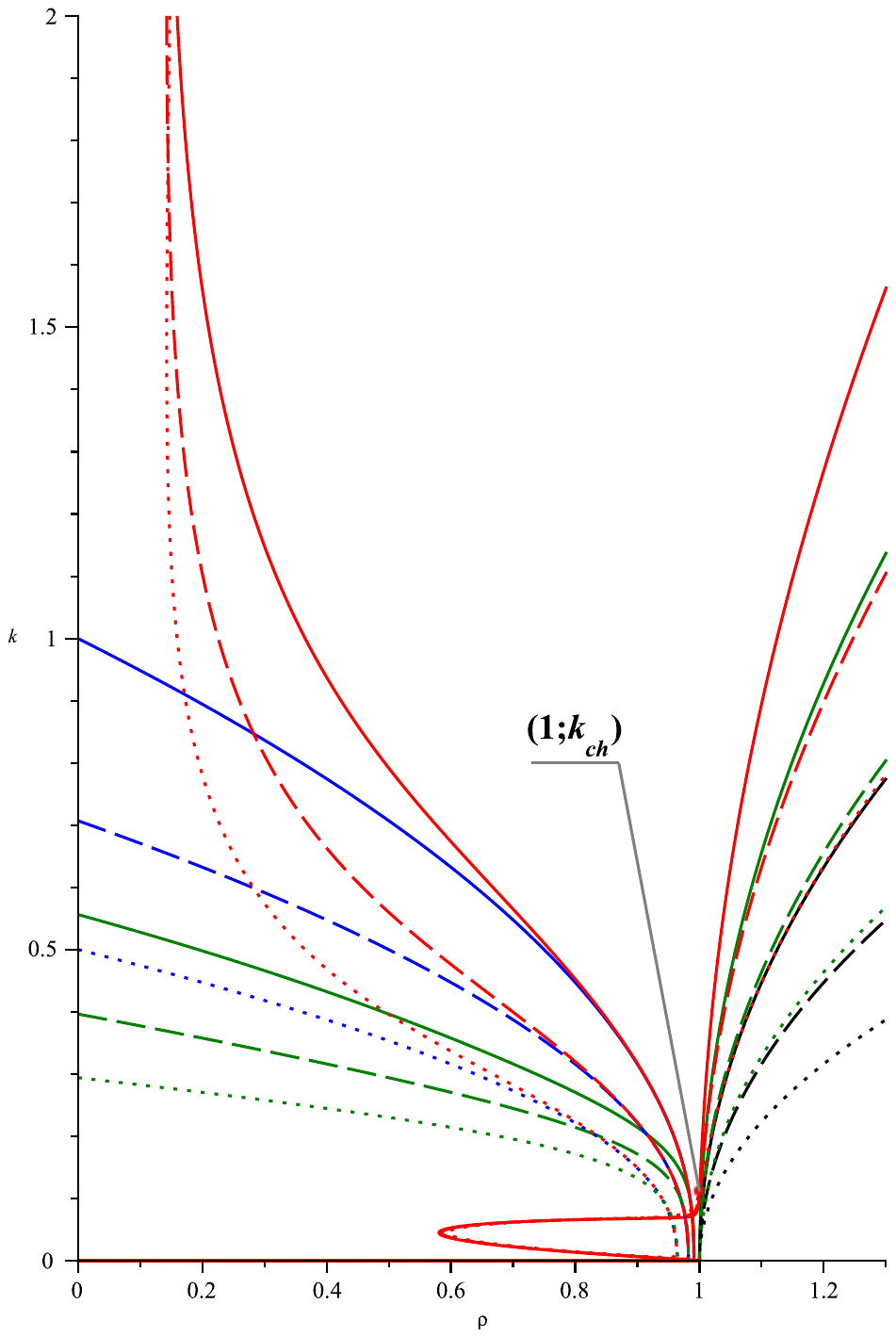}
  \caption{$h_1\!=\!13,\, h_2\!=\!13$}
  \label{fig:Fig1p}
\end{subfigure}

\caption{Modulational–stability diagrams for all combinations of
\(h_{1}, h_{2} \in \{1, 5, 9, 13\}\) for \(T=1/2, 1, 2\).}
\label{fig:Matrix}
\end{figure}

Figure~\ref{fig:Matrix} shows the modulational–stability diagrams for three representative values,
\(T={1}/{2}\), \(1\), and \(2\).
The colour convention is the same as in Part~I:
black denotes the linear–stability boundary corresponding to the
\emph{critical wavenumber}
\[
k = k_{\mathrm c} = \sqrt{(\rho - 1)/T},
\]
which separates the regions of linear stability and instability;
red indicates the curve \(J = 0\) separating focusing and defocusing nonlinearities;
blue marks the singular curve \(J \to \infty\) associated with resonance coupling;
and green represents the dispersion–curvature line \(\omega''(k)=0\),
where the sign of the group–velocity dispersion changes.
Solid, dashed, and dotted lines correspond to \(T=1/2\), \(1\), and \(2\), respectively.
This comparison highlights how an increase in surface tension
gradually modifies the topology of stable and unstable domains
across the parameter space \((\rho,k)\).

\subsection{Symmetric and asymmetric configurations of the upper unstable region}
\label{subsec_3.2}

The influence of surface tension on the upper modulationally unstable domain (bounded above by the red locus \(J=0\)) can be examined by comparing three representative values, \(T=1/2\) (solid), \(1\) (dashed), and \(2\) (dotted), across all panels of Fig.~\ref{fig:Matrix}. As shown in Part~I, deviations of \(T\) from the reference value \(T=1\) produce opposite yet comparable effects: doubling to \(T=2\) and halving to \(T=1/2\) deform the upper domain in reverse directions with nearly symmetric magnitudes.

For equal depths (\(h_1=h_2\)), the upper region remains continuous at \(\rho=1\), without a vertical cut. In Fig.~\ref{fig:Matrix}a (\(h_1=h_2=1\)) the red boundaries for different \(T\) intersect twice, the left intersection occurring at relatively large wavenumbers (\(k>2\)). For \(\rho<1\), the ordering of curves is dotted (\(T=2\)) above dashed (\(T=1\)) above solid (\(T=1/2\)), indicating that increasing surface tension narrows the upper unstable domain, whereas decreasing \(T\) broadens it. This qualitative property persists along the diagonal of the matrix (Figs.~\ref{fig:Matrix}f,\,k,\,p), where \(h_1=h_2=h\): the asymptote at \(\rho=1\) does not appear, and the intersection point of the \(T\)-dependent boundaries shifts toward smaller \(k\) as \(h\) increases. The corresponding characteristic wavenumber
\[
k_{\mathrm{ch}}(h)=\frac{1}{2h}\ln(9+4\sqrt{5}),
\]
given by formula~\eqref{eq:kstar}, decreases monotonically with \(h\) and approaches \((\rho,k)=(1,0)\) in the deep–water limit, consistent with the case of two hydrodynamic half–spaces. As can be seen in the panels corresponding to the equal–depth configurations (Figs.~\ref{fig:Matrix}a,\,f,\,k,\,p), the point
\(
(\rho,k)=\bigl(1,\,k_{\mathrm{ch}}(h)\bigr)
\)
appears as the unique location at which all red boundaries (\(J=0\)) corresponding to \(T=1/2,\,1,\) and \(2\) intersect. This common intersection is a distinctive geometric feature of the symmetric case \(h_{1}=h_{2}=h\); for unequal layer depths such a point does not exist, and instead the line \(\rho=1\) becomes a vertical asymptote of the upper unstable region, as discussed below.

When depth symmetry is broken, the topology of the upper region changes at \(\rho=1\). In the first row of Fig.~\ref{fig:Matrix}b--d (\(h_1=1\), \(h_2>1\)), the intersection at \(\rho=1\) is replaced by a vertical asymptote that divides the upper domain, although the relative ordering of the \(T\)-dependent boundaries away from \(\rho=1\) remains similar to that in Fig.~\ref{fig:Matrix}a. In contrast, Figs.~\ref{fig:Matrix}e,~i,~m (\(h_2=1\), \(h_1=5,9,13\)) exhibit a reversed response at small density ratios: the dotted curve (\(T=2\)) lies below the dashed one (\(T=1\)), while the solid curve (\(T=1/2\)) lies above both, so increasing \(T\) expands the upper unstable domain and decreasing \(T\) contracts it. For larger \(\rho\) (\(\rho\gtrsim0.3\)) this tendency reverses. In the remaining panels of Fig.~\ref{fig:Matrix} (rows 2–4, columns 2–4), the pairwise intersection points of the \(T\)-dependent boundaries cluster near \(\rho\simeq0.85\)–\(0.99\); within this range the order of curves is temporarily reversed, but just below \(\rho=1\) the usual sequence (dotted below dashed, solid above both) is restored.

The physical mechanism underlying these reorganizations is the redistribution of the balance between focusing and defocusing nonlinearities by capillarity under nearly fixed dispersion. The capillary contribution to the effective nonlinearity scales as \(T\,k^{5}(1-\rho)\): for a light upper layer (\(\rho<1\)) an increase in \(T\) stiffens the interface and shifts the \(J=0\) boundary upward in \(k\), narrowing the unstable region. Near and above density matching, particularly under pronounced depth asymmetry, the same increase in \(T\) can enhance the focusing component and expand the instability zone. Depth symmetry (\(h_1=h_2\)) removes the singular response at \(\rho=1\), producing a smooth passage of the \(J=0\) locus across the density–matching line, whereas asymmetry restores a nearly singular sensitivity manifested as a vertical asymptote. The clustering of intersection points near \(\rho\lesssim1\) shows that, for small density contrast, variations in \(T\) mainly tune the high–\(k\) capillary stiffness without significantly changing the inertial balance between layers, leading to local, topology-preserving adjustments of the upper boundary rather than qualitative alterations of its shape.

\subsection{Corridor and cut formation}
\label{subsec_3.3}

The stability structures associated with the corridor and the cut, which represent additional stable regions, are absent in Fig.~\ref{fig:Matrix}a for the symmetric case \(h_1=h_2=1\). In the subsequent panels (Figs.~\ref{fig:Matrix}b–d) a narrow cut appears, terminating at both ends on the \(\rho\)-axis. For \(h_1=1\) and \(h_2=5\) only a single small cut is visible, corresponding to \(T=1/2\) (Fig.~\ref{fig:Matrix}b). When the upper-layer thickness increases to \(h_2=9\), the cut persists for \(T=1/2\), becomes noticeably longer, and extends to higher wavenumbers (Fig.~\ref{fig:Matrix}c). At \(h_2=13\) (Fig.~\ref{fig:Matrix}d), two distinct cuts are observed: a broader one associated with \(T=1/2\) and a smaller one corresponding to \(T=1\).

\begin{figure}
\centering

\hspace{-1ex}
\begin{subfigure}[b]{0.32\textwidth}
  \includegraphics[width=\linewidth]{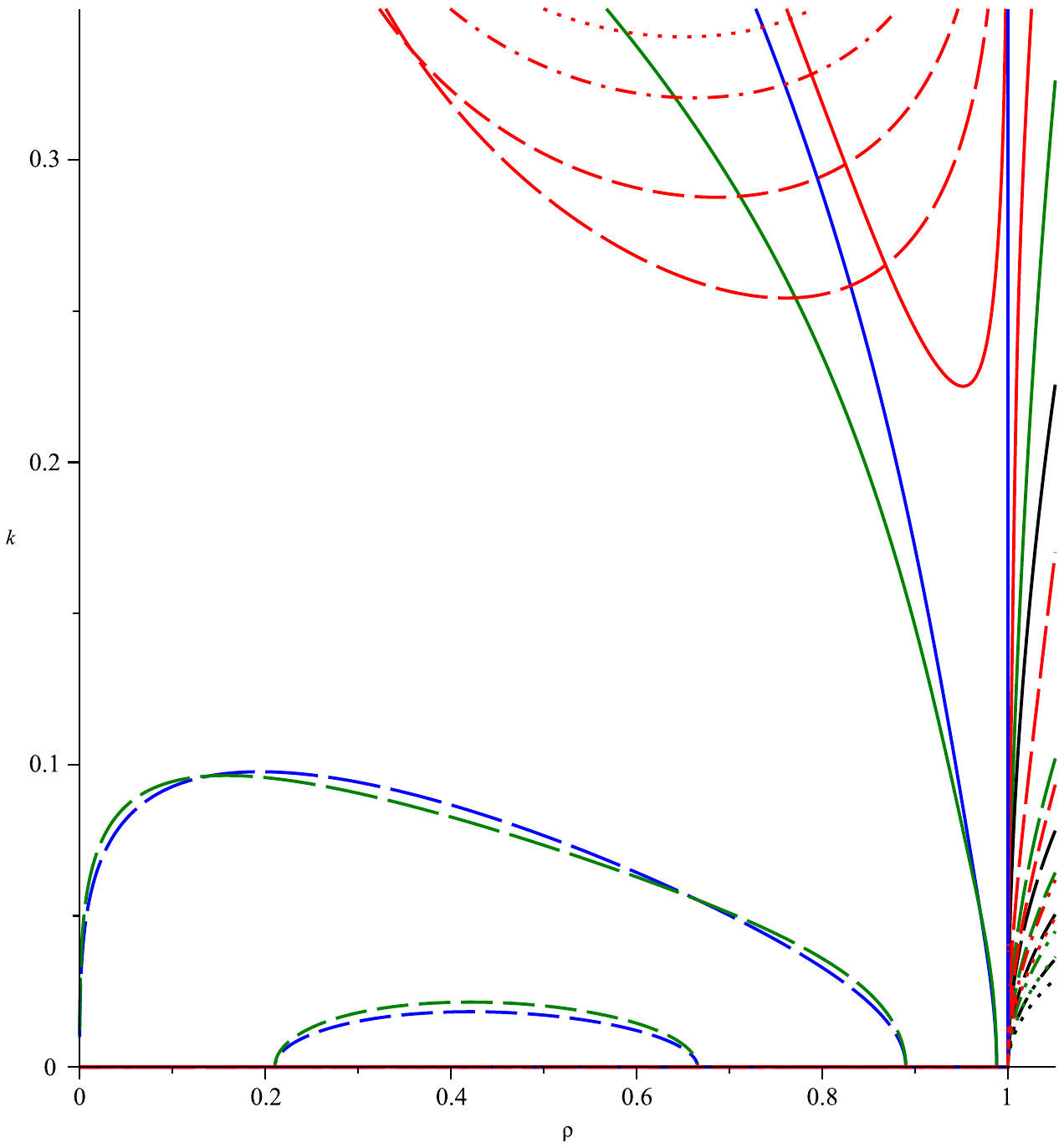}
  \caption{$h_1\!=\!5,\, h_2\!=\!50$}
  \label{fig:Fig2a}
\end{subfigure}\hspace{0ex}
\begin{subfigure}[b]{0.32\textwidth}
  \includegraphics[width=\linewidth]{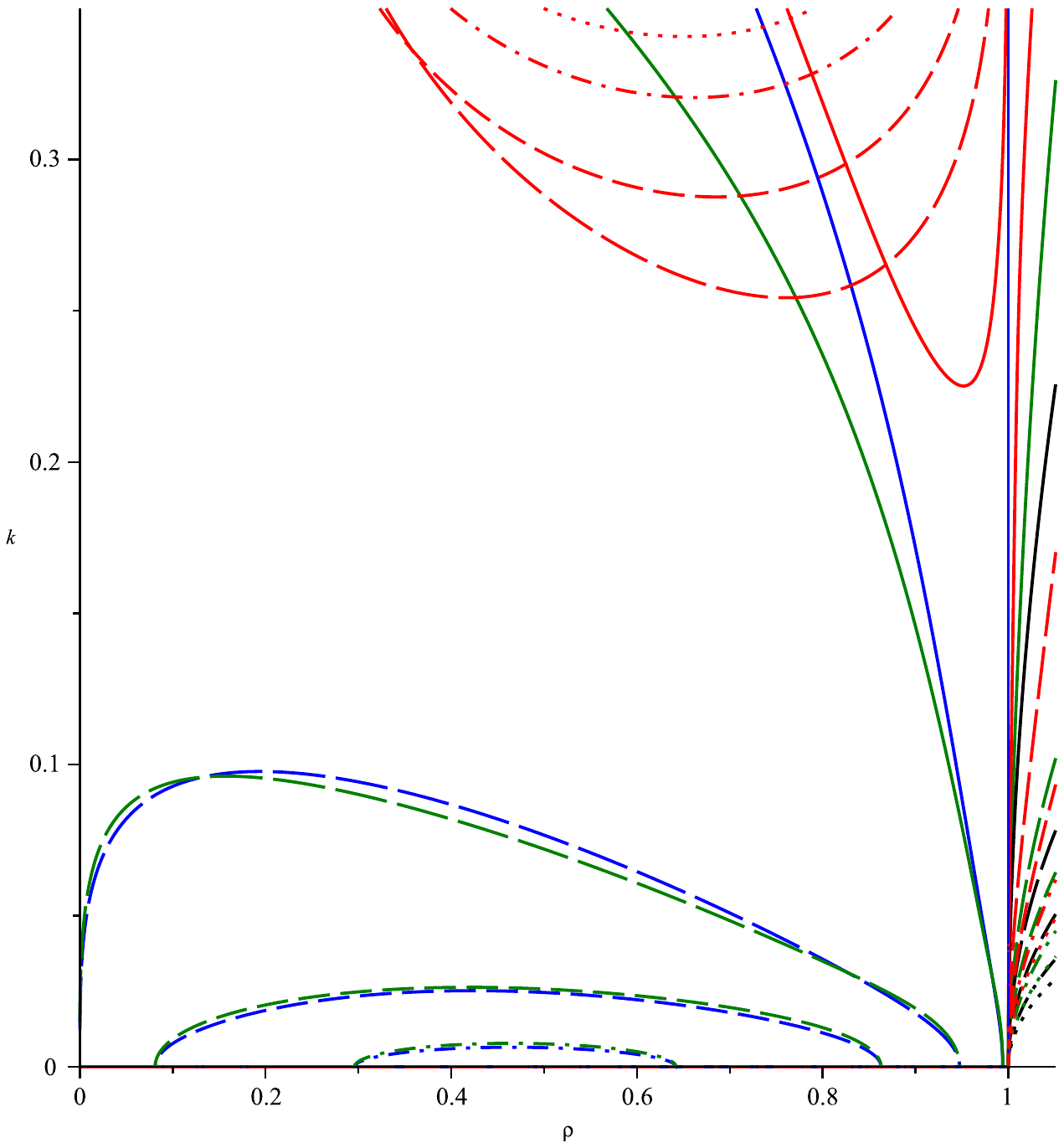}
  \caption{$h_1\!=\!5,\, h_2\!=\!100$}
  \label{fig:Fig2b}
\end{subfigure}\hspace{0ex}
\begin{subfigure}[b]{0.32\textwidth}
  \includegraphics[width=\linewidth]{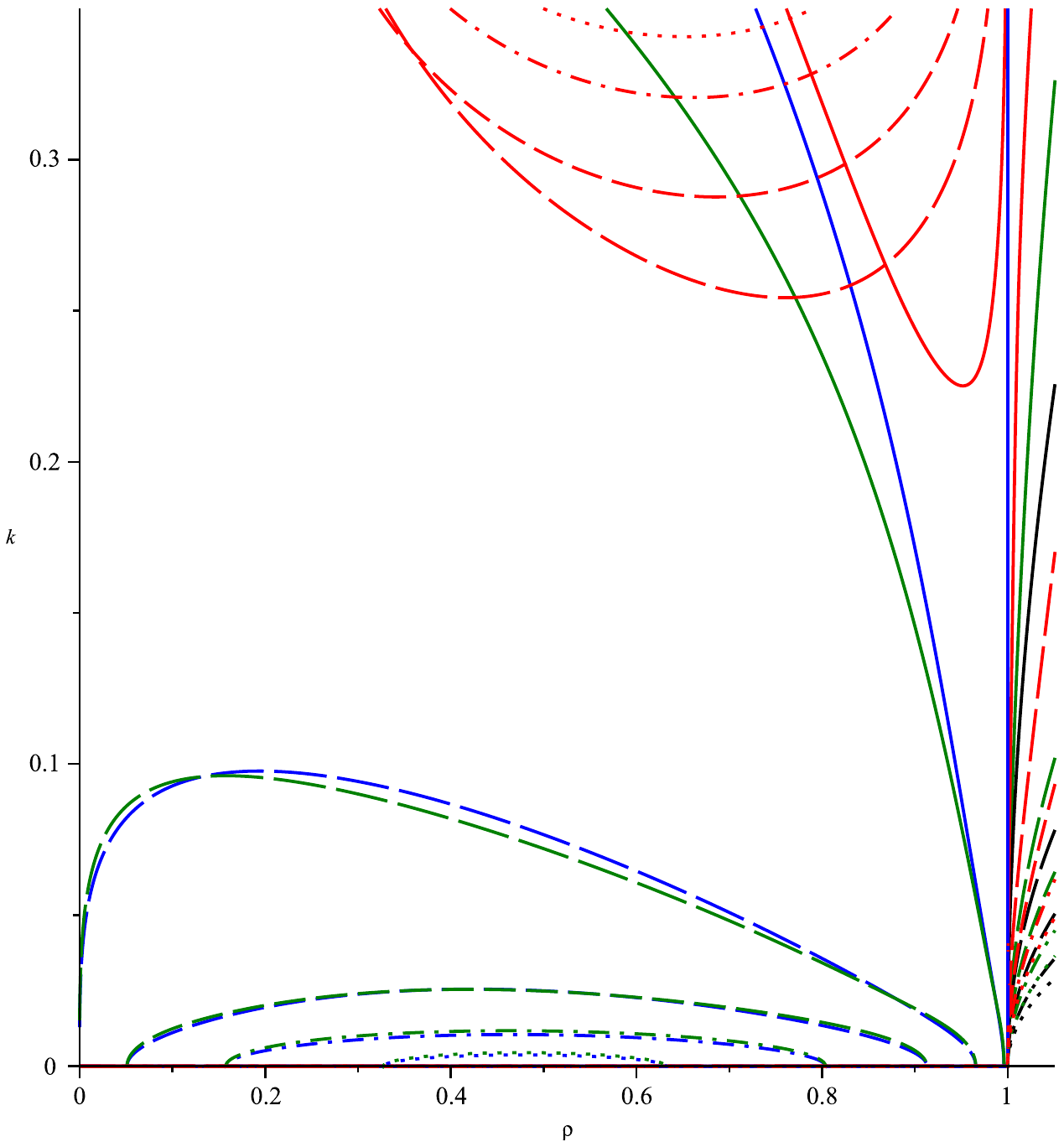}
  \caption{$h_1\!=\!5,\, h_2\!=\!150$}
  \label{fig:Fig2c}
\end{subfigure}

\caption{The cut-type corridor for \(h_1=5\) and \(h_2\in\{50,100,150\}\) for \(T=1,\,25/3,\,20,\,40,\,60\).}
\label{fig:Fig_CutKoridor}
\end{figure}

At significantly higher depth ratios, the narrow cuts characteristic of moderate asymmetry transform into broad open corridors.
For \(h_1=5\) and \(h_2=50\), \(100\), and \(150\) (Figs.~\ref{fig:Fig_CutKoridor}a–c), several distinct corridors are observed, their structure depending on the magnitude of surface tension (\(T=1\), \(25/3\), \(20\), \(40\), and \(60\)), represented by solid, long-dashed, dashed, dash-dotted, and dotted lines, respectively.
The chosen five values of \(T\) play a key methodological role.
They cover the reference case \(T=1\) used throughout Part~I, the critical long-wave threshold \eqref{eq:Tast}
\[
T^{\ast}=\frac{h_1^2}{3}=\frac{25}{3}, \qquad \mathrm{Bo}^{\ast}=\frac{1}{3},
\]
and three progressively higher values \(T=20\), \(40\), and \(60\), which extend the analysis well into the strongly capillary regime.
This set allows one to track, within a single series of diagrams, the complete transition from gravity–capillary balance to capillary dominance and to identify the scaling of corridor width, shape, and position with increasing \(T\).

For \(h_2=50\) (Fig.~\ref{fig:Fig_CutKoridor}a) only three corridors are present (\(T=1\), \(25/3\), and \(20\));
for \(h_2=100\) (Fig.~\ref{fig:Fig_CutKoridor}b) four corridors appear (\(T=1\), \(25/3\), \(20\), \(40\));
and for \(h_2=150\) (Fig.~\ref{fig:Fig_CutKoridor}c) all five corridors corresponding to \(T=1\), \(25/3\), \(20\), \(40\), and \(60\) are clearly visible.
At \(T=1\) the corridor remains broad and nearly invariant with \(h_2\), while increasing \(T\) primarily affects the region of small~\(k\), where pronounced cuts develop in the central part of the \(\rho\)-range.
As \(T\) exceeds the critical value \(T^{\ast}\), the individual corridors converge toward the origin \((\rho,k)=(0,0)\), producing a cut-type topology.

This behaviour marks the transition from a mixed gravity–capillary regime to a purely capillary one.
For moderate surface tension (\(T<T^{\ast}\)), both restoring mechanisms act simultaneously, producing multiple corridors whose positions depend on the density ratio.
As \(T\) approaches \(T^{\ast}\), the capillary stress increasingly dominates, damping long-wave modulation and localizing the instability at short wavelengths.
The cut-type corridor thus corresponds to the limit in which the interface behaves as a nearly rigid capillary sheet bounding a deep lower layer: nonlinear effects persist only within a narrow density interval, and the unstable band detaches from the \(\rho\)-axis.
In this regime, the inertia of the lower layer provides the dominant response, while dispersion is governed exclusively by surface tension, resulting in the flattening and eventual disappearance of the loop.

\subsection{Loop and its interaction with the corridor}
\label{subsec_3.4}

From the second (Figs.~\ref{fig:Matrix}e–h), third (Figs.~\ref{fig:Matrix}i–l),
and fourth (Figs.~\ref{fig:Matrix}m–p) rows of the matrix diagrams
corresponding to \(h_1=5\), \(9\), and \(13\), respectively,
it is seen that at a lower value of surface tension (\(T={1}/{2}\))
both stability structures—the loop and the corridor—expand relative to the
reference case \(T=1\), while the corridor shifts rightward along the
\(\rho\)-axis.
Its lower boundary, however, never crosses the line \(\rho=1\),
approaching the point \((\rho,k)=(1,0)\) asymptotically as the
upper–layer thickness increases.
At a higher value of surface tension (\(T=2\)), both the loop and the corridor
contract and become narrower, and the corridor moves leftward toward smaller
density ratios.
In the complete matrix of stability maps (Fig.~\ref{fig:Matrix}),
the loop is observed in Figs.~\ref{fig:Matrix}e,~f,~i--k,~m--p;
in Fig.~\ref{fig:Matrix}f it remains isolated, whereas in the
other cases it touches the \(\rho\)-axis.

\begin{figure}[ht]
\centering

\hspace{-2ex}
\begin{subfigure}[b]{0.25\textwidth}
  \includegraphics[width=\linewidth]{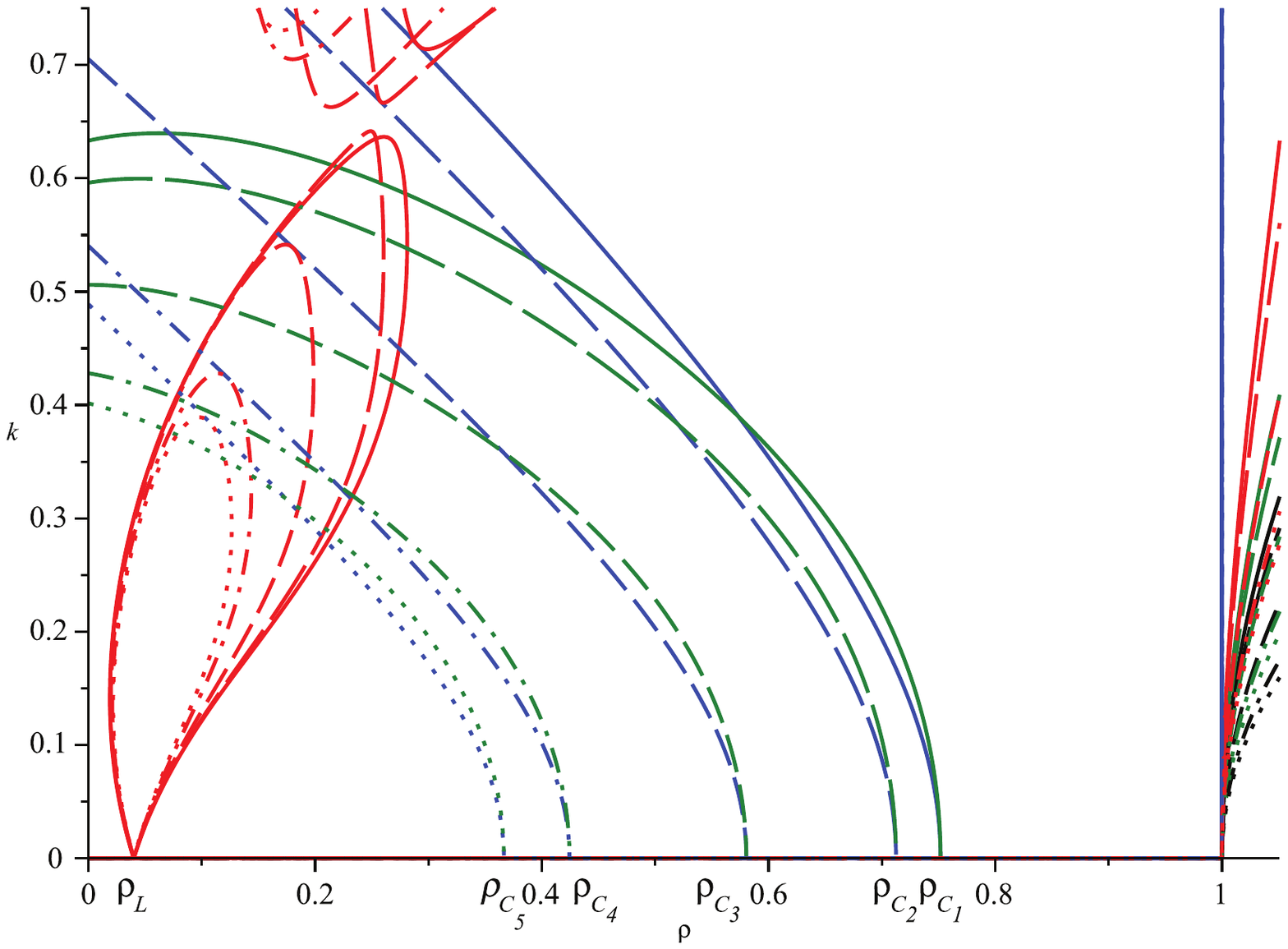}
  \caption{$h_1\!=\!5,\, h_2\!=\!1$}
  \label{fig:Fig3a}
\end{subfigure}\hspace{-1ex}
\begin{subfigure}[b]{0.25\textwidth}
  \includegraphics[width=\linewidth]{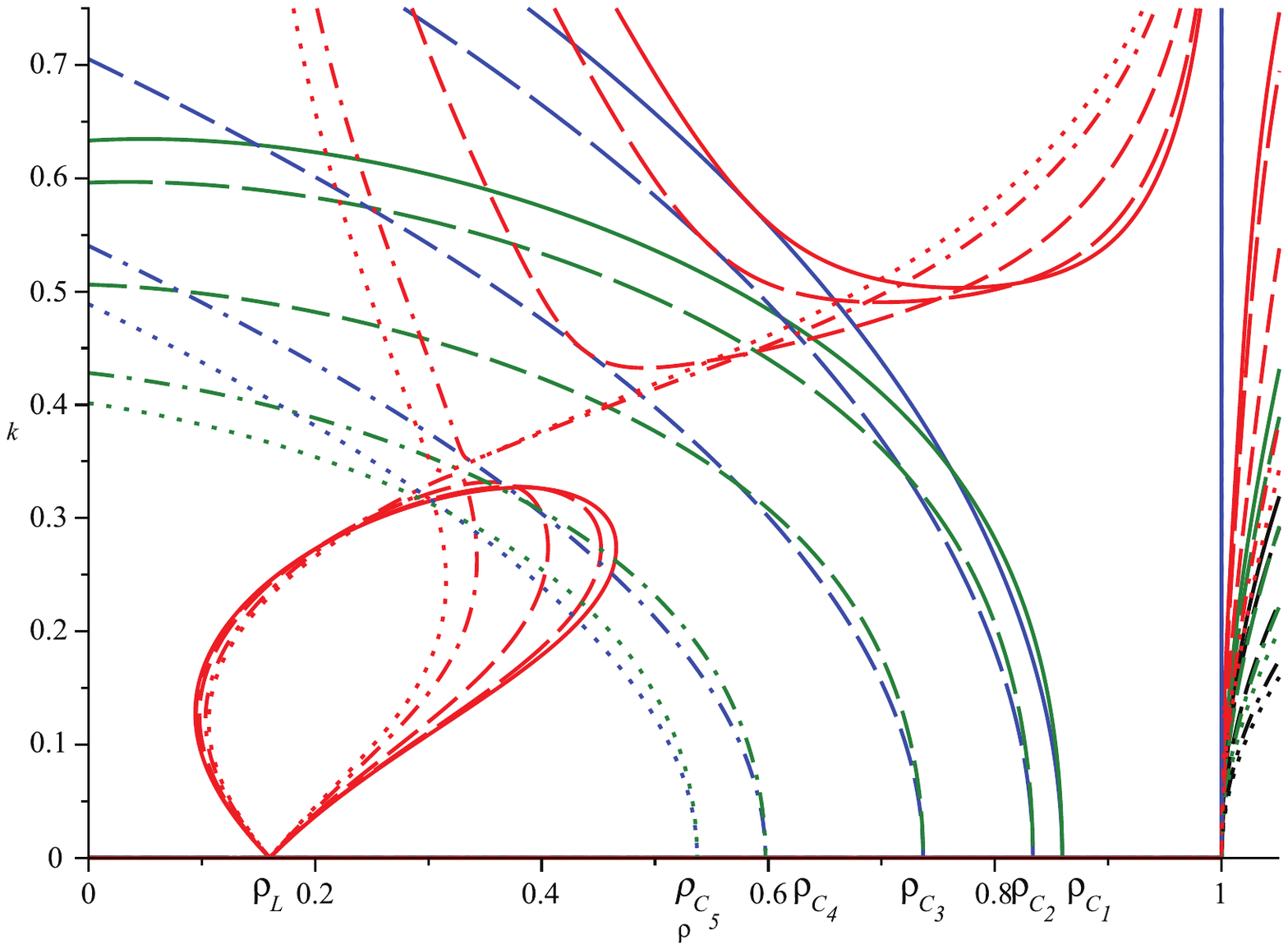}
  \caption{$h_1\!=\!5,\, h_2\!=\!2$}
  \label{fig:Fig3b}
\end{subfigure}\hspace{-1ex}
\begin{subfigure}[b]{0.25\textwidth}
  \includegraphics[width=\linewidth]{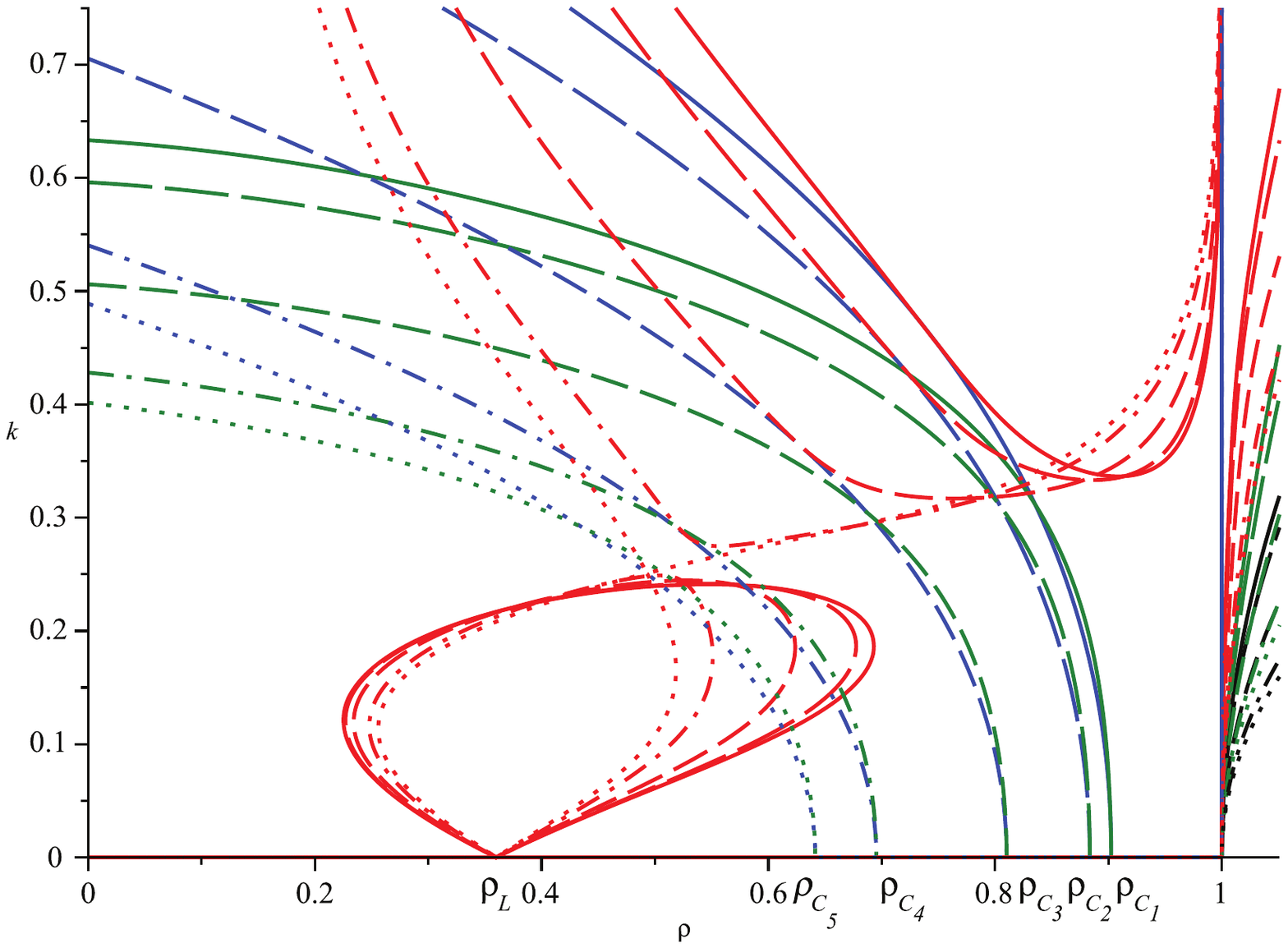}
  \caption{$h_1\!=\!5,\, h_2\!=\!3$}
  \label{fig:Fig3c}
\end{subfigure}\hspace{-1ex}
\begin{subfigure}[b]{0.25\textwidth}
  \includegraphics[width=\linewidth]{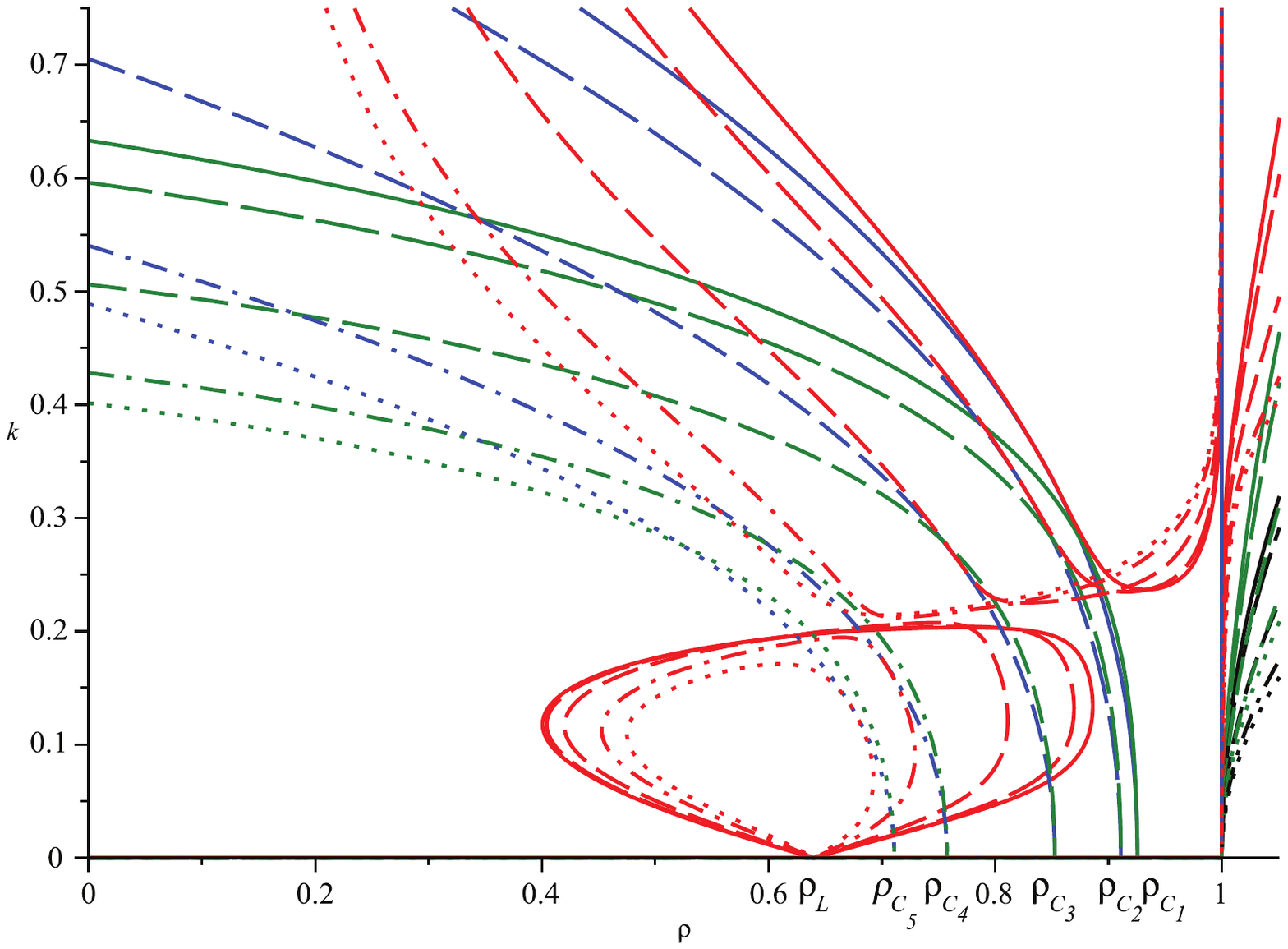}
  \caption{$h_1\!=\!5,\, h_2\!=\!4$}
  \label{fig:Fig3d}
\end{subfigure}

\vspace{0.2em}

\hspace{-2ex}
\begin{subfigure}[b]{0.25\textwidth}
  \includegraphics[width=\linewidth]{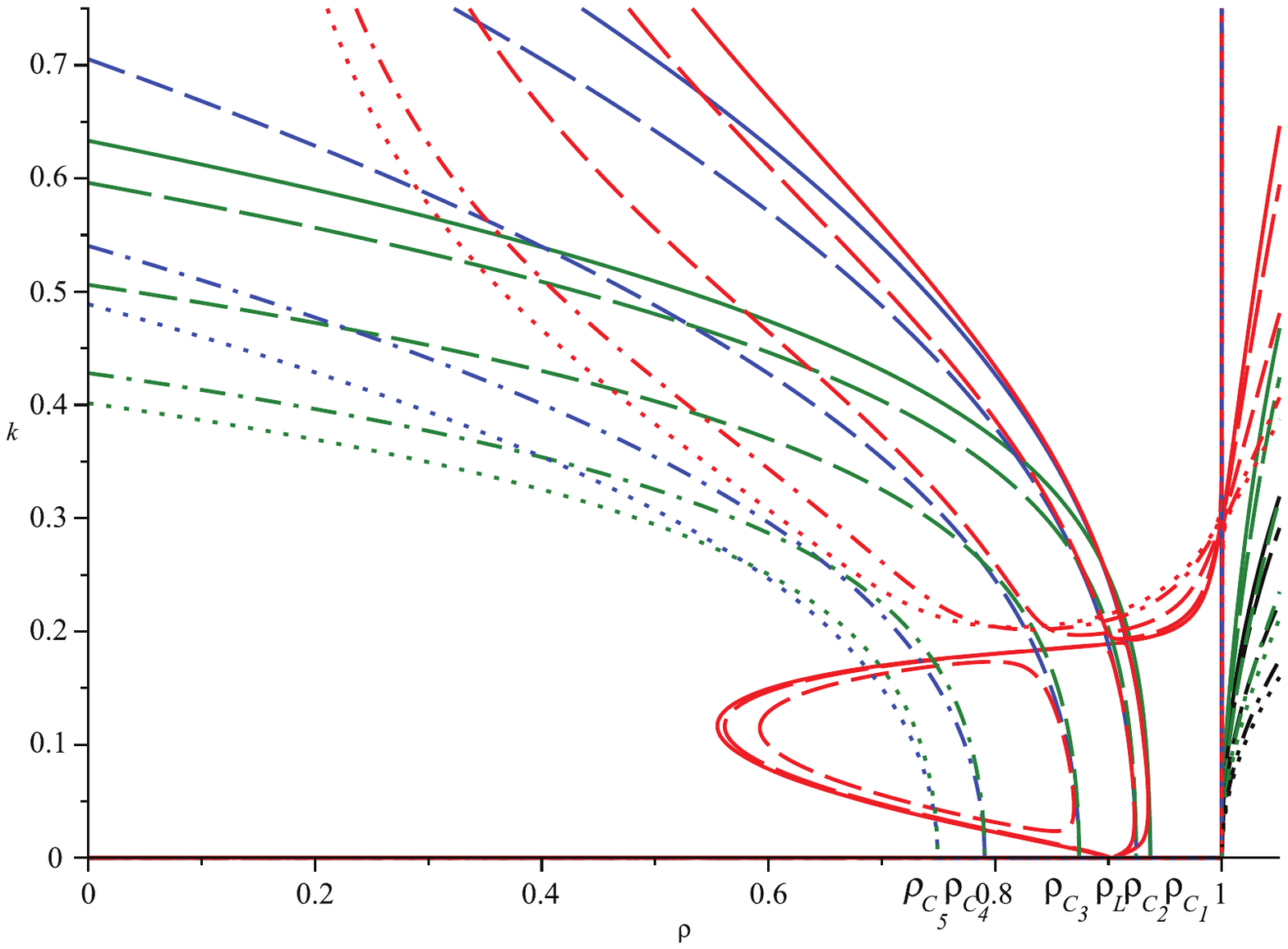}
  \caption{$h_1\!=\!5,\, h_2\!=\!4.75$}
  \label{fig:Fig3e}
\end{subfigure}\hspace{-1ex}
\begin{subfigure}[b]{0.25\textwidth}
  \includegraphics[width=\linewidth]{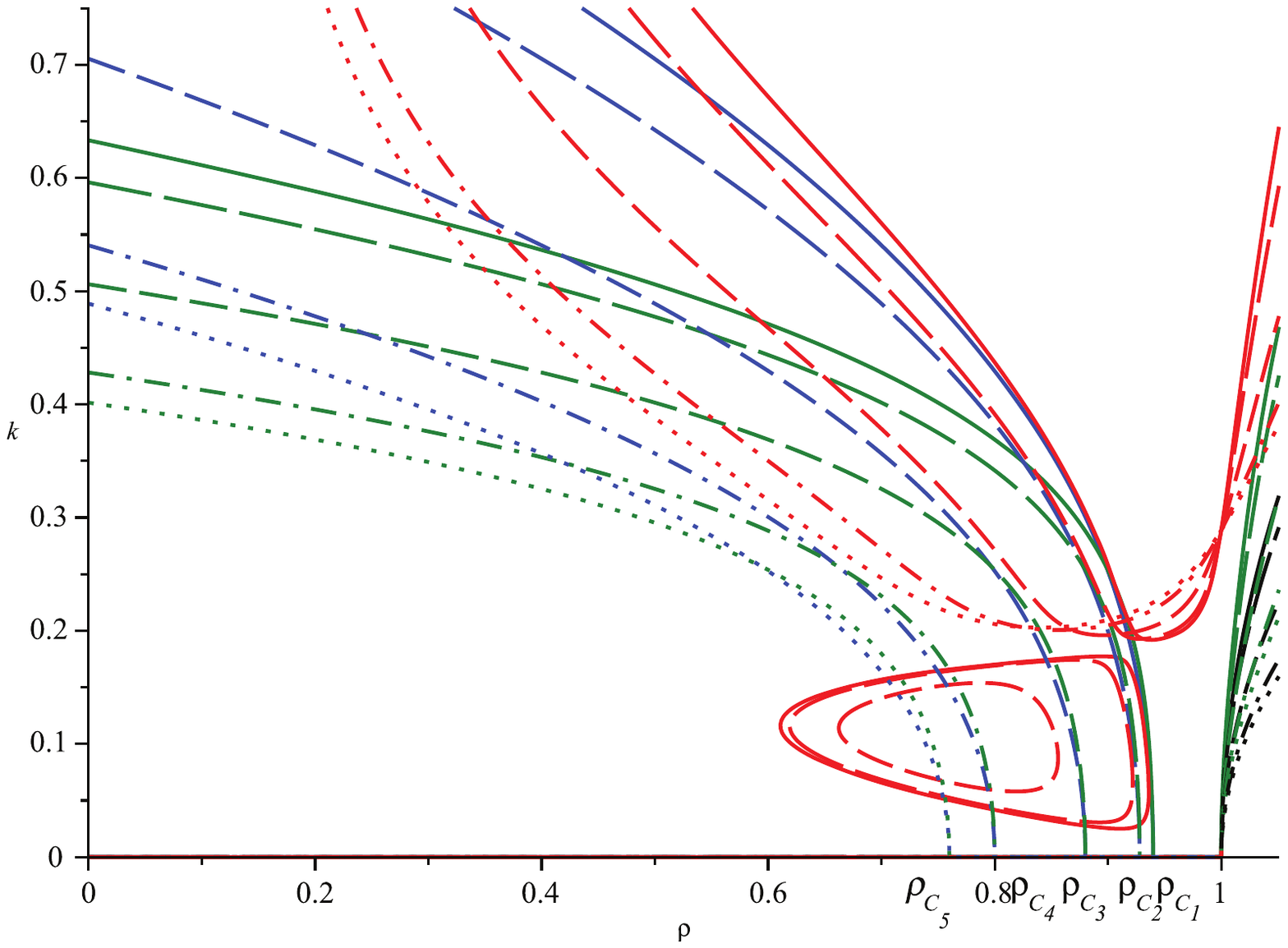}
  \caption{$h_1\!=\!5,\, h_2\!=\!5$}
  \label{fig:Fig3f}
\end{subfigure}\hspace{-1ex}
\begin{subfigure}[b]{0.25\textwidth}
  \includegraphics[width=\linewidth]{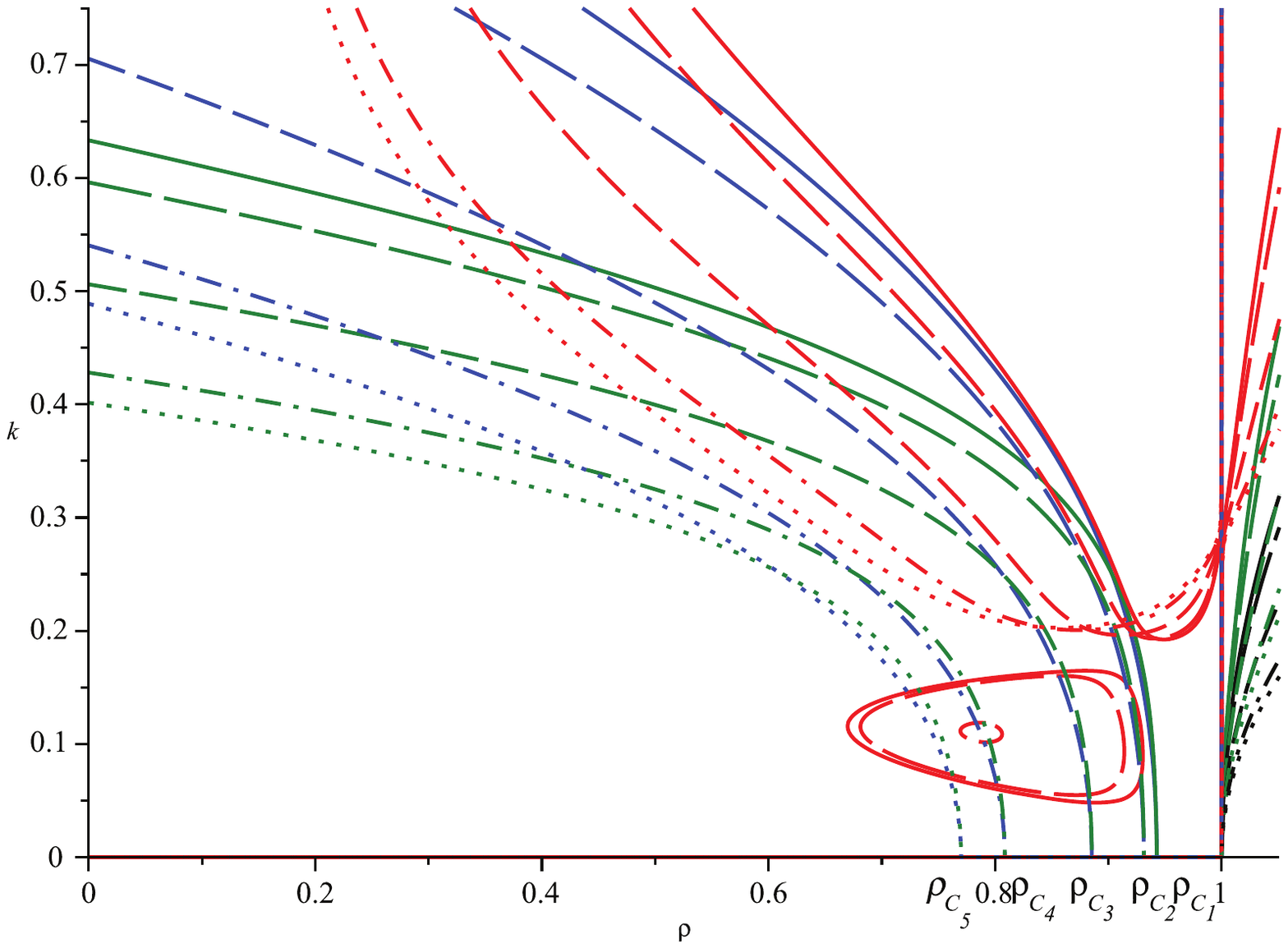}
  \caption{$h_1\!=\!5,\, h_2\!=\!5.25$}
  \label{fig:Fig3g}
\end{subfigure}\hspace{-1ex}
\begin{subfigure}[b]{0.25\textwidth}
  \includegraphics[width=\linewidth]{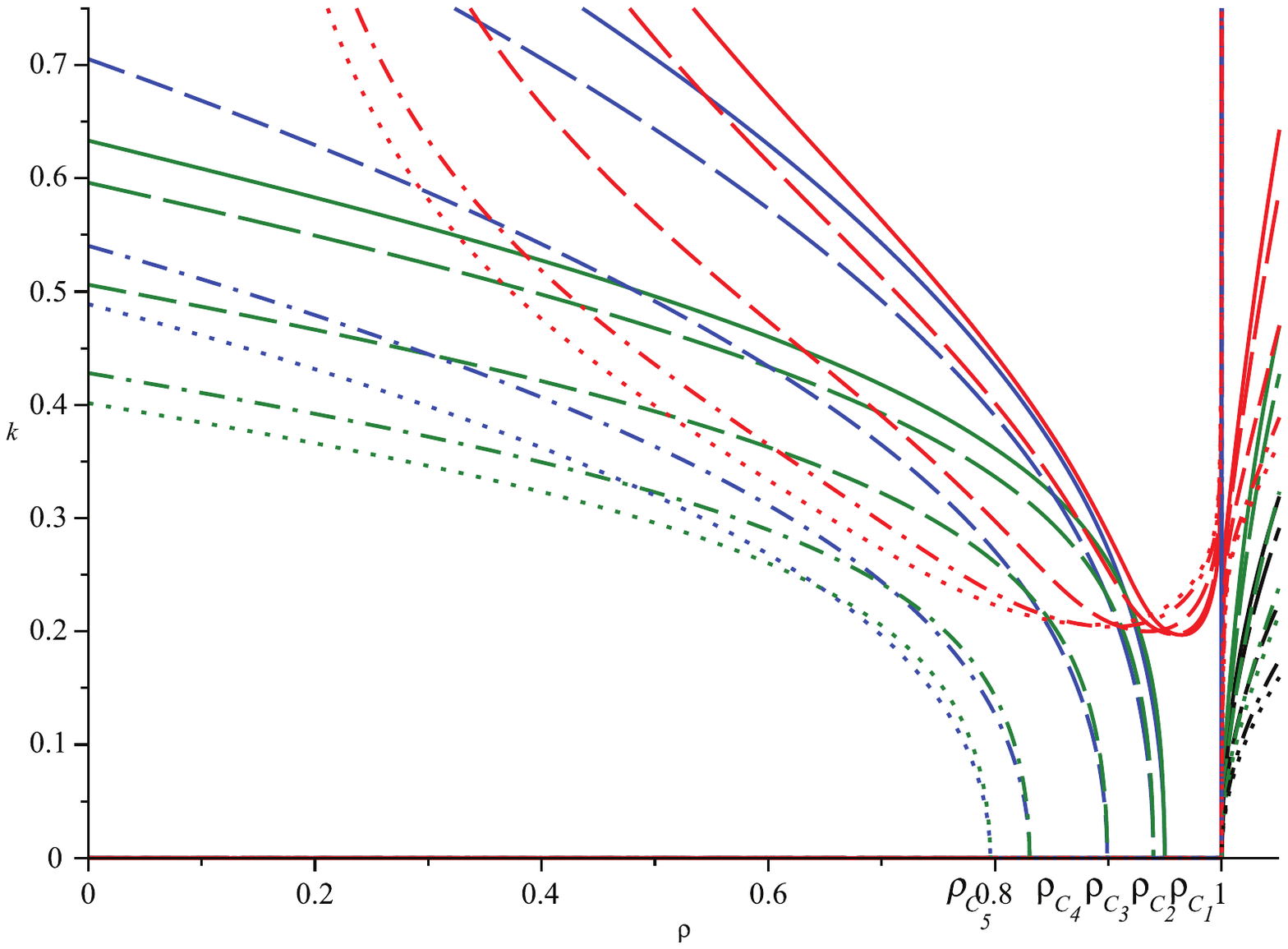}
  \caption{$h_1\!=\!5,\, h_2\!=\!6$}
  \label{fig:Fig3h}
\end{subfigure}

\caption{
Stability diagrams for surface–tension coefficients
\(T=1/2,\,3/5,\,1,\,5/3,\,2\)
for \(h_1=5\) and \(h_2\in\{1,2,3,4,4.75,5,5.25,6\}\).
}
\label{fig:Fig_LOOP}
\end{figure}

Figure~\ref{fig:Fig_LOOP} presents enlarged fragments of the
modulational–stability maps for fixed \(h_1=5\) and a sequence of
upper–layer thicknesses \(h_2=1,2,3,4,4.75,5,5.25,6\)
(Figs.~\ref{fig:Fig_LOOP}a–h), computed for five
values of surface tension,
\(T=1/2\), \(3/5\), \(1\), \(5/3\), and \(2\),
represented by solid, long-dashed, dashed, dash-dotted, and dotted lines.
Each panel shows the relative configuration of the loop and the
corridor near small wavenumbers~\(k\), illustrating how their positions
vary with both geometry and surface tension.
The loop–base point at \(k=0\) is determined solely by the depth ratio,
\(
\rho_{\mathrm L}={h_2^2}/{h_1^2},
\)
whereas the corridor–edge points
\(\rho_{\mathrm{C}_1}\)–\(\rho_{\mathrm{C}_5}\)
correspond to these five values of \(T\) and mark the roots of the
corridor branches on the horizontal axis,
that is, the intersections of the resonant (\(J\!\to\!\infty\))
and dispersive (\(\omega''\!=\!0\)) conditions at \(k\!\to\!0\).

For quantitative reference, the corresponding values of
\(\rho_{\mathrm L}\) and \(\rho_{\mathrm C}(T)\)
are summarized in Table~\ref{tab:rhoC_rhoL_LOOP}.

\begin{table}[ht]
\centering
\caption{Loop--base $\rho_{\mathrm L}$ and corridor--edge $\rho_{\mathrm C}(T)$
coordinates for five representative surface--tension values at fixed $h_1=5$.}
\label{tab:rhoC_rhoL_LOOP}
\vspace{4pt}
\begin{tabular}{c|cccccc}
\hline
$h_2$ &
$\rho_{\mathrm L}$ &
$\rho_{\mathrm{C}_1}=\rho_{\mathrm C}(1/2)$ &
$\rho_{\mathrm{C}_2}=\rho_{\mathrm C}(3/5)$ &
$\rho_{\mathrm{C}_3}=\rho_{\mathrm C}(1)$ &
$\rho_{\mathrm{C}_4}=\rho_{\mathrm C}(5/3)$ &
$\rho_{\mathrm{C}_5}=\rho_{\mathrm C}(2)$ \\
\hline
1.00  & 0.040 & 0.752 & 0.712 & 0.580 & 0.424 & 0.367 \\
2.00  & 0.160 & 0.859 & 0.833 & 0.737 & 0.597 & 0.537 \\
3.00  & 0.360 & 0.903 & 0.884 & 0.810 & 0.695 & 0.641 \\
4.00  & 0.640 & 0.926 & 0.911 & 0.853 & 0.758 & 0.711 \\
4.75  & 0.902 & 0.937 & 0.924 & 0.874 & 0.791 & 0.749 \\
5.00  & 1.000 & 0.940 & 0.928 & 0.880 & 0.800 & 0.760 \\
5.25  & 1.103 & 0.943 & 0.931 & 0.885 & 0.809 & 0.770 \\
6.00  & 1.440 & 0.950 & 0.940 & 0.899 & 0.831 & 0.796 \\
\hline
\end{tabular}
\end{table}

The inclusion of the loop–base coordinate $\rho_{\mathrm L}$ in Fig.~\ref{fig:Fig_LOOP}
clarifies the mutual placement of the loop and corridor edge points.
For all configurations, the corridor–edge coordinate $\rho_{\mathrm C}(T)$
(corresponding to $\rho_{\mathrm{C}_1}$–$\rho_{\mathrm{C}_5}$ in
Table~\ref{tab:rhoC_rhoL_LOOP}) decreases monotonically with increasing~$T$.
When $h_2<h_1$, the values $\rho_{\mathrm C}(T)$ lie strictly above the loop base
$\rho_{\mathrm L}$ and approach it from the right as $T$ increases, indicating a gradual
contraction of the corridor toward the loop.
In contrast, when $h_2>h_1$, the quantities $\rho_{\mathrm C}(T)$ remain strictly below
unity and therefore stay well separated from the loop–base coordinate
$\rho_{\mathrm L}>1$ for all admissible values of~$T$.
In this regime, the loop and corridor bases do not approach one another as the
surface tension varies; instead, their persistent separation reflects the dominance
of geometric asymmetry over the capillary adjustment of the resonant and dispersive
branches.

As shown in Figs.~\ref{fig:Fig_LOOP}(a–d),
for \(h_2=1\)–4 the loop–base coordinate
\(\rho_{\mathrm L}=h_2^2/25\) remains considerably smaller than
the corresponding corridor–edge values
\(\rho_{\mathrm{C}_1}\)–\(\rho_{\mathrm{C}_5}\)
for all surface–tension magnitudes listed in
Table~\ref{tab:rhoC_rhoL_LOOP}.
Hence, the corridor always originates to the right of the loop base,
satisfying \(\rho_{\mathrm C}(T)>\rho_{\mathrm L}\).
At small surface tension (\(T=1/2\)) the corridor widens and its
right edge extends well beyond \(\rho_{\mathrm L}\), forming a broad
composite loop–corridor domain.
As \(T\) increases, the corridor contracts and shifts leftward as
\(\rho_{\mathrm C}(T)\) approaches \(\rho_{\mathrm L}\),
and the loop correspondingly narrows.
In the most asymmetric case (\(h_2=1\)) the loop remains detached from
the \(\rho\)-axis, whereas for \(h_2=2\)–4 it touches the \(O\rho\)-axis
and gradually enlarges with increasing \(h_2\).
This behaviour indicates that a thicker upper layer increases the
effective inertia above the interface, thereby strengthening the
dispersive response and widening the interval of nonlinear–dispersive
balance at small~\(k\).

Figs.~\ref{fig:Fig_LOOP}(e,\,f) correspond to nearly symmetric
configurations (\(h_2\simeq h_1\)), where the loop–base point
\(\rho_{\mathrm L}\) approaches unity and the corridor–edge coordinates
\(\rho_{\mathrm{C}_i}\) (\(i=1\!-\!5\)) remain slightly below~1
for all values of~\(T\).
For \(h_2=4.75\), one finds \(\rho_{\mathrm L}=0.903\) and
\(\rho_{\mathrm{C}_1}=0.937\), indicating that at small~\(T\)
the corridor begins slightly to the right of the loop base, forming a
large loop resting on the \(\rho\)-axis.
As surface tension increases, the edge positions
\(\rho_{\mathrm{C}_i}\) shift left of \(\rho_{\mathrm L}\)
(e.g., for \(T=2\), \(\rho_{\mathrm{C}_5}=0.749\)),
and the loop contracts.
In the fully symmetric case \(h_2=5\), where \(\rho_{\mathrm L}=1\) and
all \(\rho_{\mathrm{C}_i}<1\), no vertical asymptote appears at
\(\rho=1\), and the loop always rests on the \(\rho\)-axis at \((1,0)\).
The corridor base lies to the left of this point and moves further left
as \(T\) increases, showing that decreasing surface tension enlarges the
corridor and enhances its overlap with the loop, whereas higher~\(T\)
suppresses this interaction.
The convergence of \(\rho_{\mathrm L}\) and \(\rho_{\mathrm{C}_i}\)
toward unity with increasing total depth explains the near coincidence
of the loop and corridor bases in these quasi–symmetric configurations.

For \(h_2>h_1\) (Figs.~\ref{fig:Fig_LOOP}(g,\,h)),
the loop–base coordinate satisfies \(\rho_{\mathrm L}>1\), while
\(\rho_{\mathrm{C}_i}<1\) for all~\(T\)
(see Table~\ref{tab:rhoC_rhoL_LOOP}).
Thus, the corridor originates to the left of the loop base, and their
interaction depends on whether the corridor’s right boundary at finite~\(k\)
extends beyond \(\rho_{\mathrm L}\).
For slightly asymmetric configurations (\(h_2\) close to \(h_1\);
Fig.~\ref{fig:Fig_LOOP}(g)), this occurs only at small~\(T\), when
\(\rho_{\mathrm{C}_1}\) is closest to unity and the corridor is widest;
under these conditions the loop persists in a degenerate form without
reaching the \(\rho\)-axis.
As the upper layer becomes thicker (\(h_2\) further increases;
Fig.~\ref{fig:Fig_LOOP}(h)), even such degenerate loops no longer
appear, and only the corridor structure remains.

These geometric dependences outline the quantitative trends governing
the interaction between the loop and the corridor under varying surface
tension. The corresponding physical interpretation is summarized in
Sec.~\ref{subsec_3.5}.

\subsection{Summary and physical interpretation}
\label{subsec_3.5}

The contrasting effects of surface tension in the two limiting geometries—
loop formation in the HS–La system and cut development in the La–HS system—
previously described in \cite{AvramenkoNarad2025,AvramenkoNarad2026},
are consistently reproduced within the present unified two-layer framework.
When the upper layer becomes thicker than the lower one, the interfacial mode
acquires a predominantly capillary character: the restoring force is governed
by surface tension rather than gravity, and increasing \(T\) enhances this dominance,
narrowing the parameter range in which nonlinearity and dispersion compensate.
Consequently, the loop shrinks and may disappear, leaving only the narrow stability corridor.

The joint consideration of \(\rho_{\mathrm L}\) and \(\rho_{\mathrm C}(T)\)
clarifies the structural transitions summarized in Fig.~\ref{fig:Fig_LOOP}.
For \(h_2<h_1\) and small \(T\), the corridor originates to the right of the loop base
(\(\rho_{\mathrm C}(T)>\rho_{\mathrm L}\)) and overlaps it, forming an extended combined domain.
As \(T\) increases, a crossing value \(T_{\times}\) (see Sec.~\ref{subsec_2.3})
is reached where \(\rho_{\mathrm C}(T_{\times})=\rho_{\mathrm L}\);
beyond this point (\(h_2\ge h_1\) or large \(T\))
the corridor shifts leftward and separates from the loop, which becomes isolated or vanishes.
Surface tension therefore redistributes the nonlinear balance at the interface:
reducing \(T\) weakens capillary stiffness and broadens the region of near-compensation,
whereas increasing \(T\) confines stability to narrow intervals of~\(\rho\) and~\(k\).
In the geometric limit \(h_2\!\to\!h_1\), \(T\!\to\!0\),
one obtains \(\rho_{\mathrm C}(T)\!\to\!1^{-}\) and \(\rho_{\mathrm L}\!\to\!1^{+}\),
and both bases merge at \((\rho,k)=(1,0)\),
corresponding to two hydrodynamic half–spaces.

A three–dimensional perspective of these transitions
is presented in Sec.~\ref{sec_3D},
which shows how loops, corridors, and cuts merge or collapse
as the surface–tension magnitude and the depth ratio \(h_2/h_1\) increase.

\section{Three-dimensional critical surfaces \(T(\rho,k)\)}
\label{sec_3D}

\subsection{Analytical framework and Bond-threshold structure}
\label{subsec_4.1}

A comprehensive understanding of the modulational–stability topology
can be achieved by extending the two-dimensional maps of
Sec.~\ref{sec_Matrix} into the full three-dimensional parameter space
\((\rho,k,T)\).
The stability diagrams discussed earlier represent planar sections of
this space at fixed \(T\), showing how the upper stability region,
the corridor, and the loop transform as the surface-tension coefficient varies.
Here these features are interpreted as intersections of continuous
critical surfaces defined by
\[
J(\rho,k;h_1,h_2,T)=0, \qquad
J(\rho,k;h_1,h_2,T)\to\infty, \qquad
\omega''(\rho,k;h_1,h_2,T)=0.
\]
The first surface (\(J=0\), red) separates focusing and defocusing
nonlinearities, the second (\(J\to\infty\), blue) corresponds to the
resonant singularity, and the third (\(\omega''=0\), green) marks the
change of sign of group-velocity dispersion.
Their mutual intersections delineate the full topology of
modulationally stable and unstable domains in the \((\rho,k,T)\) space.

It is useful to distinguish between the long-wave critical value of
surface tension and the full three-dimensional critical manifold.
The long-wave Bond threshold \(T^{\ast}\) is a scalar quantity: it
characterises the value of \(T\) at which the curvature
\(\omega''\) vanishes in the limit \(k\to0\), marking the transition
between gravity- and capillarity-dominated dispersion.
By contrast, the function \(T(\rho,k)\) defines a two-dimensional
surface in the \((\rho,k,T)\) space that collects all loci where the
nonlinear–dispersive balance is critical.
Planar cuts of this surface at fixed \(T\) reproduce the neutral curves
that bound modulationally stable and unstable regions in the
\((\rho,k)\) plane.

According to expression~\eqref{eq:Tast} in Subsec.~\ref{subsec_2.4},
the long-wave Bond threshold
\[
T^{\ast}= h_1^2/3, \qquad \mathrm{Bo}^{\ast}=1/3,
\]
is the unique value at which the resonant condition \(J\to\infty\)
and the dispersive condition \(\omega''=0\) meet at the origin
\((\rho,k)=(0,0)\).
In the three-dimensional space \((\rho,k,T)\), this coincidence
generates a nodal line along which the resonant (blue) and dispersive
(green) surfaces intersect.

Although the resonant and dispersive surfaces lie close to one another
over wide regions of parameter space for many values of \(T\),
their topology remains distinct for \(T<T^{\ast}\), and intersections
with the plane \(T=\mathrm{const}\) generate separate loop- and
corridor-type structures in the \((\rho,k)\) plane.
At the Bond-critical value \(T=T^{\ast}\), the two surfaces undergo a
local degeneracy at \((\rho,k)=(0,0)\), where they meet at the same
height and share a common tangent.
For \(T>T^{\ast}\), they separate again while remaining geometrically
close, and their intersections with planes \(T=\mathrm{const}\)
produce the characteristic cut-type configurations.

In addition to the Bond threshold \(T^{\ast}\), a second geometric
degeneracy is associated with the depth ratio. As shown in
Subsec.~\ref{subsec_2.5}, the critical surface \(T(\rho,k)\) attains a
horizontal tangent at \((\rho,k,T)=(0,0,T^{\ast})\) when the layer
depths satisfy the golden–ratio proportion \(h_{2}/h_{1}=\varphi\).
The consequences of this geometric condition become evident in the
three–dimensional structure discussed in Subsec.~\ref{subsec_4.2}, where
its impact on the critical surfaces is examined in detail.

\begin{figure}
\centering
\vspace{-10ex}

\hspace{-1ex}
\begin{subfigure}[b]{0.32\textwidth}
  \includegraphics[width=\linewidth]{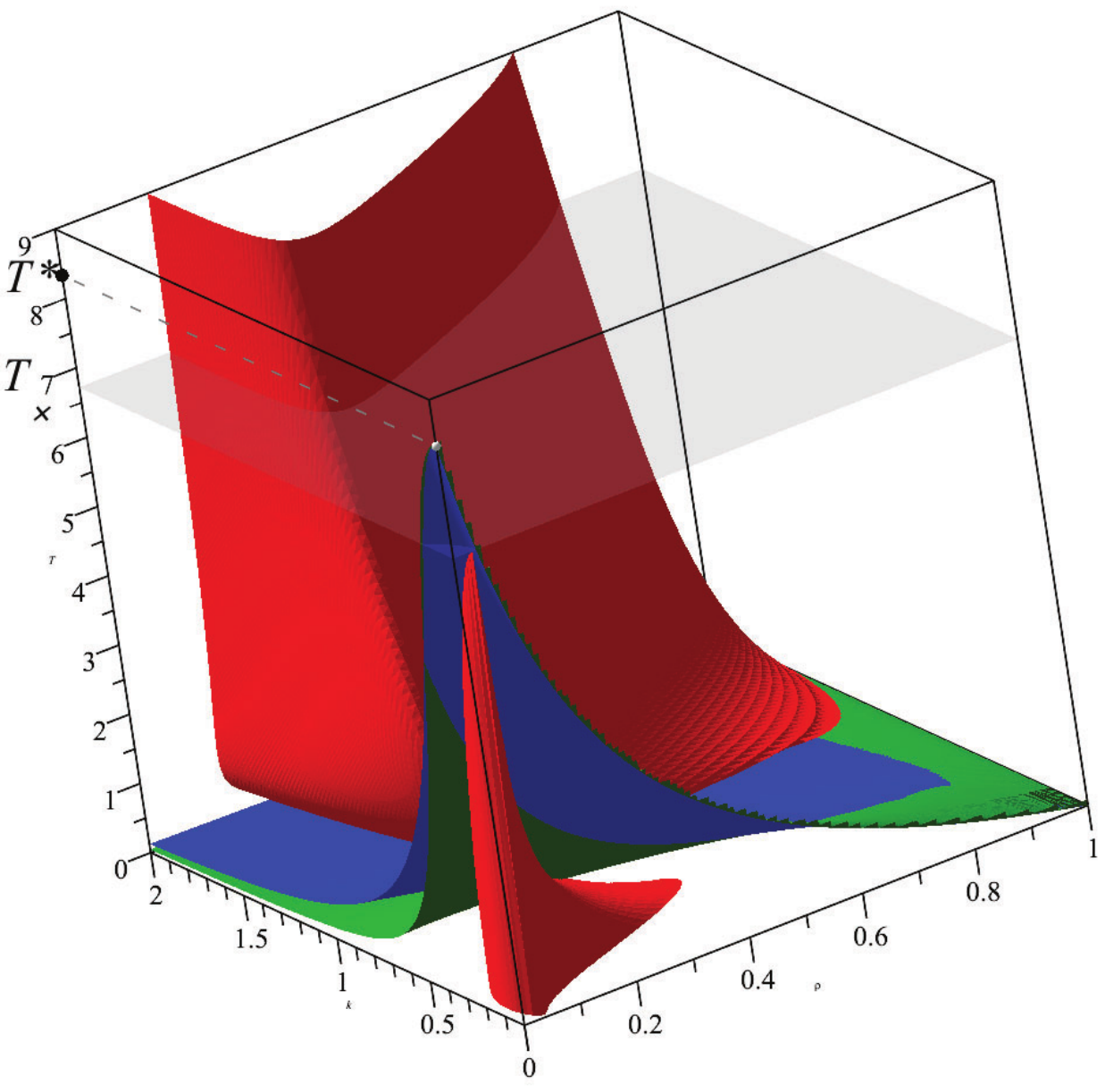}
  \caption{$h_1\!=\!5,\, h_2\!=\!1$}
  \label{fig:Fig4a}
\end{subfigure}\hspace{0ex}
\begin{subfigure}[b]{0.32\textwidth}
  \includegraphics[width=\linewidth]{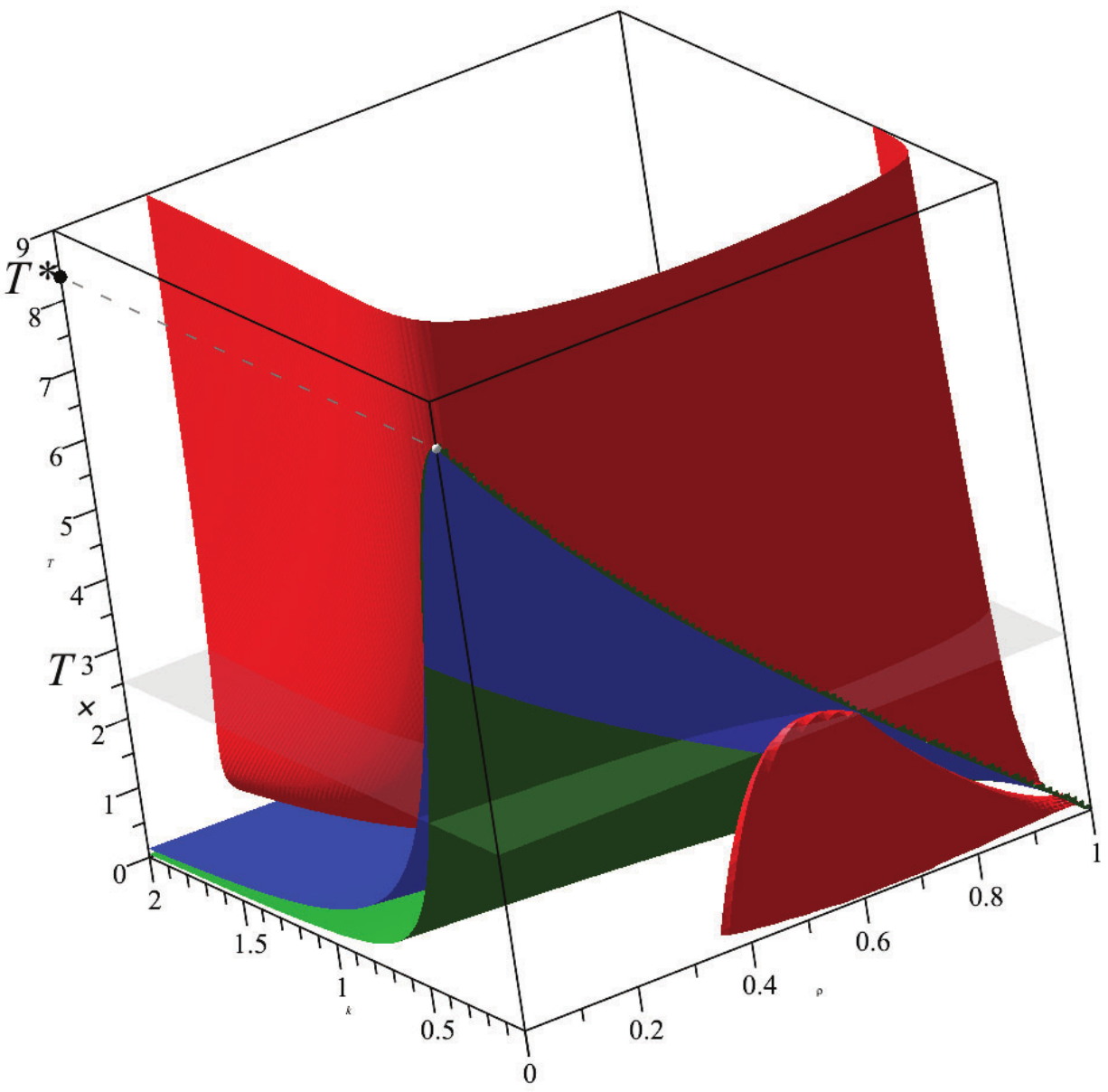}
  \caption{$h_1\!=\!5,\, h_2\!=\!4$}
  \label{fig:Fig4b}
\end{subfigure}\hspace{0ex}
\begin{subfigure}[b]{0.32\textwidth}
  \includegraphics[width=\linewidth]{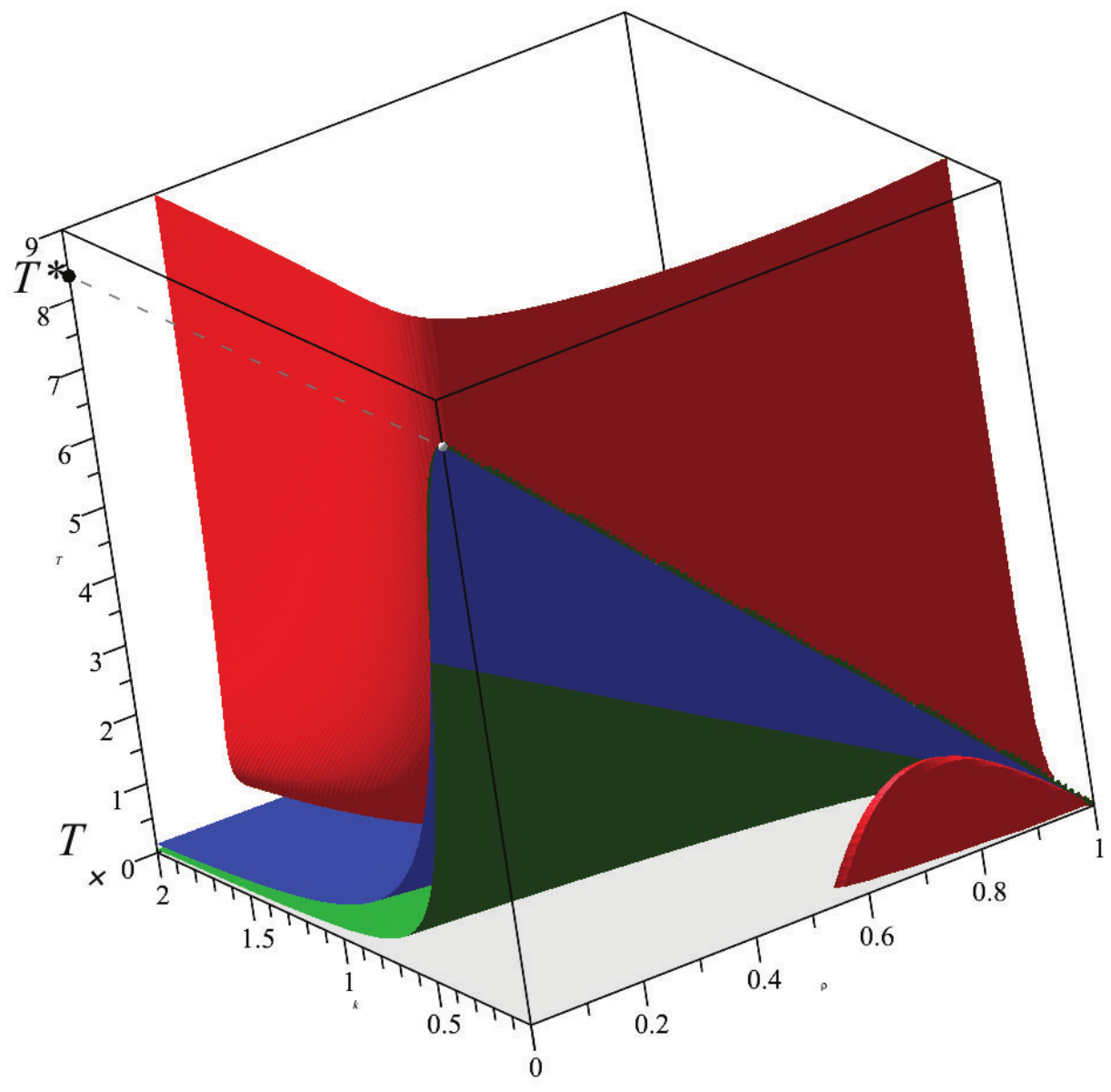}
  \caption{$h_1\!=\!5,\, h_2\!=\!5$}
  \label{fig:Fig4c}
\end{subfigure}

\vspace{-5ex}

\begin{subfigure}[b]{0.32\textwidth}
  \includegraphics[width=\linewidth]{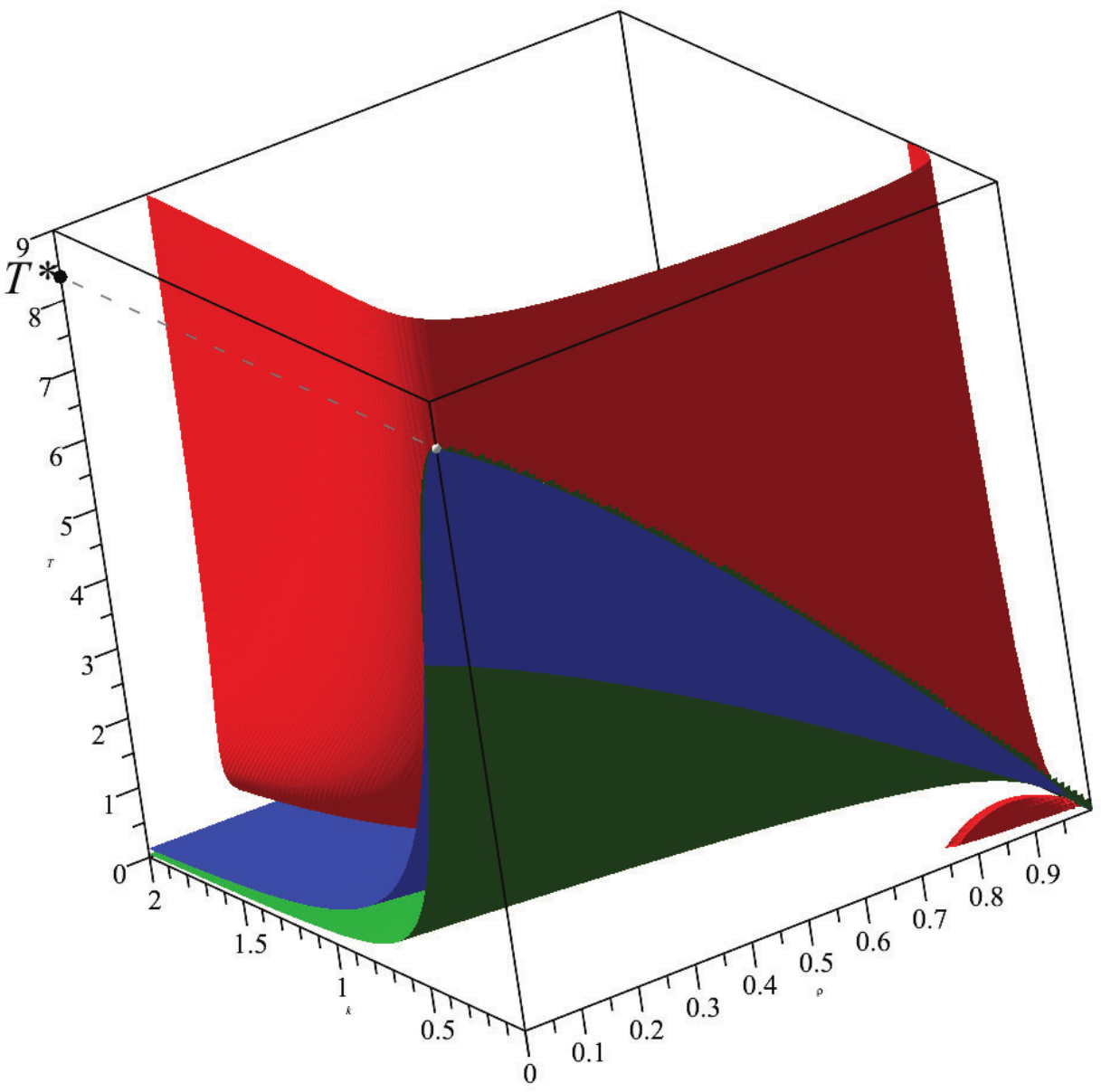}
  \caption{$h_1\!=\!5,\, h_2\!=\!6$}
  \label{fig:Fig4d}
\end{subfigure}\hspace{0ex}
\begin{subfigure}[b]{0.32\textwidth}
  \includegraphics[width=\linewidth]{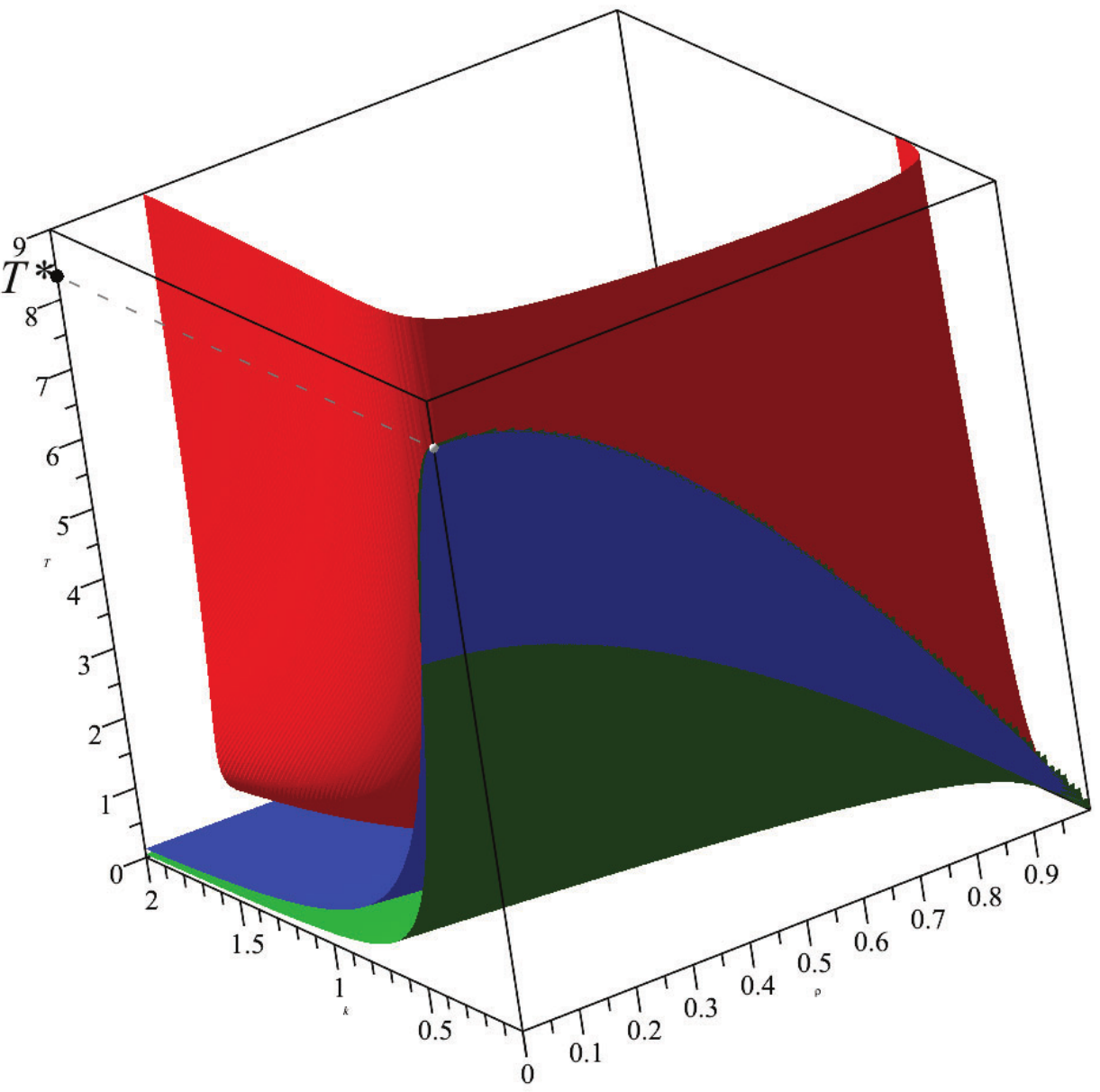}
  \caption{$h_1\!=\!5,\, h_2\!=5(1+\sqrt{5})/{2}\approx8.09\!$}
  \label{fig:Fig4e}
\end{subfigure}\hspace{0ex}
\begin{subfigure}[b]{0.32\textwidth}
  \includegraphics[width=\linewidth]{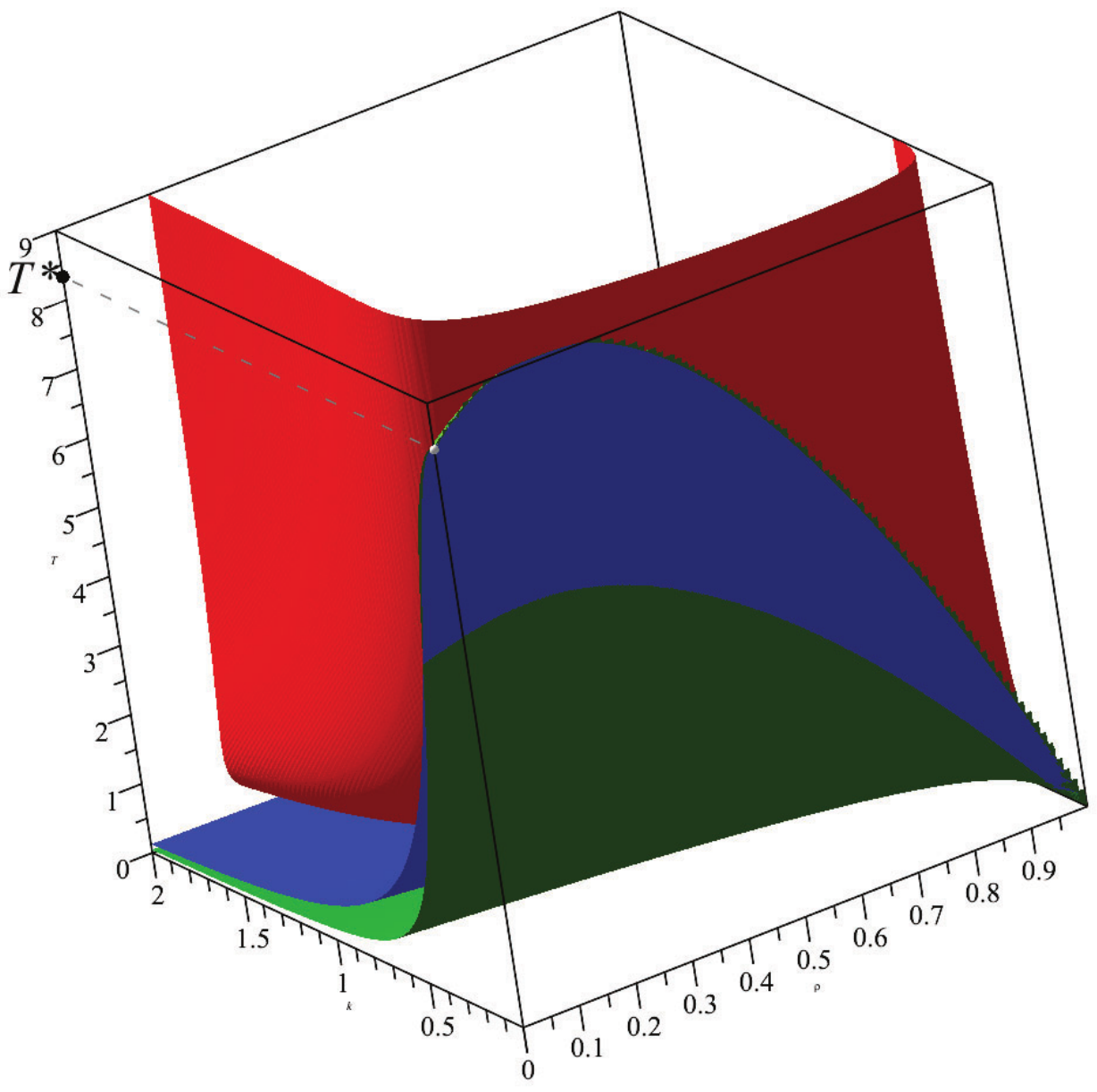}
  \caption{$h_1\!=\!5,\, h_2\!=\!12$}
  \label{fig:Fig4f}
\end{subfigure}

\hspace{-0ex}
\begin{subfigure}[b]{0.37\textwidth}
  \includegraphics[width=\linewidth]{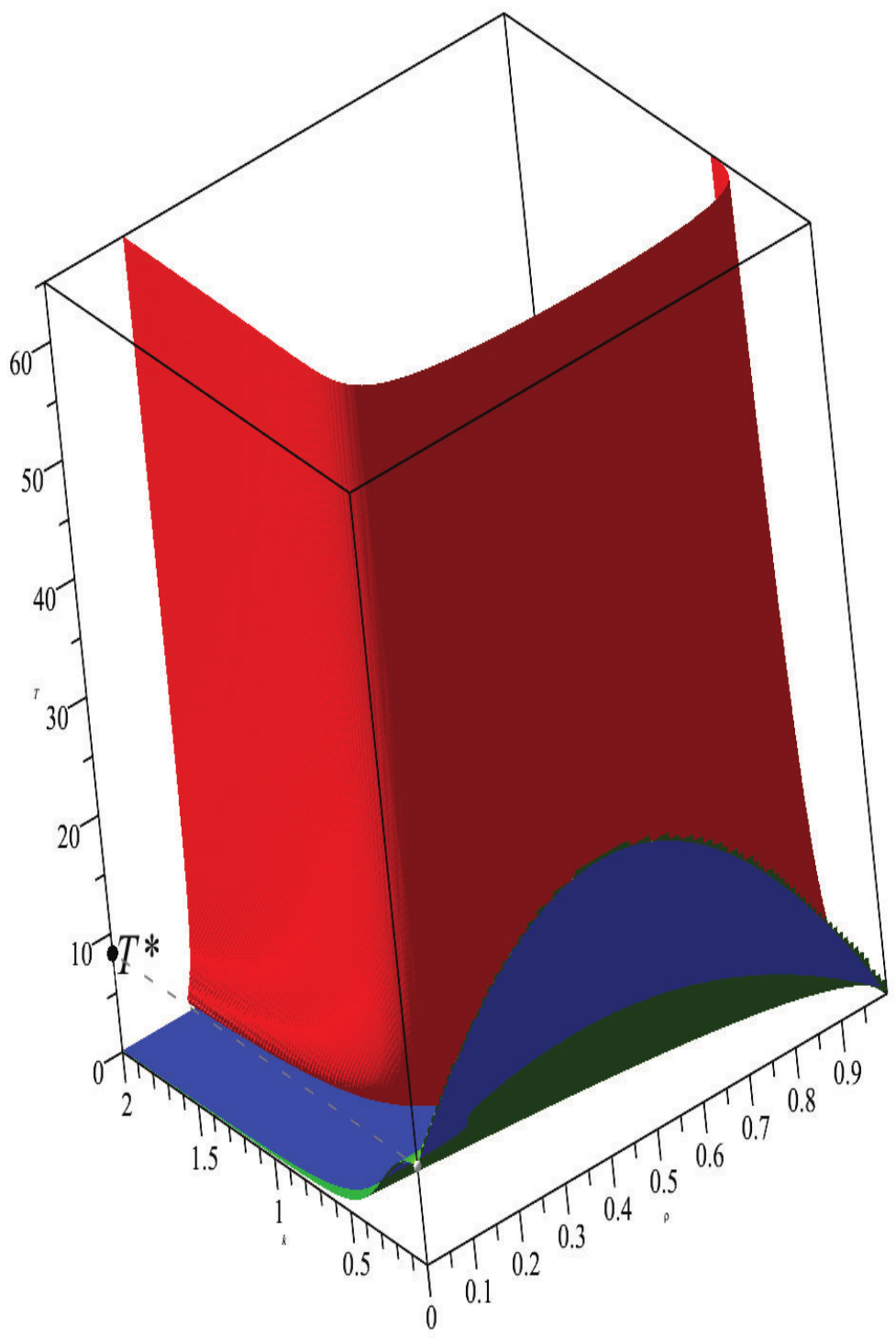}
  \caption{$h_1\!=\!5,\, h_2\!=\!50$}
  \label{fig:Fig4g}
\end{subfigure}
\hspace{-8ex}
\begin{subfigure}[b]{0.37\textwidth}
  \includegraphics[width=\linewidth]{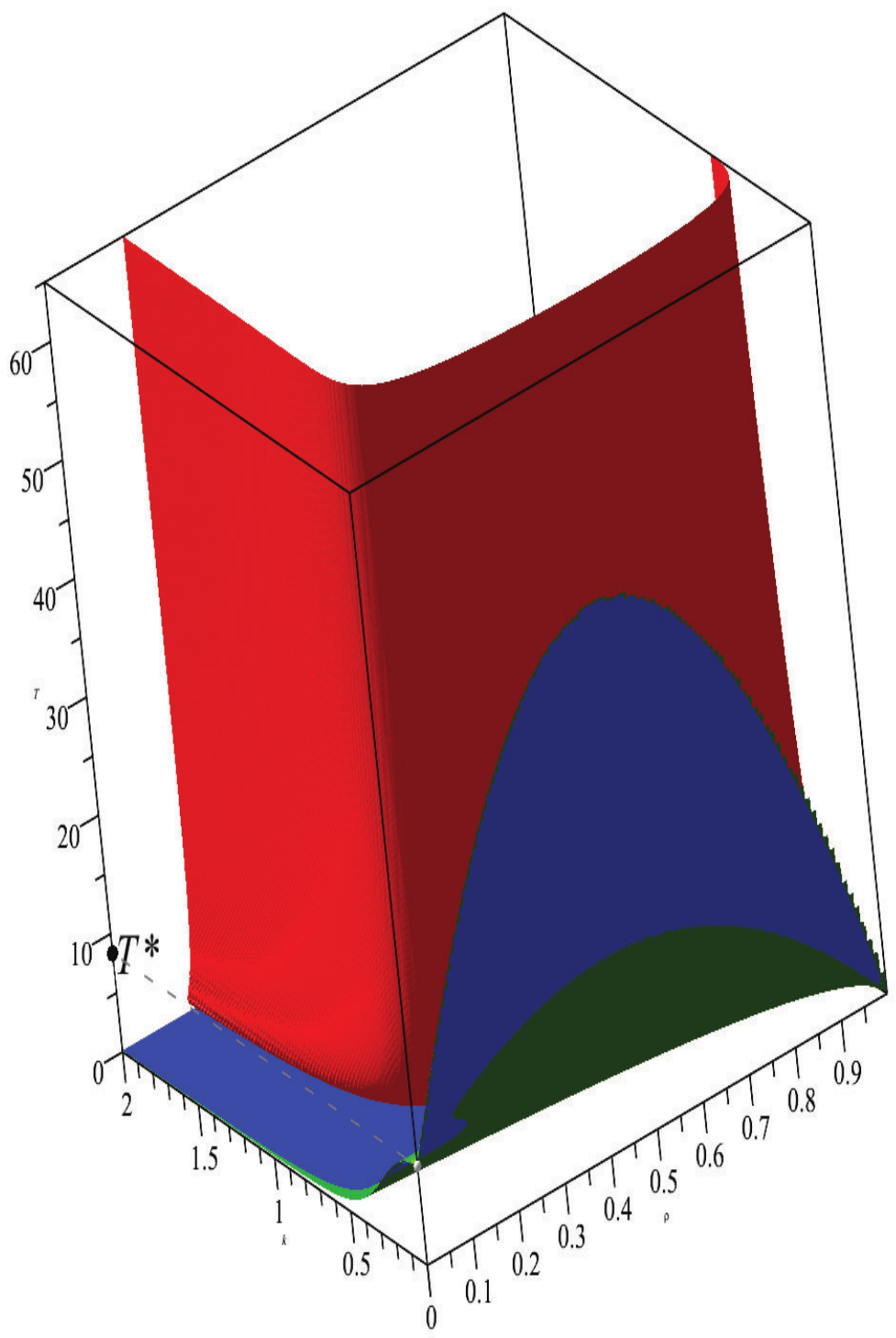}
  \caption{$h_1\!=\!5,\, h_2\!=\!100$}
  \label{fig:Fig4h}
\end{subfigure}
\hspace{-8ex}
\begin{subfigure}[b]{0.37\textwidth}
  \includegraphics[width=\linewidth]{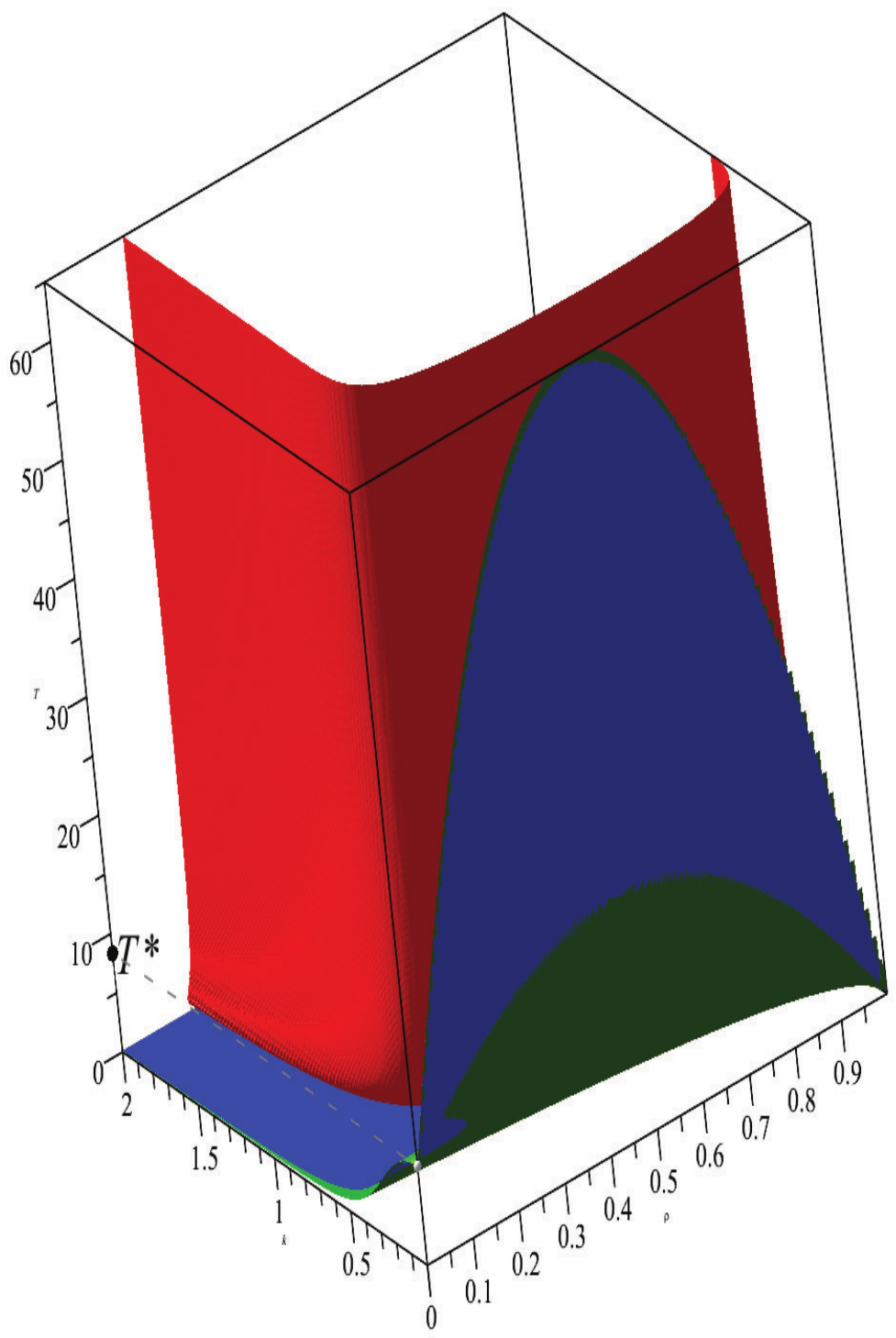}
  \caption{$h_1\!=\!5,\, h_2\!=\!150$}
  \label{fig:Fig4i}
\end{subfigure}

\caption{Critical surfaces \(T(\rho,k)\) for \(h_1=5\) and
\(h_2\in\{1,4,5,6,5(1+\sqrt{5})/{2},12,50,100,150\}\).}
\label{fig:3D}
\end{figure}

\subsection{Geometry of the critical surfaces and symmetry effects}
\label{subsec_4.2}

Figure~\ref{fig:3D} shows the critical surfaces for the fixed lower-layer
depth \(h_{1}=5\) and a sequence of upper-layer thicknesses
\(h_{2}\in\{1,4,5,6,5(1+\sqrt{5})/2,12,50,100,150\}\).
This set spans the transition from configurations with a thin upper layer
to those in which the upper layer is effectively deep.

For \(h_{2}=1\), the system corresponds to the HS–La limit, and the
nonlinear surface \(J=0\) forms a pronounced loop near the origin
\cite{AvramenkoNarad2025}.
The case \(h_{2}=4\) represents moderate asymmetry, whereas \(h_{2}=5\)
gives the symmetric configuration discussed in Subsec.~\ref{subsec_2.2}.
Increasing the upper-layer thickness to \(h_{2}=6\) introduces weak
asymmetry.
The value \(h_{2}=5(1+\sqrt{5})/2\) realises the golden–ratio depth ratio,
at which the critical surface \(T(\rho,k)\) has a horizontal tangent at
\((\rho,k,T)=(0,0,T^{\ast})\); see Subsec.~\ref{subsec_2.5}.
Larger values \(h_{2}=12,50,100,150\) correspond to progressively deeper
upper layers and approach the La–HS regime, in which the dispersive and
resonant surfaces nearly coincide
(see Subsec.~\ref{subsec_3.3} and \cite{AvramenkoNarad2026}).

In every panel the point \((\rho,k,T)=(0,0,T^{\ast})\) appears as the
intersection of the dispersive (\(\omega''=0\)) and resonant
(\(J\to\infty\)) surfaces.
For \(h_{1}=5\), its height is \(T^{\ast}=25/3\), in agreement with the
analytic Bond threshold \(T^{\ast}=h_{1}^{2}/3\) obtained in
Subsec.~\ref{subsec_2.4}.

In panels (a)–(f) of Fig.~\ref{fig:3D} the axes are scaled identically,
which facilitates comparison of curvature and elevation. Panels (g)–(i)
use an enlarged vertical scale that highlights the geometry of the
critical surfaces for large values of \(h_2\). 
These three-dimensional
plots directly extend the detailed two-dimensional sections shown in
Fig.~\ref{fig:Fig_CutKoridor} and analysed in Subsec.~\ref{subsec_3.3}.
In particular, the configurations with \(h_2=50\), \(100\), and \(150\)
correspond to planar diagrams where three, four, and five cut-type
corridors were identified for the respective sets of surface-tension
values \(T=\{20\}\), \(T=\{20,40\}\), and \(T=\{20,40,60\}\). The 3D
representation thus visualises how these cuts arise as intersections of
the critical surfaces \(J\to\infty\) and \(\omega''=0\) at the
corresponding heights in \(T\).

For small and moderate asymmetry (Figs.~\ref{fig:3D}a–d), the red surface
\(J=0\) forms a convex sheet that rises above the \((\rho,k)\)-plane at
small \(T\), while the blue and green surfaces remain well separated.
Sections at fixed \(T\) reproduce the loops and corridors identified in
the two-dimensional matrix of Subsec.~\ref{subsec_3.2}. The closed
intersections of \(J=0\) correspond to loop-type stability islands,
whereas the region between \(J\to\infty\) and \(\omega''=0\) forms the
open corridor separating upper and lower instability zones. These
surfaces therefore represent the three-dimensional extensions of the
characteristic curves forming the upper unstable region and the corridor
in the planar maps of Subsecs.~\ref{subsec_3.2}–\ref{subsec_3.4}.

In panels (a)–(c) of Fig.~\ref{fig:3D}, a grey horizontal plane marks the
level \(T=T_{\times}\) introduced in Subsec.~\ref{subsec_2.3}. This plane
bounds the upper branch of the red surface \(J=0\), separating the
region of degenerate loop configurations (\(T>T_{\times}\)) from the
regime in which the loops possess a well-defined base, as in
Fig.~\ref{fig:Fig_LOOP}. For \(h_2=1\) (panel (a)) and \(h_2=4\)
(panel (b)), the plane lies above most of the plotted region
(\(T_{\times}>0\)). In the symmetric case \(h_2=h_1=5\) (panel (c)) it
coincides with the coordinate plane \(T=0\), reflecting that
\(T_{\times}=0\). In panel (d), corresponding to \(h_2=6\), the level
\(T_{\times}\) lies entirely below the plotted domain
(\(T_{\times}<0\)), and the grey plane is therefore not visible.

A second red surface \(J=0\) is also visible in all panels of
Fig.~\ref{fig:3D} as the distant, background sheet forming the rear
boundary of the three-dimensional plots. This “back” branch of the
critical manifold undergoes its strongest visible deformation at small
upper-layer depths: from panel (a) to panel (c) it gradually straightens,
its curvature becomes weaker, and the surface approaches a more vertical
orientation. For larger depth ratios \(h_2\ge h_1\), the subsequent
changes of this background red surface become smooth and relatively
minor; although the surface continues to deform as \(h_2\) increases, the
scale and perspective of the 3D visualisation make these variations
barely noticeable in panels (d)–(i). As a result, at large asymmetry the
apparent geometry of the background red sheet remains nearly unchanged,
and the qualitative transformation of the modulational-stability
topology is governed almost entirely by the behaviour of the blue and
green surfaces.

As \(T\) approaches the critical value \(T^\ast\),
the singular surfaces \(J\to\infty\) (blue) and \(\omega''=0\) (green)
move closer together and form a line of intersection in the
three-dimensional \((\rho,k,T)\) space. The point \((\rho,k,T)=(0,0,T^\ast)\)
is the starting point of this line and corresponds to the coincidence of
the long-wave dispersive and nonlinear singularities.
The geometry of this intersection line and the shape of the upper parts
of the blue and green surfaces depend sensitively on the depth ratio
\(h_2/h_1\). For \(h_2<h_1\), the two surfaces bend downward near
\((0,0,T^\ast)\), producing a concave intersection line
(panels (a), (b)). In the symmetric case \(h_2=h_1=5\) (panel (c)), the
intersection line becomes straight, reflecting the even symmetry of the
dispersive characteristics with respect to \(k\). When \(h_2>h_1\), the
intersection line becomes convex upward, as seen starting from
panel (d).

As \(h_2\) increases further, this upward convexity becomes more
pronounced. At the golden–ratio depth ratio
\(h_2=h_1(1+\sqrt{5})/2\) (panel (e)), the upper edges of the blue and
green surfaces, as well as their intersection line, develop a horizontal
tangent at the point \((0,0,T^\ast)\). The sign of curvature does not
change: the surfaces remain convex upward both before and after this
value of \(h_2\). However, for \(h_2>h_1(1+\sqrt{5})/2\), a pronounced
local maximum appears on the upper parts of these surfaces and on their
intersection line, as clearly visible in panels (f)–(i).
The emergence of this maximum leads to a further upward bending of the
intersection curve and results in the formation of the cut–corridor
topology in the two-dimensional diagrams
(see Fig.~\ref{fig:Fig_CutKoridor}). The intersection points of the
cut–corridor boundaries correspond to the projection of the blue–green
intersection line onto the \((\rho,k)\) plane.

Overall, the evolution seen in panels (a)–(i) demonstrates that the
topological transitions in the \((\rho,k)\)-plane are governed by the
geometry of the intersection line between the resonant (\(J\to\infty\))
and dispersive (\(\omega''=0\)) surfaces. For small and moderate
\(h_2/h_1\), this line is either concave or nearly straight and produces
loop– and corridor–type structures. Once \(h_2\) exceeds the golden–ratio
value, the emergence of a local maximum on the upper parts of the
critical surfaces causes the intersection line to bend upward; its
projections on the planes \(T=\mathrm{const}\) then generate the
characteristic cut–corridor patterns described in
Subsec.~\ref{subsec_3.3}. Thus, the progressive deformation of this
intersection line provides a geometric mechanism linking the
gravity–capillary and the capillary regimes of modulational stability.

\subsection{Physical interpretation and correspondence with planar maps}
\label{subsec_4.3}

The red surface \(J = 0\) marks the boundary between the two signs of the nonlinear
NLS coefficient. Its curvature reflects how sensitively the nonlinear response of
the interface depends on the density ratio and the wavenumber. The convex portion
of this surface---clearly visible on the front side of the figure---corresponds to
the parameter range where \(J\) changes sign, and its intersection with a
constant-\(T\) plane forms a closed curve, the loop.
As the upper layer becomes thicker, the gravitational contribution to nonlinearity
is suppressed by the inertia of the upper fluid, and the capillary term in \(J\)
begins to dominate. In this regime, \(J\) loses its dependence on \(\rho\), the
\(J = 0\) surface flattens and shifts toward larger densities, and the loop
progressively shrinks and eventually disappears.

The blue surface, corresponding to the near-resonant nonlinear response
(\(J \to \infty\)), and the green surface, associated with the change of sign of
the group-velocity dispersion (\(\omega'' = 0\)), intersect for all parameter
values. However, the character of their mutual arrangement is strongly
influenced by the depth ratio of the two layers.
When the upper layer is thinner than the lower one, or only slightly thicker
so that their depths remain comparable, the intersection of the blue and
green surfaces occurs near \(\rho = 1\). Further away from this region the
surfaces diverge, leaving a finite gap between them. In planar sections
\((\rho, k)\), this gap manifests itself as a \emph{corridor}, representing a
domain of moderate modulational behaviour.
If the upper layer becomes significantly thicker than the lower one, the
situation changes. For sufficiently large surface-tension magnitudes---that is,
for values exceeding the long-wave critical threshold \(T^{\ast}\)---the blue and
green surfaces approach each other in such a way that the gap disappears.
In this regime, the corridor \emph{degenerates into a cut}, which appears as a
narrow band formed in the region of their \emph{immediate proximity}.

The curvature of the intersection line between the surfaces \(J \to \infty\) and
\(\omega'' = 0\) on the plane \(k = 0\) is determined by the distribution of
inertia between the layers in the long-wave regime and, consequently, by the
depth ratio. When the lower layer is thicker (\(h_2 < h_1\)), most of the
inertial mass is concentrated in the lower fluid, while the upper layer has only
a weak dynamical influence on the interface. As the density ratio varies, the
resonant nonlinear and dispersive mechanisms respond to this variation in
different ways, and the intersection curve becomes \emph{concave downward}
(\(\cap\)-shaped), as observed in panels~(a)–(b) of Fig.~4.
In the symmetric configuration \(h_2 = h_1\), the inertial contributions of the
two layers are exactly equivalent. In the long-wave limit the interface displaces
both layers as a single fluid column, and the dependence of the conditions
\(J \to \infty\) and \(\omega'' = 0\) on the density ratio becomes strictly
coordinated. Consequently, the intersection line reduces to a \emph{straight
line}, as shown in panel~(c).
When the upper layer becomes thicker (\(h_2 > h_1\)), the inertial dominance is
reversed, and the upper fluid governs the long-wave dynamics. The balance
between resonant nonlinearity and dispersion therefore shifts in the opposite
direction: as the density ratio varies, the two mechanisms respond
differentially, and the intersection curve becomes \emph{concave upward}
(\(\cup\)-shaped), as seen in panels~(d)–(i).
Thus, the change of convexity of the intersection line reflects the transition
from lower-layer dominance (concavity downward) through the symmetric regime
(straight line) to upper-layer dominance (concavity upward).

A second distinguished regime arises at the golden-depth ratio
\(h_2/h_1=\varphi\). In this configuration the long-wave contributions of the
two layers to the conditions \(J\to\infty\) and \(\omega''=0\) become identical,
so that the resonant and dispersive mechanisms are locally equivalent at
\((\rho,k)=(0,0)\). The corresponding critical surface satisfies
\(\partial_{\rho} T_{\mathrm{crit}}=0\), which appears as a horizontal tangent of
\(T(\rho,k)\) in the \(\rho\)-direction near the point \((0,0,T^{\ast})\). This
local degeneracy, occurring precisely when \(T=T^{\ast}\) and
\(h_2/h_1=\varphi\), enables the formation of a cut in the planar diagrams:
the critical surfaces approach each other so closely that the intermediate
region collapses, producing a cut whose lower endpoint lies on the axis
\((0,\rho)\) at \(k=0\).

Thus, the loop, corridor, and cut reflect three distinct physical–geometric
situations: local balance of nonlinearity and dispersion, gentle separation of
two dispersion mechanisms, and their near-simultaneous onset in the strongly
capillary regime.

\section{Conclusions}
\label{sec_Conclusions}

Together with Part~I, this study provides a unified geometric and asymptotic
description of modulational stability of interfacial gravity--capillary waves in
a two-layer fluid. Treating the surface-tension coefficient \(T\) as an
independent control parameter reveals how capillarity modifies both the
nonlinear coefficient \(J\) and the curvature of the dispersion relation
\(\omega''\), and how these changes reorganise the neutral boundaries in the
\((\rho,k)\)-plane.

In the long-wave limit, the base of the loop is fixed by the geometric
coordinate \(\rho_{\mathrm L}=h_{2}^{2}/h_{1}^{2}\), while the base of the
corridor is determined by \(\rho_{\mathrm C}(T)\), the density at which the
singular condition \(J\to\infty\) meets \(k=0\)
(Sec.~\ref{subsec_2.3}).
Their ordering is controlled by the threshold \(T_{\times}\), which exists only
for \(h_{1}>h_{2}\). Loops are possible while
\(\rho_{\mathrm L}<\rho_{\mathrm C}(T)\); when \(T>T_{\times}\), the ordering
reverses and the loop cannot persist. For \(h_{2}>h_{1}\), one always has
\(\rho_{\mathrm C}(T)<\rho_{\mathrm L}\) for all \(T>0\), so loop-type structures
are excluded and only corridor-type formations remain.

A second organising parameter is the long-wave Bond threshold
\(T^{\ast}=h_{1}^{2}/3\), at which the dispersive condition
\(\omega''=0\) and the nonlinear singularity \(J\to\infty\) coincide at
\((\rho,k)=(0,0)\). Below this value, gravity and surface tension act jointly,
producing loop-- and corridor--type structures with a finite region of overlap.
For \(T>T^{\ast}\), the resonant (\(J\to\infty\)) and dispersive
(\(\omega''=0\)) surfaces approach one another near the origin, and for
sufficiently large depth ratio \(h_{2}/h_{1}\) the intersection line develops a
local maximum. Its planar sections generate the capillary cut: a narrow corridor
detached from the density axis that characterises strongly capillary,
upper-layer-dominated configurations.

The three-dimensional critical surfaces \(J=0\), \(J\to\infty\), and
\(\omega''=0\) provide a single geometric framework for interpreting all
observed structures. Loops, corridors, and cuts arise as planar intersections of
the nonlinear, resonant, and dispersive surfaces with planes \(T=\mathrm{const}\),
and their evolution with varying \(T\) and depth ratio \(h_{2}/h_{1}\) follows
directly from the deformation of these surfaces. Two special geometric
configurations correspond to genuine degeneracies: equal layer depths
(\(h_{1}=h_{2}\)), where the resonant--dispersive intersection becomes a straight
line, and the golden-ratio depth ratio (\(h_{2}/h_{1}=\varphi\)), where the
critical surface is horizontally tangent at \((0,0,T^{\ast})\), marking the
transition in the curvature of the intersection line and enabling the onset of
cut-type behaviour.

Overall, the results establish a complete geometric--capillary classification of
modulational stability in two-layer fluids with variable surface tension and
extend the framework developed in Part~I. They provide a foundation for future
extensions involving shear, external forcing, flexible boundaries, or variable
bathymetry.

\subsection*{Acknowledgments}
Olga Avramenko thanks the Research Council of Lithuania for supporting this work.

\selectlanguage{ukrainian}

\authoru{Ольга Авраменко}{a,b}
\authoru{Володимир Нарадовий}{c}

\begin{center}	
\textbf{НЕСТІЙКІСТЬ БЕНДЖАМІНА–ФЕЙРА МІЖФАЗНИХ ГРАВІТАЦІЙНО-КАПІЛЯРНИХ ХВИЛЬ У ДВОШАРОВІЙ РІДИНІ. ЧАСТИНА ІІ. ВПЛИВ ПОВЕРХНЕВОГО НАТЯГУ}
\AuthorPrintu
\affiliation{a}{Національний університет ``Києво-Могилянська академія'', вул. Сковороди, 2, Київ, 04070, Україна}
\affiliation{b}{Університет Вітовта Великого, вул. К.~Донелайчо, 58, Каунас, 44248, Литва}
\affiliation{c}{Центральноукраїнський державний університет імені Володимира Винниченка, вул. Шевченка, 1, Кропивницький, 25006, Україна}
\end{center}
\vskip -0.9em
\begin{abstract}
У другій частині дослідження розроблено повний геометричний та асимптотичний
опис того, як поверхневий натяг визначає модуляційну стійкість інтерфейсних
хвиль у двошаровій рідині. Розвиваючи аналітичну схему Частини~I,
поверхневий натяг розглядається як вільний керівний параметр, що дає змогу
відстежувати нелінійні та дисперсійні властивості системи для широкого діапазону
співвідношень глибин та контрастів густин. Використовуючи зведення до
нелінійного рівняння Шредінґера разом із довгохвильовими асимптотиками,
визначено механізми, що формують межі між стабільними та нестабільними
режимами, та встановлено їхню залежність від величини поверхневого натягу.

Довгохвильова структура контролюється двома спеціальними значеннями густини,
які задають точки зародження петлі та коридору на діаграмах стійкості.
Взаємне розташування цих точок змінюється за певного порогу, що існує лише тоді,
коли нижній шар є глибшим, і саме в цьому випадку можливе існування петлі.
Другим організувальним параметром є класичний поріг Бонда, за якого
дисперсійна і нелінійна синґулярності збігаються. Коли поверхневий натяг
перевищує це значення і верхній шар є достатньо глибоким, взаємодія резонансних
та дисперсійних ефектів утворює капілярний розріз, який заміщує коридор і
характеризує режими з домінуванням капілярності.

Для об’єднання цих спостережень побудовано повні тривимірні критичні
поверхні, що розмежовують різні типи нелінійної та дисперсійної поведінки.
Петля, коридор і розріз постають як площинні перерізи цих поверхонь, а їхні
топологічні переходи безпосередньо зумовлені деформацією лінії перетину між
резонансною та дисперсійною поверхнями. Два співвідношення глибин відповідають
справжнім геометричним виродженням: рівні товщини шарів, коли лінія перетину
стає прямою, та конфігурація золотого перетину, коли критична поверхня набуває
горизонтальної дотичної при порозі Бонда.

У цілому, Частина~II завершує геометричну та фізичну класифікацію
модуляційної стійкості інтерфейсних хвиль у двошарових рідинах і формує основу
для подальших узагальнень, що враховують зсувні течії, зовнішні збурення,
гнучкі межі або змінну батиметрію.

\key{модуляційна нестійкість; міжфазні гравітаційно–капілярні хвилі;
двошарова рідина; поверхневий натяг; нестійкість Бенджаміна–Фейра}

\end{abstract}


\EndPaper


\end{document}